\documentclass[fleqn,usenatbib]{mnras}

\usepackage{newtxtext,newtxmath}

\usepackage[T1]{fontenc}

\DeclareRobustCommand{\VAN}[3]{#2}
\let\VANthebibliography\thebibliography
\def\thebibliography{\DeclareRobustCommand{\VAN}[3]{##3}\VANthebibliography}

\usepackage{graphicx}	
\usepackage{amsmath}	
\usepackage{caption} 
\usepackage{float}
\usepackage{newtxtext,newtxmath}
\usepackage{times}
\usepackage[normalem]{ulem} 
\usepackage{rotating}
\usepackage{epsfig}
\usepackage{fancyhdr}
\usepackage{epsfig}
\usepackage{graphicx}
\usepackage{amsmath}
\usepackage{theorem}
\usepackage{latexsym}
\usepackage{epic}
\usepackage{natbib} 
\usepackage[flushleft]{threeparttable}
\usepackage{tabulary}
\usepackage{tabularx}
\usepackage{pdflscape}
\usepackage{afterpage}
\usepackage{afterpage}
\usepackage{capt-of}
\usepackage{adjustbox,lipsum}
\usepackage{xcolor}
\usepackage{newtxmath}
\usepackage[normalem]{ulem} 
\usepackage{soul} 

\usepackage[utf8]{inputenc}
\DeclareUnicodeCharacter{02BC}{'}
\usepackage{textcomp}
\let\la=\lesssim
\input{epsf}
\newcommand{\emailstar}{\textcolor{blue}{\hyperlink{emailnote}{$^\star$}}}
\newcommand{\emailfootnote}{\hypertarget{emailnote}{\noindent\textsuperscript{$\star$}E-mail: \href{mailto:adlyka@ukm.edu.my}{adlyka@ukm.edu.my}}}

\title[Compton-Thick AGN in our Cosmic Backyard]{The Compton-thick AGN Population and the $N_{\rm H}$ Distribution of Low-mass AGN in our Cosmic Backyard}

\author[A. Annuar et al.]{{
A. Annuar,$^1$\emailstar}
D. M. Alexander,$^{2}$
P. Gandhi$,^{3}$
G. B. Lansbury,$^{4}$
M. N. Rosli,$^{1}$
D. Stern,$^{5}$
D. Asmus,$^{6}$
\newauthor D. R. Ballantyne,$^{7}$
M. Baloković,$^{8,9}$
F. E. Bauer,$^{10,11,12}$ 
P. G. Boorman,$^{13}$
W.N. Brandt,$^{14,15,16}$
\newauthor M. Brightman,$^{13}$
C.T.J. Chen,$^{17,18}$ 
A. Del Moro,$^{19}$
D. Farrah,$^{20,21}$
F. A. Harrison,$^{13}$
M. J. Koss,$^{22}$
\newauthor  L. Lanz,$^{23}$
S. Marchesi,$^{24,25,26}$
P. Mohanadas,$^{1}$
E. Nardini,$^{27}$
C. Ricci,$^{28,29}$ 
and L. Zappacosta$^{30}$
\\
$^{1}$Department of Applied Physics, Faculty of Science and Technology, Universiti Kebangsaan Malaysia, 43600, UKM Bangi, Selangor, Malaysia\\
$^{2}$Centre for Extragalactic Astronomy, Department of Physics, Durham University, South Road, Durham DH1 3LE, UK\\
$^{3}$School of Physics and Astronomy, University of Southampton, Southampton SO17 1BJ, UK\\
$^{4}$European Southern Observatory, Karl-Schwarzschild str. 2, D-85748 Garching bei Munchen, Germany\\
$^{5}$Jet Propulsion Laboratory, California Institute of Technology, Pasadena, CA 91109, USA\\
$^{6}$Gymnasium Schwarzenbek, 21493 Schwarzenbek, Germany\\
$^{7}$Center for Relativistic Astrophysics, School of Physics, Georgia Institute of Technology, 837 State Street, Atlanta, GA 30332-0430, USA\\
$^{8}$Yale Center for Astronomy \& Astrophysics, 52 Hillhouse Avenue, New Haven, CT 06511, USA\\
$^{9}$Department of Physics, Yale University, P.O. Box 2018120, New Haven, CT 06520, USA\\
$^{10}$Instituto de Astrofísica and Centro de Astroingeniería, Facultad de Física, Pontificia Universidad Católica de Chile, Casilla 306, Santiago 22, Chile\\
$^{11}$Millennium Institute of Astrophysics (MAS), Nuncio Monseñor Sótero Sanz 100, Providencia, Santiago, Chile \\
$^{12}$Space Science Institute, 4750 Walnut Street, Suite 205, Boulder, CO 80301, USA\\
$^{13}$Cahill Center for Astrophysics, California Institute of Technology, 1216 East California Boulevard, Pasadena, CA 91125, USA\\
$^{14}$Department of Astronomy and Astrophysics, 525 Davey Lab, The Pennsylvania State University, University Park, PA 16802, USA\\
$^{15}$Institute for Gravitation and the Cosmos, The Pennsylvania State University, University Park, PA 16802, USA\\
$^{16}$Department of Physics, 104 Davey Laboratory, The Pennsylvania State University, University Park, PA 16802, USA\\
$^{17}$Science and Technology Institute, Universities Space Research Association, Huntsville, AL 35805, USA\\
$^{18}$Astrophysics Office, NASA Marshall Space Flight Center, ST12, Huntsville, AL 35812, USA\\
$^{19}$German Aerospace Center (DLR), Space Operation and Astronaut Training, Oberpfaffenhofen, D-82234 Weßling, Germany\\
$^{20}$Institute for Astronomy, University of Hawai‘i, 2680 Woodlawn Dr., Honolulu, HI, 96822, USA\\
$^{21}$Department of Physics and Astronomy, University of Hawai‘i at Mānoa, 2505 Correa Rd., Honolulu, HI, 96822, USA\\
$^{22}$Eureka Scientific, 2452 Delmer Street Suite 100, Oakland, CA 94602-3017, USA\\
$^{23}$Department of Physics, The College of New Jersey, 2000 Pennington Road, Ewing, NJ 08628, USA\\
$^{24}$Dipartimento di Fisica e Astronomia (DIFA), Università di Bologna, via Gobetti 93/2, I-40129 Bologna, Italy \\
$^{25}$Department of Physics and Astronomy, Clemson University, Kinard Lab of Physics, Clemson, SC 29634-0978, USA \\
$^{26}$INAF - Osservatorio di Astrofisica e Scienza dello Spazio di Bologna, Via Piero Gobetti, 93/3, 40129, Bologna, Italy \\
$^{27}$INAF — Osservatorio di Astrofisica e Scienza dello Spazio di Bologna, via Gobetti 93/3, I-40129 Bologna, Italy\\
$^{28}$Instituto de Estudios Astrofísicos, Facultad de Ingeniería y Ciencias, Universidad Diego Portales, Av. Ejército Libertador 441, Santiago, Chile\\
$^{29}$Kavli Institute for Astronomy and Astrophysics, Peking University, Beijing 100871, Peopleʼs Republic of China\\
$^{30}$INAF – Osservatorio Astronomico di Roma, Via di Frascati 33, 00078 Monte Porzio Catone, Italy\\
}

\date{Accepted 2025 June 5. Received 2025 June 3; in original form 2024 December 20}

\pubyear{\the\year{}}

\begin{document}
\label{firstpage}
\pagerange{\pageref{firstpage}--\pageref{lastpage}}

\maketitle

\def\lsun{$L_\odot$}
\def\msun{$M_\odot$}
\def\micron{\mu m}

\begin{abstract}
We present a census of the Compton-thick (CT) active galactic nucleus (AGN) population and the column density ($N_{\rm{H}}$) distribution of AGN in our cosmic backyard using a mid-infrared selected AGN sample within 15 Mpc. The column densities are measured from broadband X-ray spectral analysis, mainly using data from {\it{Chandra}} and {\it{NuSTAR}}. Our sample probes AGN with intrinsic 2–10 keV luminosities of $L_{\rm 2-10, int} = 10^{37}$-$10^{43}$ erg s$^{-1}$, reaching a parameter space inaccessible to more distant samples. We directly measure a 32$^{+30}_{-18}\%$ CT AGN fraction and obtain an $N_{\rm{H}}$ distribution that agrees with that inferred by the \textsl{Swift}-BAT survey. Restricting the sample to the largely unexplored domain of low-luminosity AGN with \( L_{2-10,\mathrm{int}} \leq 10^{42} \, \mathrm{erg} \, \mathrm{s}^{-1} \), we found a CT fraction of \( 19^{+30}_{-14}\% \), consistent with those observed at higher luminosities. Comparing the host-galaxy properties between the two samples, we find consistent star formation rates, though the majority of our galaxy have lower stellar masses (by $\approx 0.3$ dex). In contrast, the two samples have very different black hole mass ($M_{\rm BH}$) distributions, with our sample having $\approx$1.5 dex lower mean mass ($M_{\rm BH}$ $\sim$ 10$^{6}$ \msun). Additionally, our sample contains a significantly higher number of LINERs and H{\sc{ii}}-type nuclei. The Eddington ratio range probed by our sample, however, is the same as \textsl{Swift}-BAT, although the latter dominates at higher accretion rates, and our sample is more evenly distributed. The majority of our sample with $\lambda_{\rm Edd} \geq$ 10$^{-3}$ tend to be CT, while those with $\lambda_{\rm Edd} <$ 10$^{-3}$ are mostly unobscured or mildly obscured.

\end{abstract}

\begin{keywords}
galaxies: active --- galaxies: nuclei --- techniques: spectroscopic --- X-rays: galaxies
\end{keywords}

\section{Introduction}

Many studies have shown that obscured active galactic nuclei (AGN) dominate the accretion energy budget of the cosmos. This has been evident from the spectral shape of the cosmic X-ray background (CXB) radiation for over three decades (e.g., \citealp{Setti89}; \citealp{Gilli07}; \citealp{Ueda14}; \citealp{Comastri15}). The obscured phase of AGN accretion is often considered to be a part of an AGN evolutionary scenario in which the central supermassive black hole (SMBH) grows rapidly due to the large amount of gas being driven to the center of the galaxy as a result of major galaxy mergers (e.g., \citealp{Martinez-Sansigre2005}; \citealp{Hopkins08}; \citealp{DraperBallantyne10}; \citealp{Treister10}; \citealp{Treister12}; \citealp{Kocevski15}; \citet{Ricci2017}). This obscured phase is mostly hidden due to enshrouding gas and dust, and is primarily characterized by significant X-ray obscuration. Once the radiation pressure (or winds) from the AGN expels this material, the central source is revealed (\citealp{Feruglio2010}; \citealp{Tombesi2015}; \citealp{Ishibashi2016}). However, this scenario might be more relevant at high redshifts.

At lower redshifts such as in the local universe, the classical AGN unification model (\citealp{Antonucci93}; \citealp{Urry95}) is more likely to be accurate. Based on this model, obscuration is attributed to a dusty, geometrically thick structure, commonly referred to as the "torus", which surrounds the central SMBH and its accretion disk, obscuring our line-of-sight toward the AGN central region depending on our the viewing angle. Having a complete census of the AGN population over a broad range of obscuration, luminosities and redshifts is therefore important to help us understand the growth of SMBHs.

To date, however, our understanding of the AGN distribution as a function of the obscuring column density ($N_{\rm H}$) remains highly uncertain, even in the nearby universe, particularly at the higher end of the distribution; i.e., the Compton-thick (CT) regime ($N_{\rm H}$ $\gtrsim$ 1.50 $\times$ 10$^{24}$ cm$^{-2}$). As a result of extreme absorption suffered by the nuclear source, direct X-ray emission from CT AGN is significantly suppressed, and the emission that we observe at $\lesssim$ 10 keV is often dominated by X-ray photons being scattered or reflected from the back-side of the torus or other circumnuclear material. This observed emission from CT AGN is typically about two orders of magnitude lower than the intrinsic AGN photons emitted in the 2--10 keV band (e.g., \citealp{Matt97}; \citealp{Balokovic14}; \citealp{Annuar17}). In extreme cases where the column density exceeds 10$^{25}$ cm$^{-2}$, the direct emission from the AGN is severely absorbed over the entire range of X-ray energy, even at the hard X-ray regime ($E >$ 10 keV; \citealp{Gilli07}). These effects make CT AGN very challenging to identify. 

Nevertheless, the CT AGN population is believed to constitute a significant fraction of the entire AGN population. For example, synthesis models of the CXB spectrum suggest that CT AGN are required to produce the CXB radiation and contribute up to 50$\%$ of the flux at the peak energy, $E \sim$ 30 keV (e.g., \citealp{Ananna2019} \citealp{Gilli07}; \citealp{Treister09} \citealp{DraperBallantyne10}; \citealp{Akylas12}; \citealp{Ueda14}; \citealp{Comastri15}). Multiwavelength studies of nearby AGN also predict that CT AGN should be numerous, accounting for $\sim$30$\%$ of the AGN population (e.g., \citealp{Risaliti99}; \citealp{Goulding11}), in agreement with predictions from CXB modelling. Interestingly, of the three AGN identified within $D =$ 4 Mpc (Circinus, NGC 4945 and NGC 5128), two are found to be CT (Circinus and NGC 4945), corresponding to a CT AGN fraction of $\sim$67$\%$ \citep{Matt00}. Yet, at larger volumes, their census seems to be far from complete. To date, only $\sim$8$\%$ of AGN  out to $z  \lesssim$ 0.055 have been \emph{directly} identified as CT on the basis of hard X-ray studies by the \textsl{Neil Gehrels Swift} Burst Alert Telescope (\textsl{Swift}-BAT) survey (\citealp{Ricci15}). This is confirmed by the \textsl{Nuclear Spectroscopic Telescope Array} (\textsl{NuSTAR}; \citealt{Harrison13}) study of \textsl{Swift}-BAT AGN that measured the same fraction within similar volume (\citealp{Torres-Alba2021}). These studies suggest that we could be missing a significant number of CT AGN, even in the local universe. A complete understanding of their population is important to help us accurately characterise the CXB radiation.

In order to form a complete census of the CT AGN population, we first need a complete AGN sample that is least limited by flux and unbiased against obscuration. A volume-limited selection within a relatively small volume is the best approach to construct an AGN sample that is least affected by flux limitations. This means that the sample will include more low-luminosity sources as compared to a purely flux-limited sample. In addition, a volume-limited sample can also be used to form representative volume-averaged statistics.

AGN identification methods at different wavelengths each have their own advantages and disadvantages. For example, AGN selected on the basis of X-ray emission (e.g., an X-ray luminosity threshold) produces a cleaner sample as it suffers relatively lower contamination by the host-galaxy (see \citealp{Brandt15}; \citealp{HickoxAlexander18} for reviews). However, such a sample is prone to be biased against finding AGN that are CT, due to extreme X-ray attenuation caused by the high column density of gas and dust. 
Although optical selection on the basis of emission-line diagnostics (e.g., \citealp{Baldwin81}; \citealp{Ho97}; \citealp{Kewley01}) is not affected by obscuration by the AGN torus, it can miss AGN that are significantly obscured by the host-galaxy (e.g., \citealp{Goulding09}). Conversely, an infrared (IR) AGN selection will be relatively unbiased against obscuration due to the lower extinction suffered at this waveband (e.g., the extinction at 12$\mu$m is $\sim$27$\times$ lower than that in the optical $V$-band for a standard dust extinction law; \citealp{Li01}). Therefore, it should be the best approach for constructing an AGN sample that is least affected by both host-galaxy and torus obscuration. However, this technique can miss AGN that are severely contaminated by host-galaxy processes such as star formation activity (e.g., \citealp{Assef13}; \citealp{Kirkpatrick13}). 

A multiwavelength selection approach would of course be the best technique to yield the most complete AGN sample that is independent of AGN diagnostics at any one wavelength. However, the advantage of forming a sample using a single waveband selection is that the selection effect is simpler and better understood. Therefore in this paper, we use a sample of local, mid-IR selected AGN within $D \leq$ 15 Mpc to form a census of the CT AGN population and the $N_{\rm H}$ distribution of AGN in our cosmic backyard. The $N_{\rm{H}}$ values and intrinsic luminosities for each AGN were directly measured via broadband X-ray spectroscopy. In  most cases, we used data from multiple focusing X-ray observatories, primarily \textsl{Chandra} in combination with the \textit{Nuclear Spectroscopic Telescope Array}  (\textsl{NuSTAR}; \citealp{Harrison13}). We describe our AGN sample in Section 2. In Section 3, we detail the X-ray observations and analysis, including data that were specifically obtained for this program. The multiwavelength properties of the AGN are discussed in Section 4 to complement our X-ray results. The CT AGN fraction and $N_{\rm H}$ distribution of our sample are presented in Section 5. This is followed by a discussion of the AGN Eddington ratio and host-galaxy properties in comparison with the Swift-BAT AGN sample, in Section 6. Finally, we summarize our results in Section 7. 

\vspace{2em}
\emailfootnote

\begin{table*}
\caption{Complete list of AGN at $D$ $\leq$ 15 Mpc and their basic properties.}
\centering
\begin{adjustbox}{width=\textwidth,totalheight=\textheight,keepaspectratio}
\begin{tabular}{cccccccccccc}
  \hline 
   Name & D & Hubble & Spectral & \textsl{Swift-BAT} & log $M_{\rm BH}$ & $\log{L_{[\rm OIV]}}$ &  $\log{L_{[\rm NeV]}}$ &  $\log{L_{\rm 12\mu m}}$ & $\log{L_{[\rm O {\sc III}],corr}}$ & $\log{L_{\rm IR}}$ & \textsl{NuSTAR} \\
             &  [Mpc] & Type   & Class & AGN? & [\msun] & [erg s$^{-1}$] & [erg s$^{-1}$] & [erg s$^{-1}$] & [erg s$^{-1}$] & [\lsun] & obs.? \\
   (1) & (2)  & (3) & (4) & (5) & (6) & (7) & (8) & (9) & (10) & (11) & (12) \\
   \hline
   Circinus & 4.0 & Sb & S2 & Yes &6.23$^{[2]}$ & 40.11$^{[4]}$& 39.62$^{[5]}$& 42.60 & 40.52$^{[8]}$ & 10.15$^{[10,11]}$ & Yes \\
   ESO121-G6 & 14.5  & Sc & H {\sc{ii}} & No & 6.10 & 39.04& 38.21 & 40.28$^{[6]}$ &- & 9.70 & Yes$^{\ast}$  \\  
   NGC 0613 & 15.0  & Sbc & H {\sc{ii}} & No & 7.34 & 39.38 & 38.25& 41.27& 39.67 & 10.37 & No \\  
   NGC 0660 & 12.3  & Sa & L & No & 7.35 & 39.71& 38.85& 41.18$^{[6]}$ & 40.03 & 10.49 & Yes$^{\ast}$  \\
   NGC 1068 & 13.7  & Sb & S2 & Yes & 7.20 & 41.66& 41.30 & 43.76& 42.61 & 11.27 & Yes \\
   NGC 1448 & 11.5  & Scd & S2$^{[1]}$ & No & 5.99 & 39.40& 38.76 & 40.60$^{[1]}$ & 38.84$^{[1]}$ & 9.78 & Yes$^{\ast}$  \\ 
   NGC 1792 & 12.5  & Sbc & H {\sc{ii}} & No & 6.83 & 38.26& 37.88 &$<$42.03$^{[7]}$ & -  & 10.33 & Yes$^{\ast}$  \\ 
   NGC 3621 & 6.6 & Sd & S2 & No & 6.50 & 38.18& 37.18&$<$40.75$^{[7]}$ & 37.19 & 9.74 & Yes$^{\ast}$  \\ 
   NGC 3627 & 10.0 & Sb & L & No & 7.30 & 38.38& 37.55 & 40.59&39.01 & 10.38 & Yes$^{\ast}$  \\ 
   NGC 3628 & 10.0  & Sb & L & No & 6.53 & 38.81& 38.05& $<$ 40.47& 36.97 & 10.25 & Yes$^{\ast}$  \\ 
   NGC 4051 & 13.1 & Sbc & S1 & Yes & 6.15 & 39.88&39.36& 42.38 & 40.25 & 9.90 & Yes \\ 
   NGC 4565 & 10.0 & Sb  & S2 & No & 7.56$^{[3]}$ & 38.40 & 37.58$^{[5]}$  &$<$41.35$^{[7]}$ & 38.84$^{[9]}$ & 9.66 & No \\ 
   NGC 4945 & 3.9 & Scd & S2 & Yes & 6.04 & 38.72& 38.11 & 40.00 &35.04 & 10.48 & Yes \\ 
   NGC 5033 & 13.8  & Sc & S1 & Yes & 7.62 & 39.08&37.98 & 40.96 &39.78 & 10.13 & Yes \\ 
   NGC 5128 & 4.0 & S0 & H {\sc{ii}} & Yes & 8.38 & 39.38& 38.62& 41.86 & 38.09 & 10.11 & Yes \\ 
   NGC 5194 & 8.6 & Sbc & S2 & Yes & 6.88 & 38.85& 37.81& 40.80 & 40.18 & 10.42 & Yes \\  
   NGC 5195 & 8.3 & Irr & L & No & 7.31 & 37.89& 37.22& $<$41.90$^{[7]}$ & 37.12 & 9.50 & Yes \\ 
   NGC 5643 & 13.9 & Sc & S2 & Yes & 6.44 & 40.43& 39.75& 42.17 & 41.32 & 10.24 & Yes$^{\ast}$   \\ 
   NGC 6300 & 13.1 & Sb & S2 & Yes & 6.80 & 39.78 & 39.41& 42.45 & 40.02 & 10.09 & Yes \\ 
   \hline
\end{tabular}
\end{adjustbox}
  \begin{tablenotes}
   \footnotesize
   \item \emph{Notes.} Column (1) Galaxy name; (2) Distance in Mpc; (3) Hubble classification of the host-galaxy; (4) Optical spectral class on the basis of emission-line diagnostic diagrams (\citealp{Kewley01}; \citealp{Kauffmann03}); (5) Whether the AGN is detected in the 105-month \textsl{Swift}-BAT survey \citep{Oh18}; (6) Logarithm of the black hole mass  relative to the mass of the Sun, \msun measured from various techniques (e.g., maser mapping, velocity dispersion and bulge luminosity; see GA09 for more details); (7)-(10) [O {\sc{iv}}], [Ne {\sc{v}}], 12 $\mu$m continuum, and [O {\sc{iii}}] (corrected for the Balmer decrement) luminosities in logarithmic scale and expressed in erg s$^{-1}$, respectively; (11) Logarithm of the total IR luminosity in solar luminosities, \lsun; (12) Whether the AGN has been observed by \textsl{NuSTAR}. Most data are taken from GA09 or \cite{Goulding10}, unless indicated otherwise. The spectral classes for the \textsl{Swift}-BAT detected AGN were obtained from \cite{Oh22}. 
    \item $^{\ast}$ observed as part of our program.
   \item \emph{References.} [1] \cite{Annuar17}; [2] \cite{Greenhill03}; [3] \cite{McKernan10}; [4] \cite{DiamondStanic09}; [5] \cite{Pereira-Santaella10}; [6] \cite{annuar20}; [7] {\textsl{WISE}} 12{$\mu$m} luminosity used as upper limit [8] \cite{Bianchi02}; [9] \cite{Ho97}; [10] \cite{Helou88}; [11] \cite{Sanders96}.
   \end{tablenotes}
\end{table*} 

\section{The $D \leq$ 15 Mpc AGN Sample}

The parent AGN sample that we use for this work was constructed by \cite{Goulding09} (hereafter GA09; see also \citealp{Goulding10}). Here, we briefly describe their AGN selection, and refer the reader to GA09 and its subsequent paper, \cite{Goulding10}, for further details on the sample. Their sample consists of 17 IR-selected AGN within $D =$ 15 Mpc. It was derived using the \textsl{Infrared Astronomical Satellite} (\textsl{IRAS}) Revised Bright Galaxy Sample (RBGS; \citealp{Sanders03}) which provides the most complete census of IR-bright galaxies ($f$$_{60\,\mathrm{\mu m}}$ $>$ 5.24 Jy) in the local universe above a Galactic latitude of $|b|$ $=$ 5$^{\circ}$. The constraint of 15 Mpc on the distance was placed to avoid the Virgo cluster at $\approx$16 Mpc so that the AGN sample is representative of the field-galaxy population, as demonstrated in \cite{Goulding10}. The distances were calculated using the \cite{Mould00} cosmic attractor model which adjusts heliocentric redshifts to the centroid of the Local Group, taking into account the gravitational attraction toward the Virgo Cluster, the Great Attractor and the Shapley supercluster. In addition, GA09 apply a total IR (8--1000$\mu$m) luminosity cut-off of $L_{\rm IR} >$ 3 $\times$ 10$^{9}$ \lsun\ to their sample to match the flux-sensitivity limit of RBGS, and exclude low-luminosity dwarf galaxies as well as relatively inactive galaxies with low star formation rate. In total, GA09 found that there are 67 galaxies detected by IRAS with $L_{\rm IR} >$ 3 $\times$ 10$^{9}$ \lsun\ within $D \leq$ 15 Mpc.\footnote{GA09 presented 68 galaxies in their sample. However, we find that one of the galaxies; NGC 3486, has an IR luminosity below the threshold value ($L_{\rm IR} =$ 2.04 $\times$ 10$^{9}$ \lsun). Therefore, we exclude NGC 3486 from the sample in this paper.} 

GA09 then used the detection of the high-ionization [Ne {\sc v}]$\uplambda$14.3$\mu$m emission-line by the \emph{Spitzer Space Telescope}  \citep{Werner04} high resolution IR Spectrographs (IRS) Short-High (SH) module ($\lambda =$ 9.9--19.6$\mu$m; spectral resolution $R$ $\sim$ 600; aperture size $=$ 4.7 $\times$ 11.3 arcsec$^{2}$; \citealp{Houck04}) to identify the presence of AGN in these galaxies. Out of the 67 galaxies, 64 have \emph{Spitzer}-IRS SH data ($\sim$94$\%$ complete). The [Ne {\sc v}] line is primarily produced in the NLR of AGN through ionization by the primary emission of the nuclear source. Because the line is produced in the NLR, it is not strongly affected by obscuration by the AGN torus, unlike the primary emission. In addition, given that it is emitted at mid-IR wavelength (i.e., $\lambda =$ 14.32 $\mu$m), it does not suffer from significant absorption by the host-galaxy, as opposed the NLR emission-lines that are produced at optical wavelengths (the extinction at 14.32 $\mu$m is $\sim$50$\times$ lower than that in the optical $V$-band; \citealp{Li01}). The energy required to ionise this line is also relatively large; i.e. 97.1 eV, meaning that it can only be produced by extremely energetic phenomena such as AGN activity. The detection of this line therefore provides an almost unambiguous identifier of AGN (\citealp{Weedman2005}; \citealp{Iwasawa2011}; \citealp{Negus2023}). Although the line is predicted to be produced by a dense population of Wolf-Rayet stars \citep{Schaerer99} and extremely high velocity shocks caused by a starburst, these were not observed by \emph{Spitzer}, and do not appear to be the case for the AGN in the GA09 sample (see GA09 for further details).

Based on this technique, GA09 found 17/64 galaxies (27$^{+13}_{-10}\%$) have significant [Ne {\sc v}] line detections, and thus are identified as AGN. \footnote{The uncertainties were calculated using the approximate algebraic expression for small number Poisson statistics, based on the 90\% confidence double-sided interval (95\% single-sided) given in \cite{Gehrels86}, with the upper limit capped at 100\%.} To further extend the GA09 sample, we included two other known AGN within 15 Mpc that match the GA09 selection criteria, but were not originally selected. These are Circinus and NGC 4565. Circinus is identified as an AGN in optical and X-ray data (e.g., \citealp{Moorwood84}; \citealp{Baumgartner13}), but was not selected in the original RBGS sample (therefore GA09) due to its low Galactic latitude; i.e., $\sim4^{\rm o}$ below the Galactic plane (see also Section 2.1). NGC 4565 was one of the galaxies in GA09 that lacked high-resolution \emph{Spitzer}-IRS spectroscopic data at the time of that study but has since been observed with the high-resolution spectrograph. \cite{Pereira-Santaella10} present the high-resolution \emph{Spitzer}-IRS data for both Circinus and NGC 4565, and detected [Ne {\sc v}] emission-lines in both galaxies. These galaxies are also identified as Seyferts on the basis of their optical emission-line ratios \citep{Ho97} and have [O {\sc iv}]$\uplambda$25.89$\mu$m line detections \citep{DiamondStanic09}.\footnote{The [O {\sc iv}]$\uplambda$25.89$\mu$m emission-line is also often used for AGN identification due to its high ionization energy (59.4 eV). However, it is a more ambiguous AGN indicator than the [Ne {\sc v}] line since energetic starbursts can also produce this line.} Our final sample thus consists of 19 AGN within $D =$ 15 Mpc. The complete list of the AGN and their basic properties are presented in Table 1. 

Out of these 19 AGN, 8 (42$^{+34}_{-21}\%$) are not identified as AGN using the optical emission-line diagnostic diagram (see Table 1) due to significant dilution by the host galaxies (e.g., highly inclined, presence of dust lanes, strong star formation activity; GA09). Furthermore, 10/19 (53$^{+37}_{-23}\%$) are found to be unidentified as AGN in X-rays on the basis of high X-ray energy non detection by the \textsl{Swift}-BAT 105-month all sky survey \citep{Oh18}.\footnote{The \textsl{Swift}-BAT telescope provides a hard X-ray all sky survey in the 14--195 keV band. The 105-month survey is sensitive down to an X-ray flux of $f_{\rm 14-195, obs} \sim$ 8 $\times$ 10$^{-12}$ erg s$^{-1}$ cm$^{-2}$ \citep{Oh18}. This means that at $D =$ 15 Mpc, it is able to detect an AGN down to an observed X-ray luminosity of $L_{\rm 14-195, obs} \sim$ 2 $\times$ 10$^{41}$ erg s$^{-1}$, or $L_{\rm 2-10, obs} \sim$ 8 $\times$ 10$^{40}$ erg s$^{-1}$ on the basis of the scaling relation derived by \cite{Rigby09}.} This demonstrates the relative inefficiency of identifying AGN using optical and X-ray wavelengths as compared to IR spectroscopy (i.e., $\lesssim$50$\%$ in this case). 

In addition, only 1/19 (5.0$^{+21}_{-4.8}\%$) and 7/19 (37$^{+33}_{-20}\%$) of our sources overlapped with the \textsl{NuSTAR} Local AGN $N_{\rm H}$ Distribution Survey (NuLANDS; \citealp{Boorman2024-arXiv}) and the Local AGN Survey (LASr; \citealp{Asmus2020}), respectively, which used infrared colours AGN selection method. This corresponds to inefficiencies of 95$^{+5.0}_{-34}\%$ and 63$^{+37}_{-26}\%$, respectively, when using this technique in AGN identification, particularly for those with low luminosity and when the AGN contribute only a small fraction of the total galaxy emission.

\begin{figure}
\begin{center}
\includegraphics[scale=0.55]{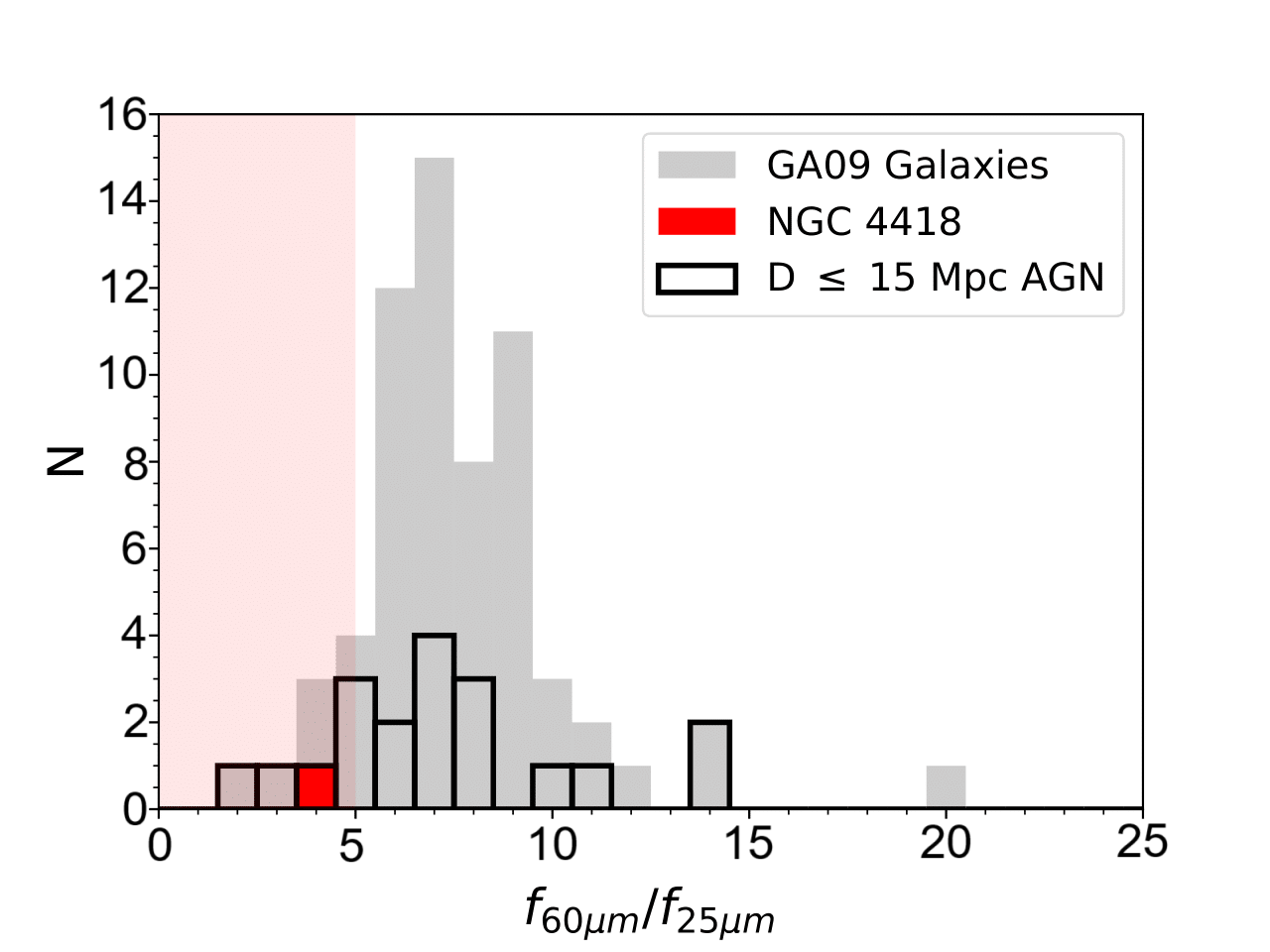}
\includegraphics[scale=0.24]{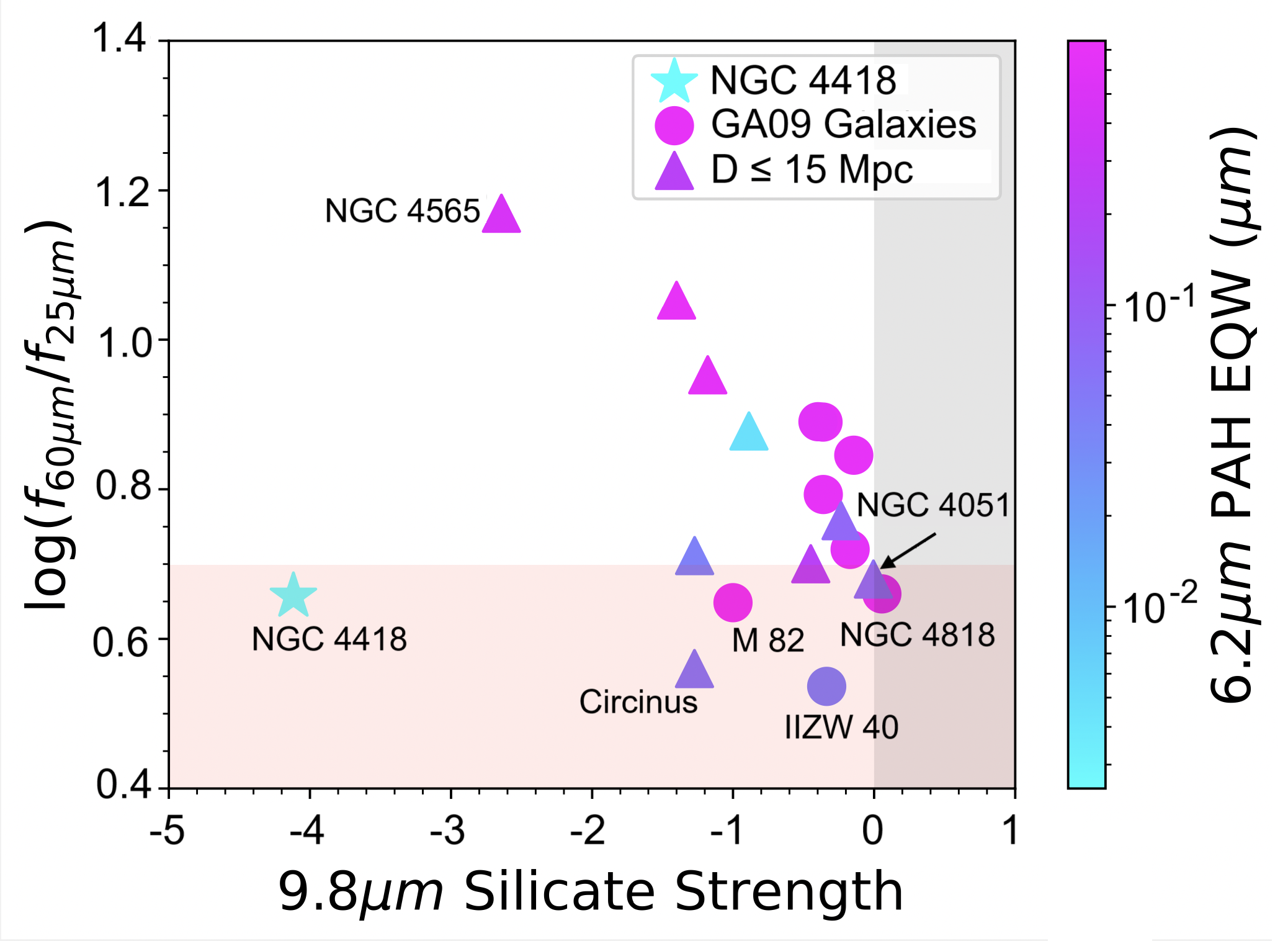} 
\caption[
  IRAS flux-ratio distribution and silicate strength
]{
  \emph{Top}: Distribution of IRAS $f_{\rm 60\micron}/f_{\rm 25\micron}$ ratio
  for the galaxies in GA09 (grey), our AGN sample (black solid line), and
  NGC 4418 (red). The pink shaded region marks a flux ratio of $\le 5$,
  indicating an AGN-dominated SED.
  \emph{Bottom}: IRAS $f_{\rm 60\micron}/f_{\rm 25\micron}$ versus
  $9.8\micron$ silicate strength for GA09 galaxies (circles), our AGN sample
  (triangles), and NGC 4418 (star); silicate and $6.2\micron$ PAH equivalent-width
  measurements are taken from \protect\cite{Spoon22}.
  Marker colours denote different PAH equivalent-width strengths (a star-formation
  indicator). The pink region marks $f_{\rm 60\micron}/f_{\rm 25\micron}\le 5$
  (AGN-dominated spectra); the grey region marks silicate emission instead of
  absorption.
}

\end{center}
\end{figure} 

\subsection{Sample Completeness}

In extreme cases where the nuclei of the galaxies are highly obscured such as in merging galaxies and Compact Obscured Nuclei (CONs; eg \citealp{Sakamoto2013}; \citealp{Martin2016}; \citealp{aalto2019}), even the [Ne {\sc v}] line can be extinguished and undetected. An example of this is the local prototype of a deeply buried nucleus in NGC 4418 (e.g. \citealp{Roche86}; \citealp{Gonzalez-Alfonso19}; \citealp{Wethers2024}). This galaxy is located at a distance of 31.9 Mpc \citep{Sanders03} and therefore lies beyond of our sample's distance threshold. It is prominent for its unusual deep silicate absorption feature at 9.8$\mu$m (e.g., \citealp{Rieke85}; \citealp{Roche86}; \citealp{Spoon07}; \citealp{Stierwalt13}), indicating extreme extinction. The presence of an AGN has long been suggested by numerous studies based on multiple pieces of evidence (e.g., \citealp{Roche86}; \citealp{Spoon01}; \citealp{Sakamoto21}), including the presence of a warm IR spectral energy distribution (SED), indicating a hidden hot source, likely an AGN accretion disk. However, the [Ne {\sc v}] emission-line was not detected in the galaxy by \emph{Spitzer}  \citep{Spoon22}, which could be due to the extreme absorption suffered by the nucleus.  

We can therefore take NGC 4418 as a model galaxy in order  to investigate if GA09 may have missed more AGN within their galaxy sample due to similar cause. Given the distinctive IR properties of NGC 4418, we search for evidence of a warm IR SED and deep silicate absorption. The IRAS $f_{\rm 60\micron}/f_{\rm 25\micron}$ flux ratio can be used as an indicator for a warm IR SED.  Galaxies with an AGN would have warmer IRAS colours and therefore smaller $f_{\rm 60\micron}/f_{\rm 25\micron}$ ratios. Figure 1 (top) shows the distribution of $f_{\rm 60\micron}/f_{\rm 25\micron}$ ratio for the galaxies in GA09, AGN in our sample and NGC 4418. We use a $f_{\rm 60\micron}/f_{\rm 25\micron}$ threshold value of $\leq 5$ to indicate AGN-dominated SED based on \cite{Alexander01}, and which is sufficiently "warm" to include NGC 4418. Based on the figure, we see that only three AGN in our sample (Circinus, NGC 1068, and NGC 4051) are identified using this technique. Interestingly, there are three additional galaxies in the GA09 sample with undetected [Ne {\sc v}] line emission that falls below our $f_{\rm 60\micron}/f_{\rm 25\micron}$ threshold: IIZW 40, M 82, NGC 4818. 

For these three galaxies however, there has been no convincing evidence for AGN in earlier studies. These galaxies also do not have extreme silicate absorption like NGC 4418, which implies that the extinction towards their mid-IR emission is not sufficiently high to detect [Ne {\sc{v}}] emission from the central source, if any. In fact, their silicate absorption is comparable to most of the AGN in our sample shown in Figure 1 (bottom), which means that GA09 should have been able to detect the line if a significant AGN existed in these galaxies. In addition, their 6.2 $\mu$m polycyclic aromatic hydrocarbon (PAH) equivalent widths are relatively high and are typical for starburst galaxies \citep{Spoon22}, except for IIZW 40 which shows a hot-dust dominated spectrum, typical for an AGN on the basis of its silicate strength and PAH feature \citep{Spoon22}. However, we argue that this system is a metal-poor dwarf galaxy and thus may not follow the same trends and features of more massive galaxies. Because of this and due to the lack of other evidence for the presence of AGN in the galaxy  \citep{Leitherer18}, we therefore infer that it does not host an AGN. 

Based on the  \textsl{Swift}-BAT 105-month survey (\citealp{Oh22}), there are five additional AGN that are detected which lies within our distance threshold and not selected in GA09, and therefore our sample (see Figure 2). They are M81, M106, NGC 1566, NGC 4151 and NGC 4395. However, only three of these are found to be located within 15 Mpc if we were to use the same distance calculation method by \cite{Mould00} as in GA09. These are M81, M106 and NGC 4395. NGC 4395 is a dwarf galaxy with $f$$_{60\,\mathrm{\mu m}}$ $<$ 5.24 Jy, and therefore was excluded in the RBGS sample \citep{Sanders03}. M81 has an IR luminosity of $L_{\rm IR} =$ 2.95 $\times$ 10$^{9}$ \lsun \citep{Sanders03}, which is just below GA09 selection criteria. M106 is not in the RBGS sample, although it would have matched all the galaxy selection criteria in GA09. However, \textsl{Spitzer} did not detect [Ne {\sc{v}}] emission from the galaxy \citep{Spoon22}.

Finally, based on the {\it{Swift}}-BAT survey, Circinus is the only AGN that is found within the Galactic latitude. Based on all these, we deduce that GA09 did not clearly miss any AGN in their galaxy sample on the basis of their selection criteria. Hence, we conclude that our AGN sample can be assumed to be near complete. 

\begin{figure}
\begin{center}
\includegraphics[scale=0.33]{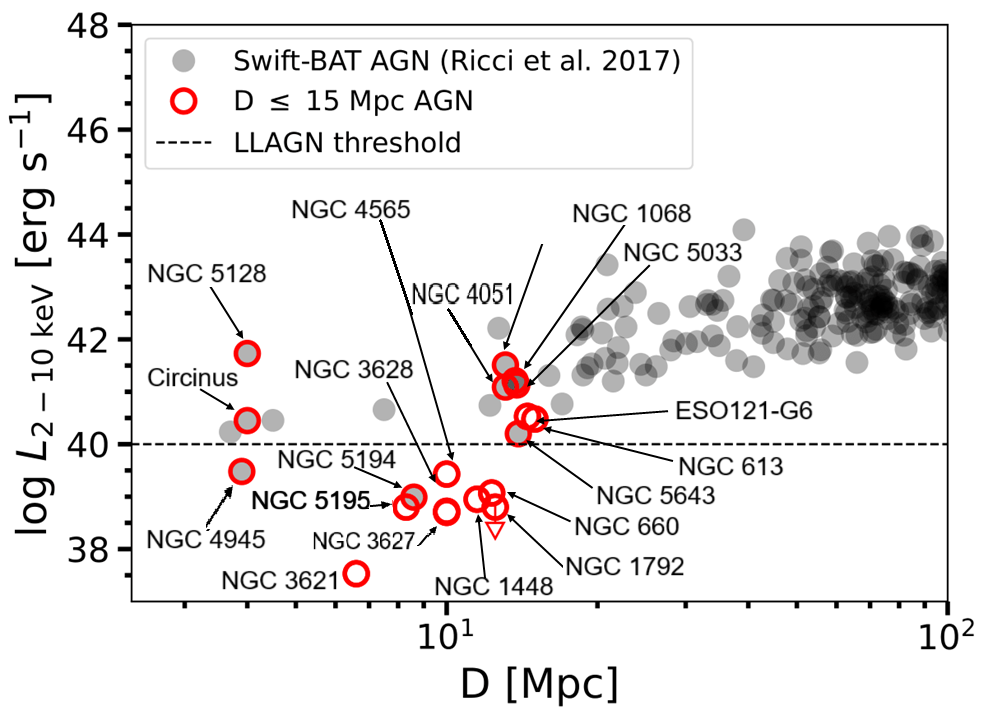}
\caption{2--10 keV luminosity versus distance for the AGN in our sample (observed luminosity; red), in comparison with the {\it{Swift}}-BAT AGN detected in the 70-month survey (intrinsic luminosity; grey; \citealp{Ricci15}). Our AGN sample extends approximately two orders of magnitude beyond the fainter end of the {\it{Swift}}-BAT AGN sample (dashed line). {\it{Swift}}-BAT AGN within 15 Mpc that are not in our sample are discussed in Section 2.1.}
\end{center}
\end{figure} 

\section{X-ray Observations and Data Reduction}

\begin{table*}
\centering
\caption{Log of additional X-ray observations used in this work. }
\begin{adjustbox}{width=\textwidth,totalheight=\textheight,keepaspectratio}
\begin{tabular}{ccccccccccc}
  \hline 
   Name & R.A. & Dec. & Observatory & ObsID & Date & Energy Band & $t_{\rm exp}$  & Extraction & Count Rate & Previous Work \\
   &&&&&&&& Region & \\
             &   &  &  &    &  &[keV] & [ks] & [$\arcsec$] & [10$^{-3}$ cts s$^{-1}$] &   \\
   (1) & (2)  & (3) & (4) & (5) & (6) & (7) & (8) & (9) & (10) & (11)\\
   \hline
      NGC 613 & 1:34:18.23 & $-$29:25:06.35 &\textsl{Chandra} & 16351  & 2014-08-21 & 0.5--8 & 48.9 & 8 & 45.14 $\pm$ 0.97 & \cite{daSilva20a}  \\  
     NGC 1792 & 5:05:14.58 & $-$37:58:50.85  & \textsl{Chandra} & 19524 & 2016-11-23 & 0.5--8 & 19.8 & 20 & 6.42 $\pm$ 1.22 & ...  \\  
   		      & & & \textsl{NuSTAR} & 60371001002  & 2018-05-13 & 3--24 & 45.7 & 20 & $<$0.49 & ...  \\
    NGC 3621 & 11:18:16.51 & $-$32:48:50.78  & \textsl{Chandra} & 9278 & 2008-03-06 & 0.5--8 & 21.2 & 3 & 1.06 $\pm$ 0.30 & \cite{Gliozzi09}   \\  
   	 & & & \textsl{XMM-Newton} & 0795660101  & 2017-12-16 & 0.5--10 & 56.4 & 30 & 602 $\pm$ 5 & ... \\
     & & & \textsl{NuSTAR} & 60371002002  & 2017-12-15 & 3--24 & 61.4 & 30 & 10.90 $\pm$ 0.50  
 & \cite{Osorio22} \\
    NGC 3627 & 11:20:15.04 & $+$12:59:29.92  & \textsl{Chandra} & 394 & 1999-11-03 & 0.5--8 & 1.8 & 2  & 6.17 $\pm$ 1.98  & \cite{Panessa06}  \\  
     & & & ...  & 9548  & 2008-03-31 & 0.5--8 & 49.6 & 2 & 5.09 $\pm$ 0.33 & \cite{Cisternas13}\\ 
     &  &  & ... & ... & ... & ... & ... & ... & ... &  \cite{Saade22} \\
          & & & \textsl{NuSTAR} & 60371003002  & 2017-12-20 & 3-24 & 101.3 & 30 & $<$2.19 & \cite{Esparza20}\\
          &  &  & ... & ... & ... & ... & ... & ... & ... &   \cite{Saade22} \\
    NGC 3628 & 11:20:16.92 & $+$13:35:20.59  & \textsl{Chandra} & 395 & 1999-11-03 & 0.5--8 & 1.8 & 5 & $<$3.60 & ...  \\  
     &  &   & ... & ... & ... &...  & ... &  20 & 131.5 $\pm$ 10.3  & ...  \\ 
     & & & ... & 2039  & 2000-12-02 & 0.5--8 & 58.0  & 5 & 5.35 $\pm$ 0.34 & \cite{GonzalezMartin09}  \\
     & & & ... & ...  & ... & ... & ... & 20 & 71.36 $\pm$ 1.25 & \cite{GonzalezMartin09}  \\
          & & & \textsl{NuSTAR} & 60371004002  & 2017-12-23 & 3-24 & 100.5 & 20 & 4.15 $\pm$ 0.25 & \cite{Esparza20} \\
          &  &  & ... & ... & ... & ... & ... & ... & ... &  \cite{Osorio22}  \\
      NGC 4565 & 12:36:20.78 & $+$25:59:15.58 &\textsl{Chandra} & 404  & 2000-06-30 & 0.5--8 & 2.8 & 3 & 50.83 $\pm$ 4.26 & \cite{Wu02}    \\  
    & & & ... &  3950  & 2003-02-08 & 0.5--8 & 59.2 & 3 & 37.83 $\pm$ 0.81 & \cite{Chiaberge06} \\
   \hline
\end{tabular}
\end{adjustbox} 
  \begin{tablenotes}
   \footnotesize
   \item \emph{Notes.} (1) Galaxy name; (2)-(3) AGN position that was used to extract the spectra, mostly from \emph{Chandra}, expect for NGC 1792; (4) observatory; (5) observation identification number; (6) observation UT start date; (7) energy band in keV; (8) the net (clean) exposure time in ks; (9) radius of circular region used to extract the spectra; (10) net count rate within the extraction region in the given energy band in units of 10$^{-3}$ cts s$^{-1}$. The net exposure times and count rates for \textsl{NuSTAR} and \textsl{XMM-Newton} are the total values from the FPMs, and EPIC cameras, respectively. (11) Previous work that has analyzed the same data. 
   \end{tablenotes}  
\end{table*}

Throughout our studies on the $D <$ 15 Mpc sample, \emph{Chandra} data is critical as it provides us with high resolution X-ray images of the galaxies which is important in isolating the AGN emission from off-nuclear sources and accounting for contaminants in the AGN emission in data obtained by other telescopes (e.g., \citealp{Annuar17, annuar20}). This is especially crucial for our AGN sample where about half of our sources have very low observed X-ray luminosities; i.e., $L_{\rm 2-10, obs} \leq$ 10$^{39}$ erg s$^{-1}$ (see Figure 2), comparable to ultraluminous X-ray sources. This makes them prone to significant contamination by host-galaxy emission. Sensitive high-energy X-ray data from \emph{NuSTAR} have also been demonstrated to be important to provide good quality high-energy data for these low-luminosity sources which are not detected by \emph{Swift}-BAT. Broadband X-ray data are essential for characterizing the AGN spectra accurately in order to obtain reliable measurements of their properties. 

Among the 19 AGN, broadband X-ray spectral analyses for 13 sources ($\sim$68$\%$) have been performed in detail by past studies utilising low-energy data from \emph{Chandra} and/or \emph{XMM-Newton}, and high-energy data from \emph{NuSTAR} and/or \emph{Swift}-BAT, mostly using physically motivated torus models by e.g., \cite{Murphy12}, \cite{Brightman08} and/or \cite{Balokovic18}, to measure the X-ray properties of the AGN, including their torus column densities and intrinsic luminosities (see Table 3). The analyses for five of these sources were published as part of our work in \cite{Annuar15, Annuar17, annuar20}. To improve the X-ray completeness, we acquired a total of 8 \emph{NuSTAR} observations, of which five were coordinated with \emph{Chandra} (three), \emph{XMM-Newton} (one) or \emph{Swift}-XRT (one). This boosted the \emph{Chandra} and \emph{NuSTAR} data for our sample from $\sim$89$\%$ to 100$\%$, and $\sim$47$\%$ to $\sim$89$\%$ complete, respectively. We did not propose \emph{NuSTAR} observations for the remaining two ($\sim$11$\%$) AGN (i.e., NGC 613 and NGC 4565)  since we believe that the archival low-energy X-ray data for those two sources already provide reliable measurements on their column densities and AGN properties (see Table 1).

In this section, we present the X-ray observations and analyses for these remaining six objects in our sample. We note that most of the data for these AGN have been analyzed and published in past papers. However, we re-analyzed them here to ensure our analyses are consistent with our previous studies. The X-ray observations used in this paper are detailed in Table 2.

As mentioned earlier, we prioritize low-energy X-ray data from \textsl{Chandra} in our analysis. The \textsl{Chandra} data were reprocessed to create event files with updated calibration modifications using the {\sc{ciao}} pipeline \citep{Fruscione06}, following standard procedures. We then used the {\sc{dmcopy}} task to produce X-ray images of each source in different energy bands, and extracted the source spectra using the {\sc{specextract}} task in {\sc{ciao}}. One of our sources (NGC 3621) has coordinated \textsl{NuSTAR} and \textsl{XMM-Newton} observations, which we present in this paper. We analyzed the Pipeline Processing System (PPS) data products using the Science Analysis Software (SAS), with standard filter flags. Background flares were subtracted from the data by visually examining the source light curves, and the X-ray spectra from the three EPIC cameras were then extracted using the {\sc{evselect}} task in SAS.

In addition to these, we also used high-energy X-ray observations from \textsl{NuSTAR} where available to facilitate our X-ray spectral analysis of the AGN at high energies. We processed the {\emph{NuSTAR}} data for our sources with the \textsl{NuSTAR} Data Analysis Software ({\sc{nustardas}}) within {\sc{heasoft}}. The {\sc{nupipeline}} script was used to produce the calibrated and cleaned event files using standard filter flags. We extracted the spectra and response files from each of the \textsl{NuSTAR} focal plane modules, named A and B (FPM A and FPM B), using the {\sc{nuproducts}} task. In addition to the spectral extraction, we also combined the \textsl{NuSTAR} event files from the two FPMs using {\sc{xselect}} to produce the total event file. The total image counts at different energy bands were then produced from the resultant event files using the {\sc{dmcopy}} task in {\sc{ciao}}.

In all cases, the spectra and response files from each \textsl{NuSTAR} FPM are combined using the {\sc{addascaspec}} script to increase the overall signal-to-noise ratio of the data in our spectral fitting (see Section 4).\footnote{More details on the {\sc{addascaspec}} script can be found at https://heasarc.gsfc.nasa.gov/docs/asca/adspecinfo.html .} For \textsl{XMM-Newton} data (NGC 3621), spectra extracted for the EPIC MOS1 and MOS2 cameras were combined using the {\sc{epicspeccombine}} task in SAS. In most cases, we binned our spectra to a minimum of 20 counts per bin to allow the use of $\chi^{2}$ statistics. However, for cases in which the count rate is relatively low ($<$200 counts), we grouped the spectra to a minimum of 5 counts per bin for the \textsl{NuSTAR} and \textsl{XMM-Newton} data, and 1 count per bin for the \textsl{Chandra} data, and optimised our fitting using the Poisson C-statistic \citep{Cash79}. This was done using the {\sc{grppha}} task in {\sc{heasoft}}.

\section {Xray Spectral Analyses}

The X-ray spectral analysis was performed using {\sc{xspec}}. We included a fixed Galactic absorption component for each source \citep{Kalberla05} using the {\sc{xspec}} model ``{\sc{phabs}}" in all spectral fits, and assumed solar abundances for all models. Redshifts for all sources were obtained from the NASA/IPAC Extragalactic Database (NED).

Due to the modest quality of most of our data, we also fixed the cross-calibration uncertainties of each observatory with respect to \textsl{NuSTAR} to the values found by \citet{Madsen15} using a constant parameter, \textsl{C}. In general, we started our analysis using a simple absorbed power-law model to provide an initial indication of the level of obscuration suffered by the AGN. If the model indicated that the sources were obscured with $N_{\rm H} \geq$ 10$^{22}$ cm$^{-2}$, we then utilised the physically-motivated torus models by \cite{Murphy12} (MY{\sc{torus}}) and \cite{Balokovic18} (B{\sc{orus}}). We did not use these models for the unobscured sources due to the model limitations, which could only measure column densities down to $N_{\rm H} =$ 10$^{22}$ cm$^{-2}$. In all cases, we used the most simple version of the MY{\sc{torus}} and B{\sc{orus}} models, with all the AGN emission components tied together and most non-crucial parameters (e.g., torus inclination angle, iron abundance and high-energy cut-off) fixed to the recommended values. We refer the reader to \cite{Murphy12}, \cite{Balokovic18} and our previous works (\citealp{Annuar15,Annuar17,annuar20}) for more detailed information on the modelling technique used. We summarize the main results of our spectral analysis in Table 3. All errors are quoted at 90$\%$ confidence. Below we discuss the individual sources and their spectral analysis results.

\subsection{NGC 613}

\begin{figure*}
\begin{center}
\includegraphics[scale=0.27]{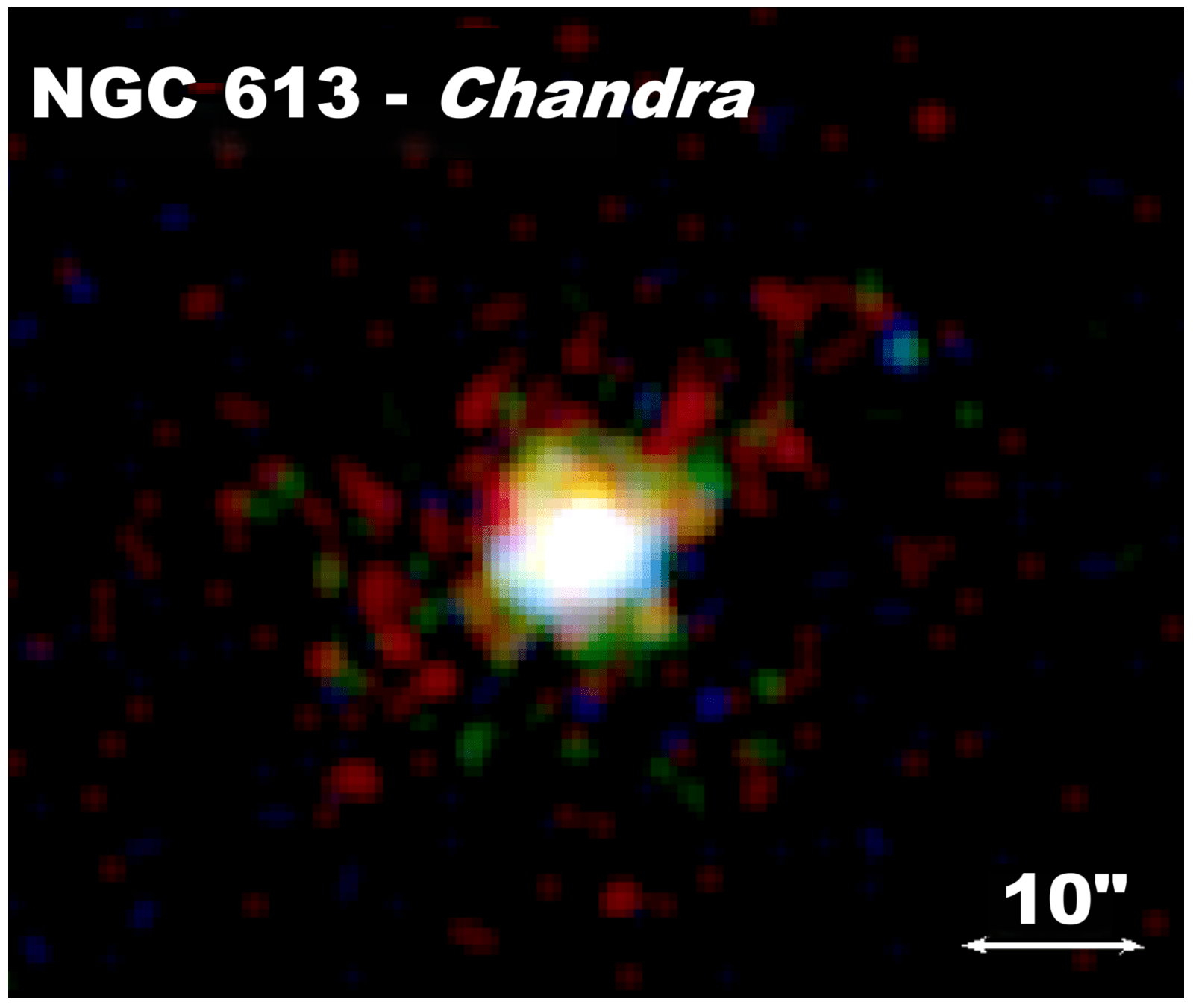} \\
\includegraphics[scale=0.31]{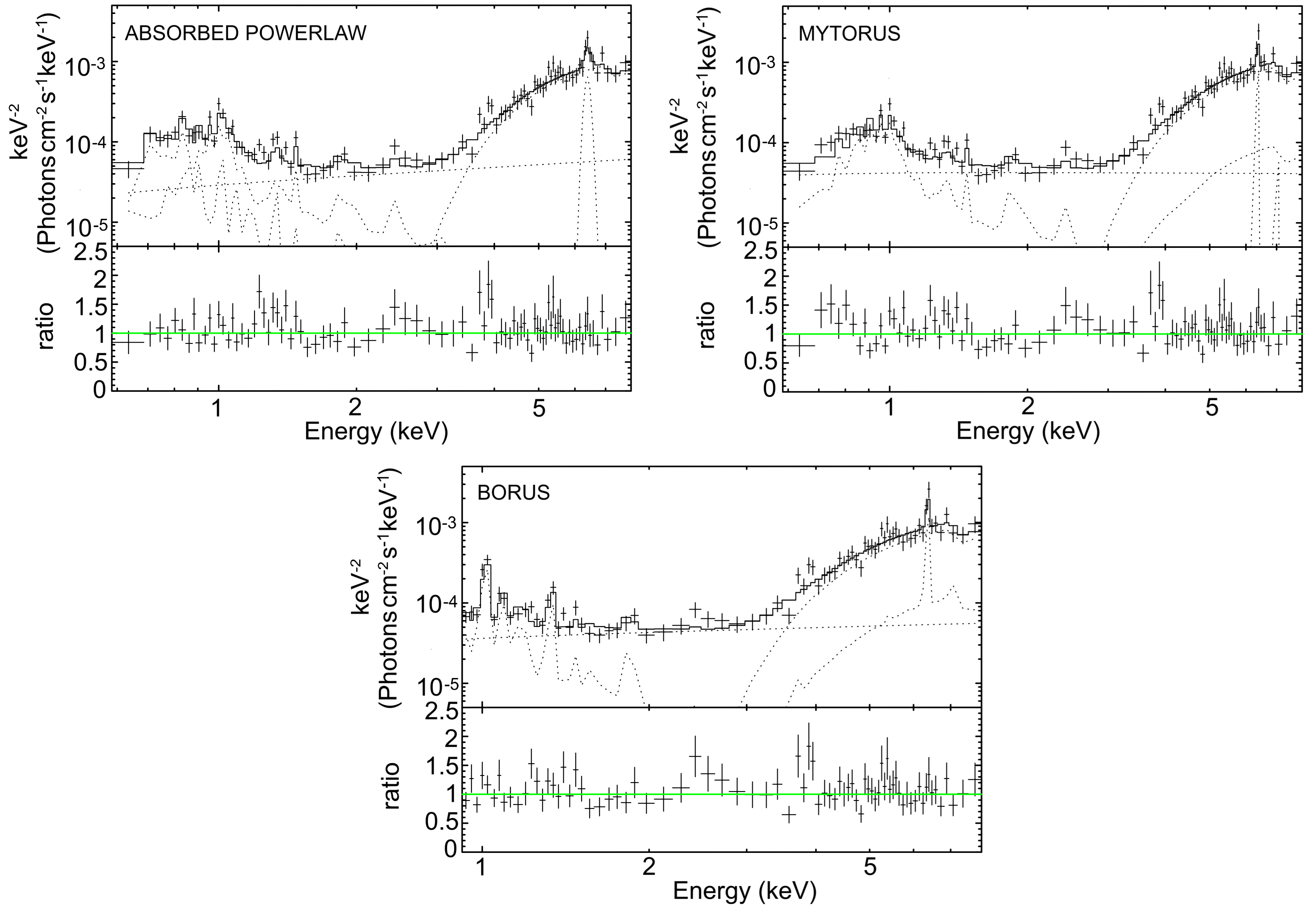}
\caption[
  NGC613
]{
 \emph{Top}: \textsl{Chandra} RGB images of NGC 613 (Red: 0.5--1 keV, Green: 1--2 keV, Blue: 2--8 keV). The image is smoothed with a Gaussian function of radius 3 pixels, corresponding to 1.5$\arcsec$. \emph{Bottom}: Best-fitting absorbed power-law model (top left), MY{\sc{torus}} (top right) and B{\sc{orus}} (bottom) models to the spectra. The top panels show the data and unfolded model in $E^{2}F_{E}$ units, whilst the bottom panels show the ratio between the data and the folded model. The model components fitted to the data are shown as dotted curves, and the combined model fits are shown as solid curves.
}
\end{center}
\end{figure*}
  
NGC 613 is an Sbc galaxy located at a distance of $D =$ 15 Mpc. Evidence for the presence of an AGN in the galaxy was provided by GA09 on the basis of [Ne {\sc v}] emission and the presence of nuclear radio jet (\citealp{Hummel87}; \citealp{Hummel92}; \citealp{Miyamoto17}). Nuclear water maser emission has also been detected in the galaxy \citep{Kondratko06}. The AGN is unidentified in the optical wavebands (i.e., classified as H{\sc{II}} in optical; GA09) and it is not detected in the \emph{Swift}-BAT survey \citep{Oh18}.

The AGN is surrounded by a starburst ring (\citealp{Hummel92}; \citealp{Falcon-Barroso14}; \citealp{Audibert19}), but using high angular resolution mid-IR observations ($\sim$0.4 arcsec) by Gemini T-ReCS, \cite{Asmus14} managed to resolve the compact nucleus from this circumnuclear ring. The nuclear region was recently studied in great detail by \cite{daSilva20a, daSilva20b} using multiwavelength data in order to understand its complexity. The nucleus has been observed in X-rays by \emph{XMM-Newton} and \emph{Chandra} in 2010 and 2014, respectively. \cite{Castangia13} analysed the \emph{XMM-Newton} data and measured a column density of $N_{\rm H} =$ (36.0 $\pm$ 5.0) $\times$ 10$^{23}$ cm$^{-2}$, indicating that it is heavily obscured, but Compton-thin. The \emph{Chandra} data are analyzed in \cite{daSilva20a}, though no column density measurement was presented.  

\subsubsection{X-ray observations and spectral fitting}

We re-analyse the \emph{Chandra} data of NGC 613 to obtain the column density value from this higher resolution X-ray observation. We did not try to obtain a \emph{NuSTAR} observation for this object as there were no indications of it being a CT AGN based on multiwavelength diagnostics (see Figures 12 and 13), and thus the current X-ray data are sufficient to provide us with a reliable column density measurement. The AGN is clearly detected in the \textsl{Chandra} data, located at RA $=$ 1:34:18.23, Dec. $=$ $-$29:25:06.35 as determined by the {\sc{wavdetect}} tool within {\sc{ciao}} in the 2--8 keV energy band. In Figure 3, we show the combined \textsl{Chandra} RGB image of NGC 613. We extracted the spectra of the AGN using an 8$\arcsec$-radius circular region to incorporate all the source emission. The total net count rate obtained from this extraction region is 45.1 $\times$ 10$^{-3}$ counts s$^{-1}$ ($\sim$2200 counts), allowing us to perform a relatively detailed modelling of the spectrum. 

We modelled the spectrum  using  three models: an absorbed power-law, MY{\sc{torus}} and B{\sc{orus}}. There is an excess of emission at $\sim$6.4 keV, suggesting the presence of an Fe K$\alpha$ line, indicating heavy obscuration. We therefore added a {\sc{gaussian}} component to our power-law model in order to simulate this emission-line. In addition, we also found that the spectrum required a soft power-law and two {\sc{apec}} components to simulate the thermal emission at low energy. Based on our analyses, we found that all three models provide comparably good fits to the data, with the B{\sc{orus}} model having the lowest reduced $\chi^{2}$ value of $\sim$ 1.1. However, we had to fix the photon index for this model to 1.8 \citep{Ricci17} as it was unconstrained. The photon indices measured by the other two models are consistent with the typical intrinsic value found for AGN (e.g., \citealp{Burlon11}; \citealp{Ricci17}). All three models provided an $N_{\rm H}$ value of $\approx$3.0 $\times$ 10$^{23}$ cm$^{-2}$, indicating that the AGN is heavily obscured but not CT (see Table 3). This is in agreement with was what found by \cite{Castangia13} using \textsl{XMM-Newton} data. The plasma temperatures measured by the two {\sc{apec}} components in the power-law model are $kT_{\rm 1} = 0.51_{-0.13}^{+0.14}$ keV and $kT_{\rm 2} = 1.14_{-0.11}^{+0.12}$ keV, respectively, consistent with \cite{Castangia13}. We show the modelled spectrum of NGC 613 fitted by all three models in Figure 3.

\subsection{NGC 1792}

NGC 1792, located at a distance of 12.5 Mpc, is classified as an H{\sc{ii}} galaxy in the optical \citep{Veron86}. The source has been observed at high spatial resolution at 12$\mu m$ by Gemini/T-ReCS ($t_{\rm exp} =$ 319 s), but was not detected. In X-rays, it has only been previously observed by \textsl{XMM-Newton} (2007-03-29; ObsID 0403070301; $t_{\rm exp} =$ 23.3 ks). The \textsl{XMM-Newton} observation revealed diffuse soft emission at the nuclear position of the galaxy, with no clear indication of a point source. However, a [Ne {\sc v}] emission-line is clearly detected at the central part of the galaxy in GA09, indicating that it hosts an AGN. 

\begin{figure}
\begin{center}
\includegraphics[scale=0.29]{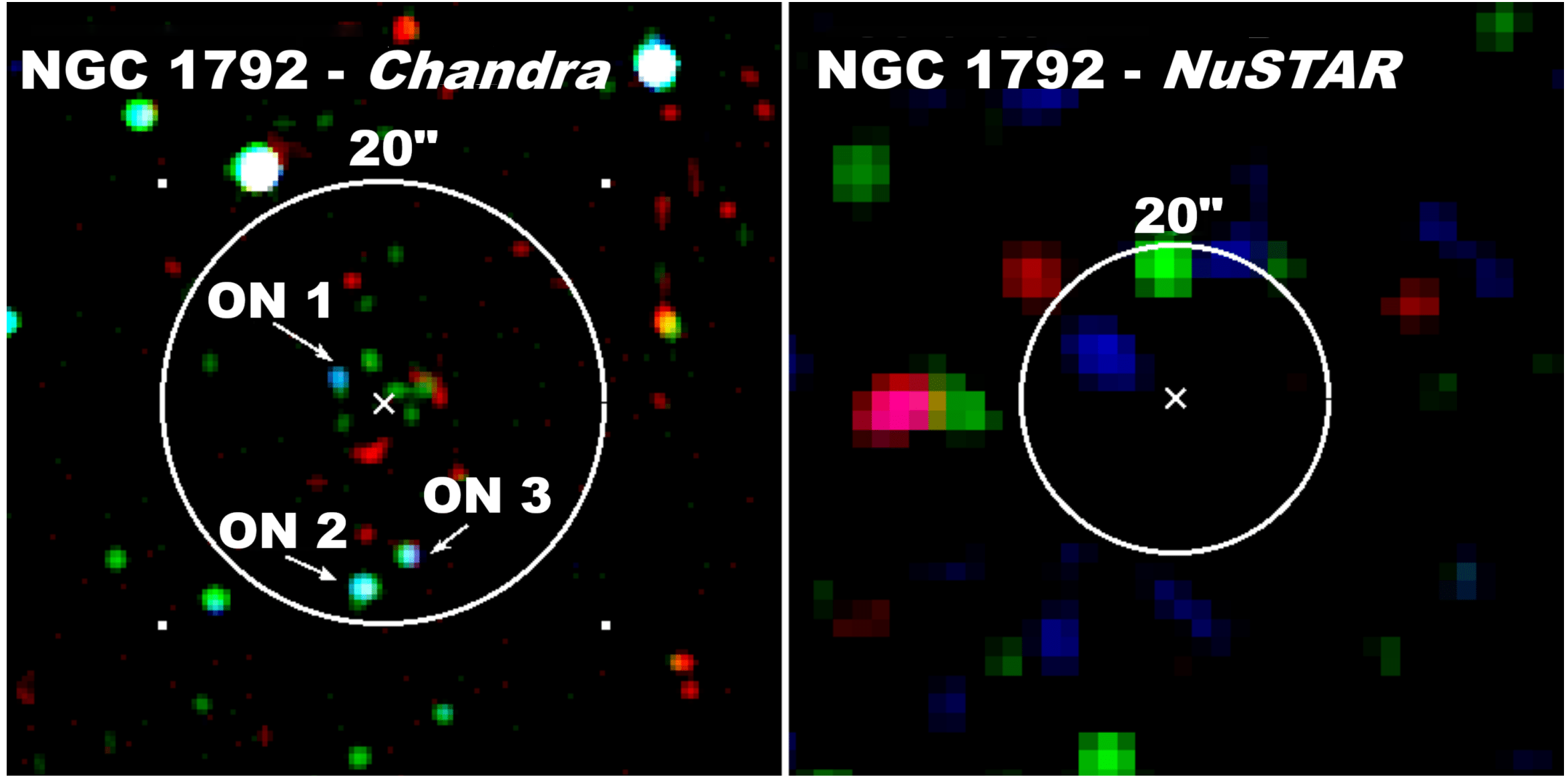}
\includegraphics[scale=0.31]{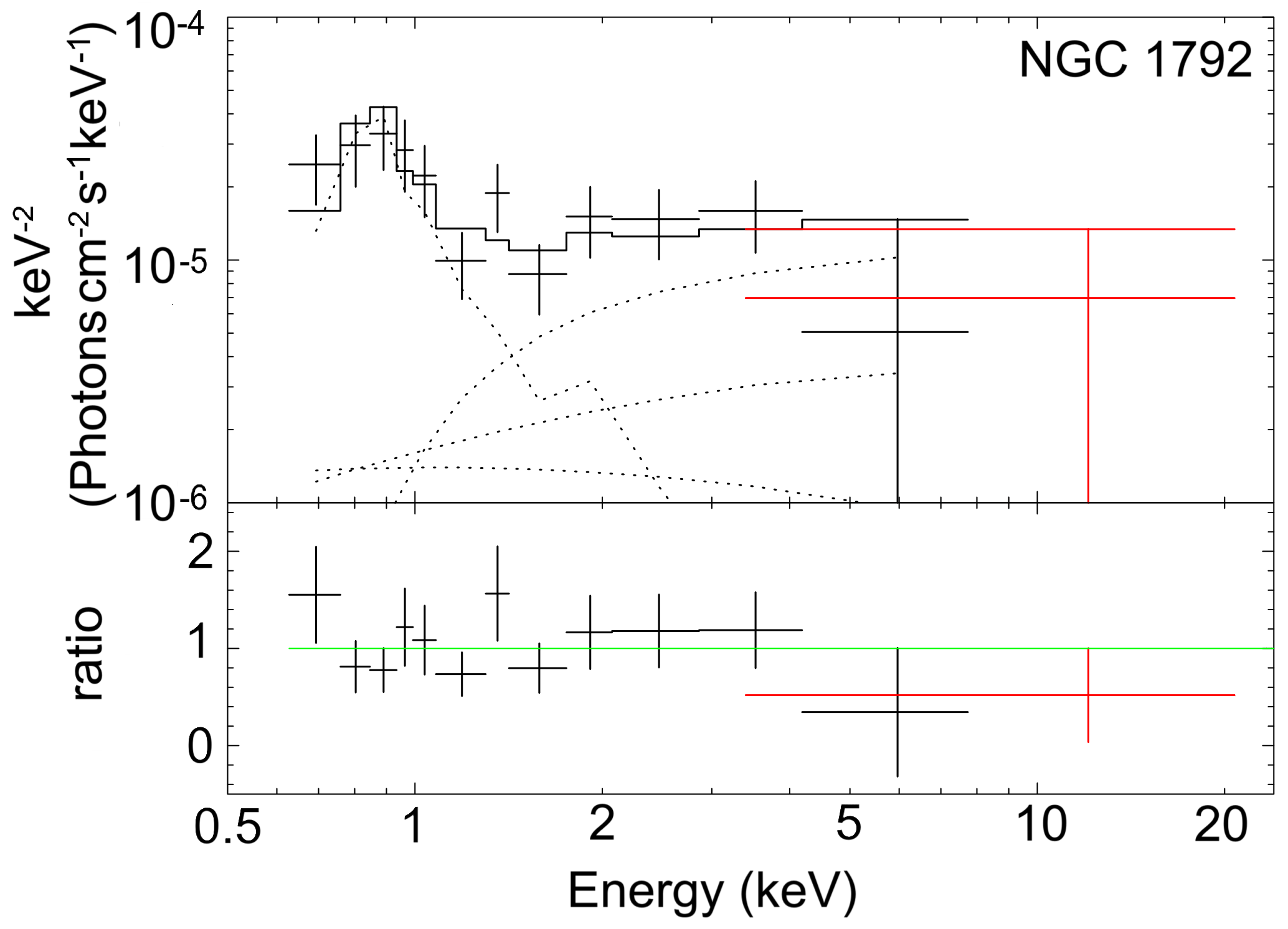}
\caption[
  NGC1792
]{
 \emph{Top}: \textsl{Chandra} and \textsl{NuSTAR} RGB images of NGC 1792 (\textsl{Chandra} - Red: 0.5--1 keV, Green: 1--2 keV, Blue: 2--8 keV; \textsl{NuSTAR} - Red: 3--8 keV, Green: 8--24 keV, Blue: 24--79 keV). The off-nuclear sources detected within the 20$\arcsec$-radius extraction region in \textsl{Chandra} are labelled ON 1, ON 2 and ON 3. The images are smoothed with a Gaussian function of radius 3 pixels, corresponding to 1.5$\arcsec$ and and 7.4$\arcsec$ for \textsl{Chandra} and \textsl{NuSTAR}, respectively. \emph{Bottom}: Best-fitting model to the combined \textsl{Chandra} (black) and \textsl{NuSTAR} (red) data. The data have been rebinned to a minimum of 3$\sigma$ significance with a maximum of 500 bins for visual clarity. The top panel shows the data and unfolded model in $E^{2}F_{E}$ units, whilst the bottom panel shows the ratio between the data and the folded model. The spectra were fitted using an absorbed power-law model to simulate the AGN emission, and two cut-off power law components to model the off-nuclear sources, ON 2 and ON 3. ON 1 was not included in the spectral fitting as its contribution was found to be insignificant. The model components fitted to the data are shown as dotted curves, and the combined model fit is shown as a solid curve.
}

\end{center}
\end{figure}

\subsubsection{X-ray observations and spectral analysis}

We observed NGC 1792 with \textsl{Chandra} in 2016 for 19.8 ks (2016-11-23; ObsID 19524), to provide complete \textsl{Chandra} coverage for our sample. However, an X-ray source associated with the nuclear position of NGC 1792 was not detected. The nearest source to the 2MASS nuclear position of the galaxy is detected $\sim$5$\arcsec$ away. The upper limit fluxes measured at this central position are 7.6 $\times$ 10$^{-15}$ erg s$^{-1}$ cm$^{-2}$ and 8.54 $\times$ 10$^{-15}$ erg s$^{-1}$ cm$^{-2}$ at 0.5--2 keV and 2--8 keV, respectively. We also managed to obtain \textsl{NuSTAR} observation of the source in 2018 for 22.9 ks (2018-05-13; ObsID 60371001002). However, no strong emission was detected near the central position of the galaxy using the detection technique adopted in other \textsl{NuSTAR} studies of faint sources (significance $\lesssim$2.6$\sigma$; e.g. \citealp{Luo13}; \citealp{Lansbury14}; \citealp{Stern14}). We show \textsl{Chandra} and \textsl{NuSTAR} RGB images of NGC 1792 in Figure 4. 

We extracted X-ray spectra of the source anyway to obtain a measurement of $N_{\rm H}$ that could be used as an estimation for the AGN column density.  The spectra was extracted using a circular extraction region with 20$\arcsec$ radius, centred on the 2MASS position of the galaxy. This region size corresponds to $\sim$30$\%$ of the \textsl{NuSTAR} encircled energy fraction, and the size was chosen to minimize contamination of off-nuclear sources. There are no significant differences between the \textsl{Chandra} and \textsl{NuSTAR} spectra. We therefore fitted the two spectra simultaneously using a simple absorbed cutoff power-law model to simulate the AGN emission, with an additional {\sc{apec}} component to model the thermal emission at low-energy. In addition, there are three off-nuclear sources detected within the extraction region in the full \textsl{Chandra} 0.5--8 keV band. We included the power-law components for two of these sources (ON 2 and 3) into our model. The component for ON 1 was not included as its contribution was relatively insignificant, although it is more significant than the AGN emission, which is completely undetected. Under this assumption, the observe AGN luminosity would be much lower than the measured value.

\begin{figure*}
\begin{center}
\includegraphics[scale=0.45]{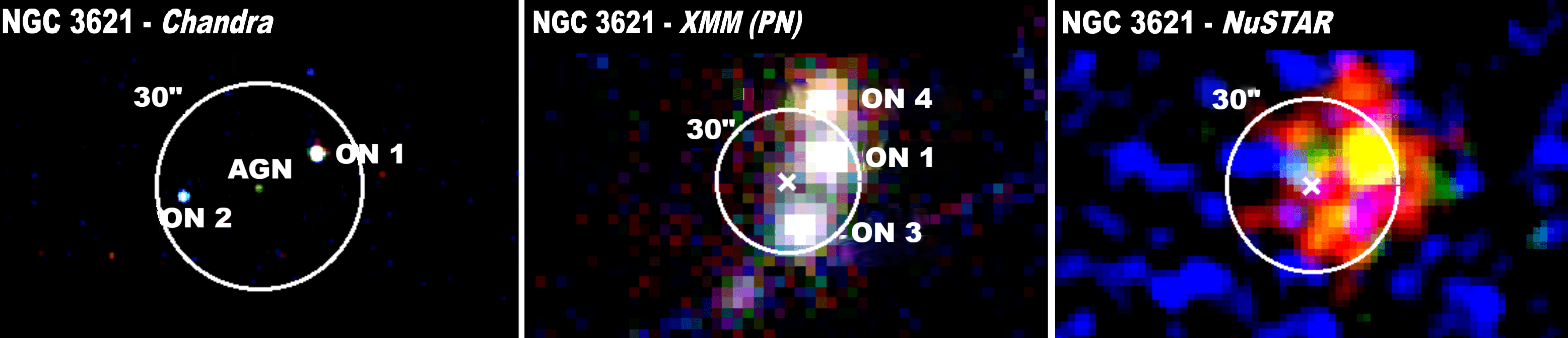}
\includegraphics[scale=0.32]{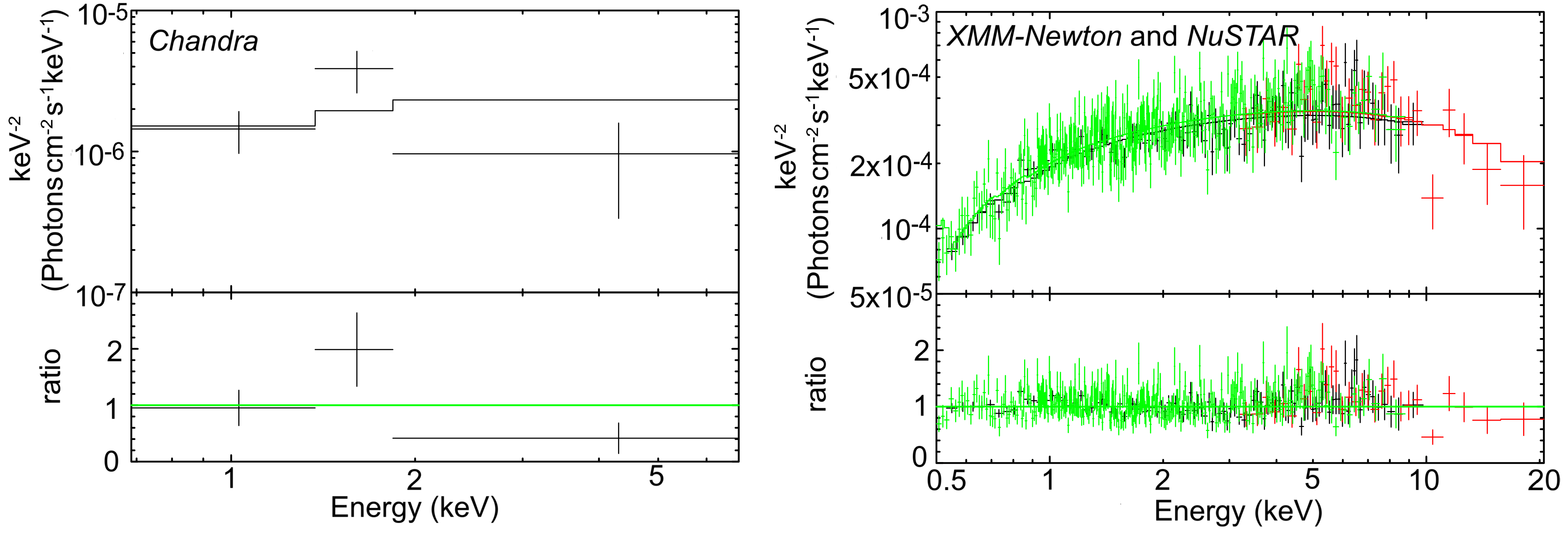}
\caption[
  NGC3621
]{
 \emph{Top}: \textsl{Chandra}, \textsl{XMM-Newton} and \textsl{NuSTAR} RGB images of NGC 3621 (\textsl{Chandra} - Red: 0.5--1 keV, Green: 1--2 keV, Blue: 2--8 keV; \textsl{XMM-Newton} - Red: 0.5--1 keV, Green: 1--2 keV, Blue: 2--10 keV; \textsl{NuSTAR} - Red: 3--8 keV, Green: 8--24 keV, Blue: 24--79 keV). The off-nuclear sources which are detected within the 30$\arcsec$-radius extraction region in \textsl{Chandra} and \textsl{XMM-Newton} are labelled ON 1, ON 2, ON 3 and ON 4. \emph{Bottom}: Best-fitting absorbed power-law model to the \textsl{Chandra} spectrum extracted from a small 3$\arcsec$ region corresponding to the AGN (left), and to the combined \textsl{NuSTAR} (red) and \textsl{XMM-Newton} data (black - PN; MOS - green). Figure description is the same as Figure 4.
}

\end{center}
\end{figure*}

The photon index was fixed to 1.8 as we were not able to constrain it. We obtained a good fit to the data (reduced $\chi^{2} \sim$ 1.3) that provided an upper limit of $N_{\rm H} \leq$ 2.5 $\times$ 10$^{22}$ cm$^{-2}$, suggesting that it is just mildy obscured at most. Note that, the AGN could also be extremely CT, causing it to not be detected in our X-ray data. The observed luminosity measured for the AGN using this model is 6.30 $\times$ 10$^{38}$ erg s$^{-1}$, which should be taken as an upper limit. The intrinsic luminosity of the source however, could be higher than this observed luminosity if it is indeed heavily CT. The plasma temperature measured by the {\sc{apec}} component is $kT = 0.75^{+0.15}_{-0.41}$ keV. The best-fit spectra for the source are shown in Figure 4.

\subsection{NGC 3621}
 
NGC 3621 is a late-type (Sd) bulgeless spiral galaxy located at a distance of 6.6 Mpc. The presence of an AGN in the galaxy was initially discovered by the detection of the [Ne {\sc v}] emission-line from \emph{Spitzer} spectroscopic observation \citep{Satyapal07}, which was also later confirmed by GA09. Optical spectroscopy later identified the presence of a Seyfert 2 nucleus \citep{Barth09}. A \emph{Chandra} observation also detected a weak X-ray point source coincident with the nucleus of the galaxy, adding to the growing evidence that black holes can in fact form and grow in a bulgeless disk galaxy \citep{Gliozzi09}. \emph{Hubble Space Telescope (HST)} images for this object reveal a bright and compact nuclear star cluster, providing evidence that black holes can be found inside some nuclear star clusters \citep{Barth09}. The \emph{Chandra} observation also reveals the presence of two potential ULXs located almost symmetrically 20 arcsec away from the centre. However, \cite{Gliozzi09} did not perform X-ray spectral fitting on the AGN due to low count rates. Despite this, they concluded that the collective evidence from optical and infrared spectroscopic results provides strong support that NGC 3621 harbours a buried AGN. The AGN is not detected in the \emph{Swift}-BAT survey \citep{Oh18}, and has not been observed at high angular resolution at mid-IR wavelengths.

\subsubsection{X-ray observations and spectral fitting}

Prior to our work, NGC 3621 had only been observed in X-rays by \emph{Chandra} (see above). We obtained simultaneous \emph{NuSTAR} (2017-12-15; ObsID 60371002002) and \emph{XMM-Newton} (2017-12-16; ObsID 0795660101) observations for the galaxy as we expect the AGN and the off-nuclear sources to be spatially resolved by \emph{XMM-Newton}, and the observation could provide higher quality low-energy data to complement the \emph{NuSTAR} data. \cite{Osorio22} have analyzed the \emph{NuSTAR} data alone using the simple reflection models {\sc{pexrav}} \citep{Magdziarz95} and {\sc{pexmon}} \citep{Nandra2007}. They did not detect any reflection signatures in the spectrum, and measure a column density of $N_{\rm H} =$ (5.4 $\pm$ 3.6) $\times$ 10$^{21}$ cm$^{-2}$, indicating that the AGN is just mildy obscured. In this paper, we present the \emph{XMM-Newton} observation of the source, and re-analyze the \emph{Chandra} and \emph{NuSTAR} data, taking into consideration contamination from the relatively bright off-nuclear sources in the \emph{NuSTAR} data that could significantly affect the results presented by \cite{Osorio22} due to the relatively large extraction region used in that study (2$\arcmin$ radius).

\begin{figure*}
\begin{center}
\includegraphics[scale=0.34]{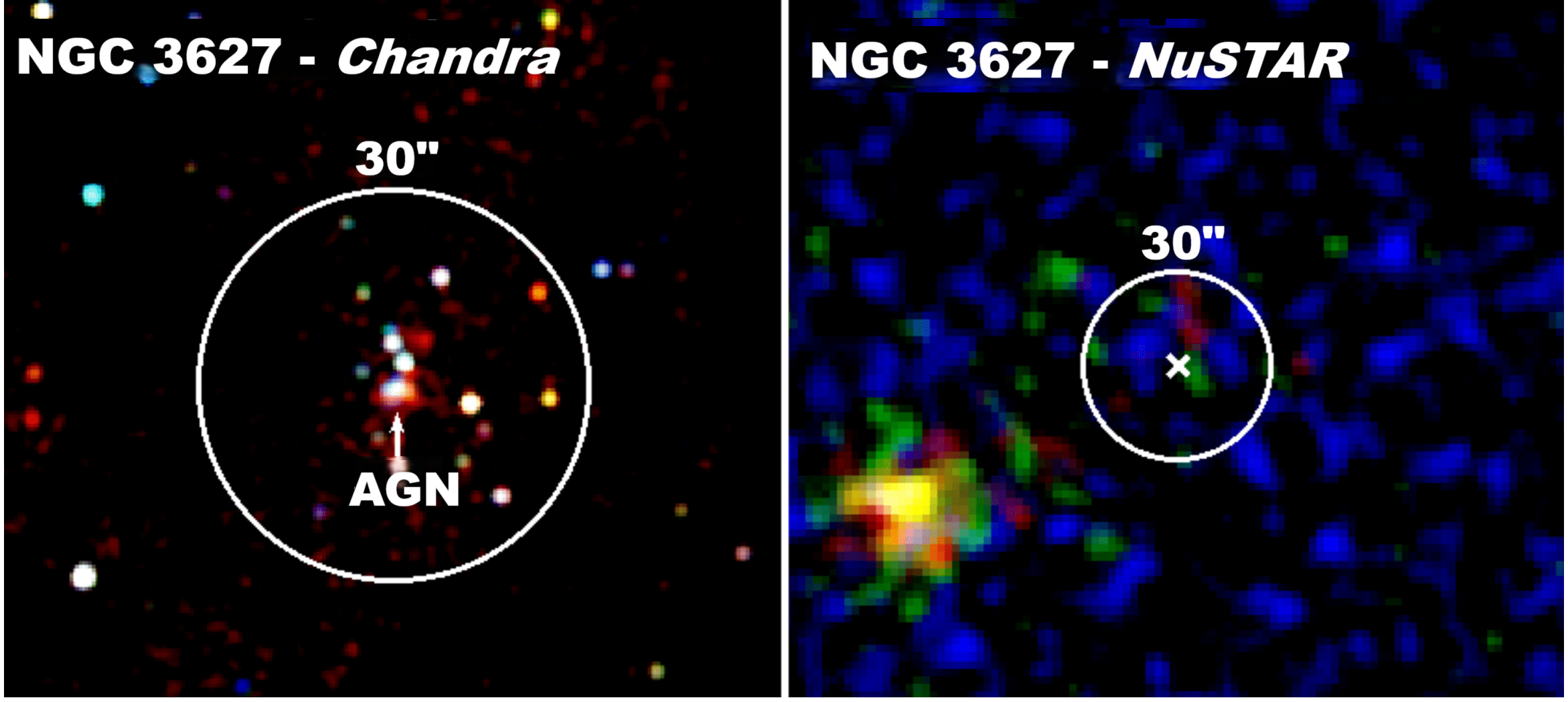}\\
\includegraphics[scale=0.32]{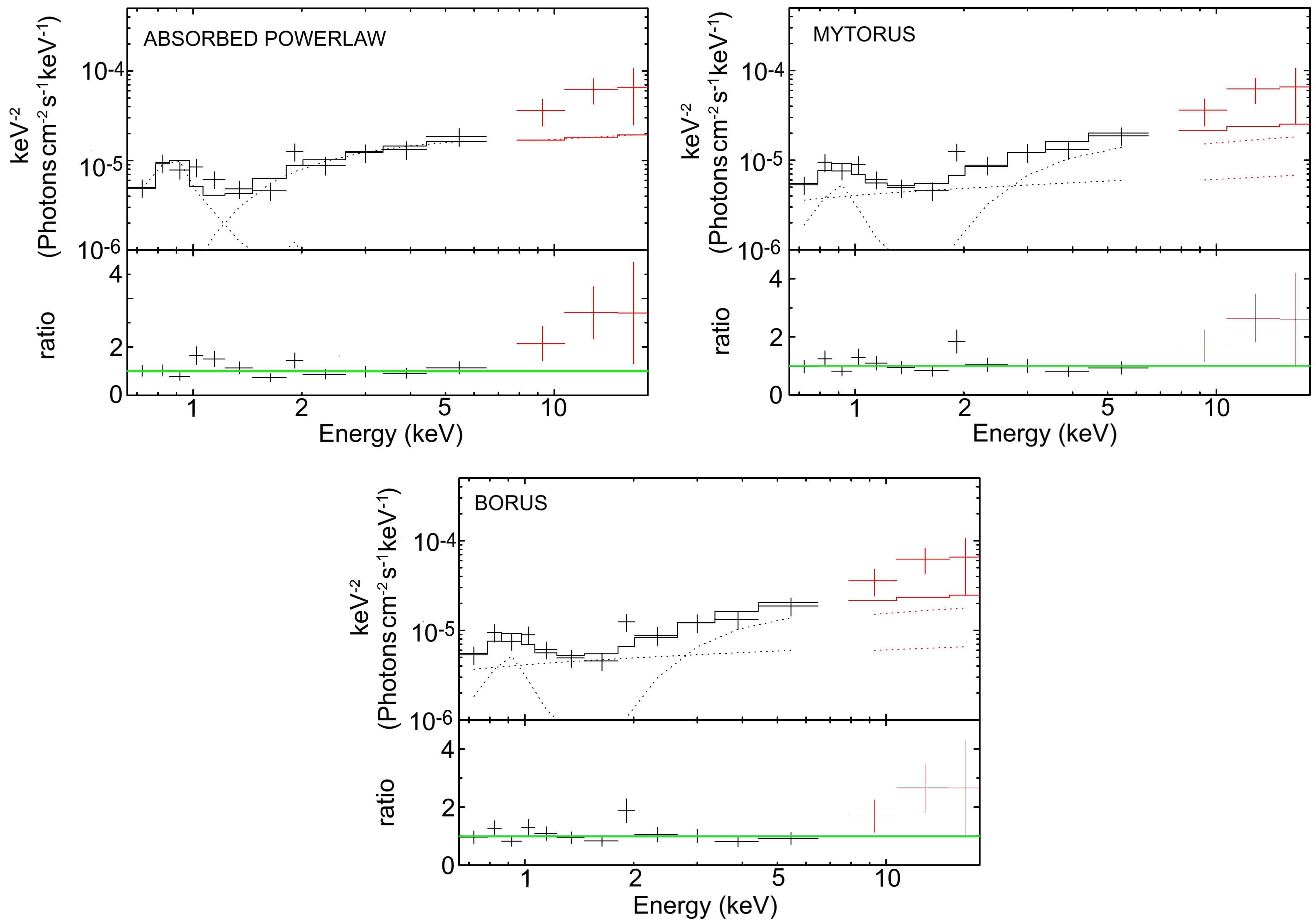}
\caption[
  NGC3627
]{
 \emph{Top}: \textsl{Chandra} and \textsl{NuSTAR} RGB images of NGC 3627. \emph{Bottom}: Best-fitting absorbed power-law (top left), MY{\sc{torus}} (top right) and B{\sc{orus}} (bottom) models to the combined \textsl{NuSTAR} (red) and \textsl{Chandra} (black) data. Figure description is the same as Figure 4.
}

\end{center}
\end{figure*}

The \emph{XMM-Newton} data reveal three sources within 30$\arcsec$ of the centroid position of the galaxy (Figure 5). However, none is consistent with the \emph{Chandra} position of the AGN (i.e., RA $=$ 11:18:16.51 and Dec. $=$ $-$32:48:50.78 in the 0.5--8 keV band). Comparing the \emph{Chandra} and \emph{XMM-Newton} images of the galaxy, we found that one of the two off-nuclear sources detected in the older \emph{Chandra} data (ON 2) was not detected in the more recent \emph{XMM-Newton} observation. However, two new off-nuclear sources (ON 3 and 4) emerge in the \emph{XMM-Newton} observation that were not detected in the \emph{Chandra} data. Source ON 1 is clearly visible in the \emph{NuSTAR} data, whilst weak and diffuse emission can be seen at the ON 3 and 4 positions.

\begin{figure*}
\begin{center}
\includegraphics[scale=0.32]{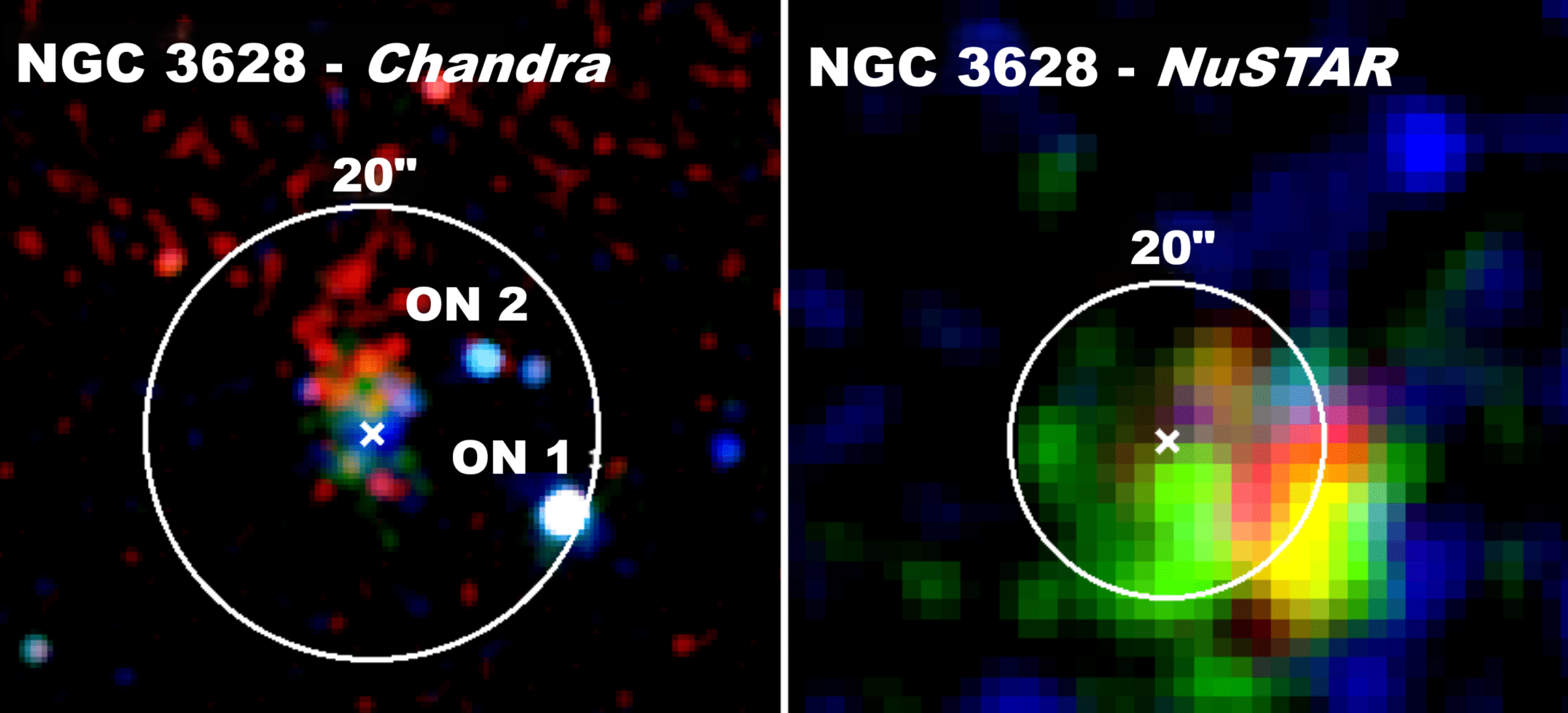}\\
\includegraphics[scale=0.31]{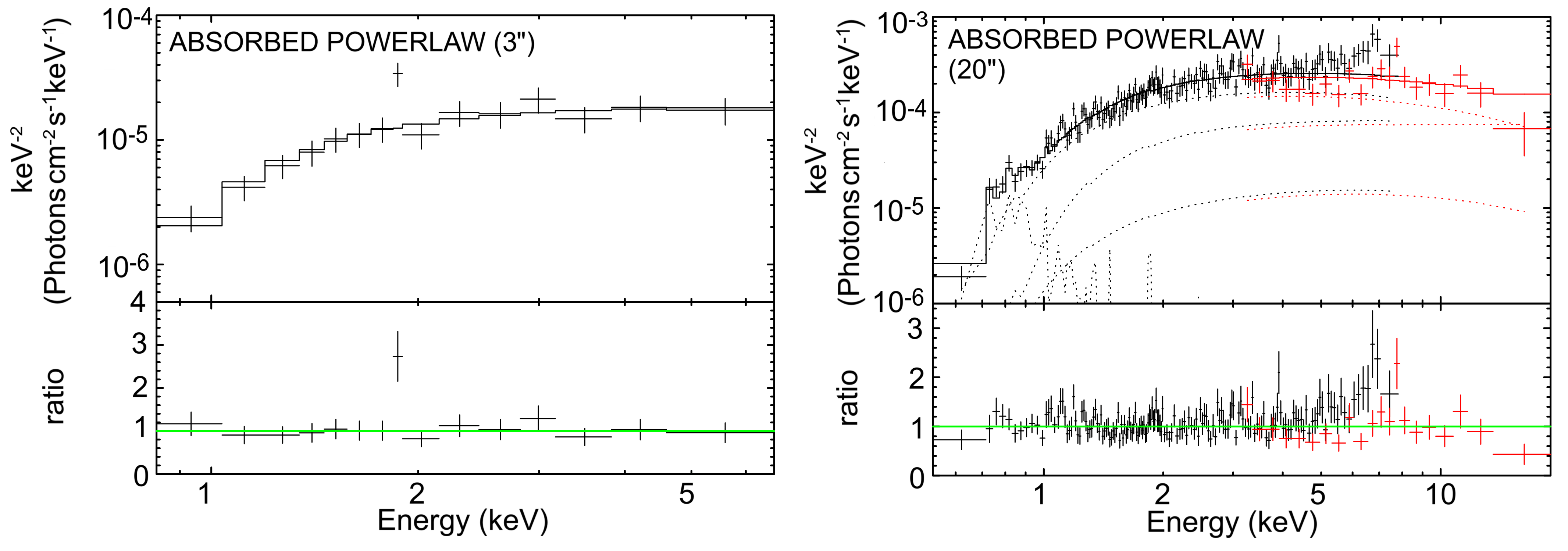}
\caption[
  NGC3628
]{
 \emph{Top}: \textsl{Chandra} and \textsl{NuSTAR} RGB images of NGC 3628. The off-nuclear sources which were detected significantly detected within the 20$\arcsec$-radius extraction region in \textsl{Chandra} are labelled as ON 1 and ON 2. \emph{Bottom}: Best-fitting absorbed power-law model to the \textsl{Chandra} spectrum extracted from a small 3$\arcsec$ radius region (left), and to the combined \textsl{NuSTAR} (red) and \textsl{Chandra} data (black) extracted with a larger 20$\arcsec$ {radius} region. The spectra were fitted using an absorbed power-law model to simulate the AGN emission, and two cut-off power law components to model the off-nuclear sources, ON 1 and ON 2. Figure description is the same as Figure 4.
}
\end{center}
\end{figure*}

Similar to NGC 1792, we performed X-ray spectral analysis on the NGC 3621 data even though it is not significantly detected in both of our \emph{XMM-Newton} and \emph{NuSTAR} data. For this source however, we chose a larger extraction region of 30$\arcsec$ to fully include the off-nuclear source ON 1 emission. Based on our modelling of the off-nuclear sources, we found that a simple absorbed power-law can best describe ON 1 and ON 4, whilst an absorbed cut-off power-law is the best-fitted model for ON 3. However, when we tried to incorporate these components into our modelling to decompose the AGN emission using a simple absorbed power-law model, we could not constrain the parameters. We therefore just modelled the total spectra using an absorbed cutoff power-law model as it clearly turns down at high energy. Indeed, our model measured a cut-off energy of $E \sim$ 12 keV. The photon index and column density measured towards the AGN are $\Gamma =$ 1.60 $\pm$ 0.10 and $N_{\rm H} =$ (3.5 $\pm$ 1.9) $\times$ 10$^{20}$ cm$^{-2}$, respectively, suggesting that the AGN is unobscured, with a 2--10 keV intrinsic luminosity of 4.37 $\times$ 10$^{39}$ erg s$^{-1}$.

The measured $N_{\rm{H}}$ value is consistent with that constrained by the \emph{Chandra} observation using a much smaller aperture region of 3$\arcsec$, clearly isolating the AGN emission from any off-nuclear sources. Based on our analysis of the \emph{Chandra} data using a simple power-law model with a photon index 1.8, we measured a column density upper limit towards the AGN of $N_{\rm H} \leq$ 1.8 $\times$ 10$^{21}$ cm$^{-2}$ with 2--10 keV intrinsic luminosity of 3.39 $\times$ 10$^{37}$ erg s$^{-1}$. As we believe that the results from the \emph{Chandra} data are more reliable (due to actual detection of the AGN and lack of contamination), we therefore used the results from these data throughout this paper (see Table~3). In Figure 5, we show the best-fit spectra using the \emph{Chandra}, \emph{XMM-Newton} and \emph{NuSTAR} data.

\subsection{NGC 3627}

NGC 3627 is a a spiral galaxy (Sb) located at a distance of 10 Mpc. It is a member of the Leo triplet galaxies and is in tidal interaction with NGC 3623 and NGC 3628. In the optical, the AGN has been variously identified as a LINER (e.g., \citealp{Veron06}; GA09), transition object (e.g., \citealp{Dudik05}) and Seyfert 2 (e.g., \citealp{Brightman11}), depending on the diagnostics used. In X-rays, the presence of an AGN has also been in debate, mainly due to the lack of a clearly resolved nuclear point source and faint flux emission in \emph{Chandra} data (e.g., \citealp{Panessa06}; \citealp{Cisternas13}). It is also not detected in the \emph{Swift}-BAT survey \citep{Oh18}. However, the high ionization [Ne {\sc{v}}] line was detected by GA09, providing strong evidence for the presence of an AGN in the galaxy. Using high angular resolution mid-IR observations by VLT-VISIR, complex extended emission was detected in the nuclear region, in which a compact source cannot be clearly identified \citep{Asmus14}.

\cite{GonzalezMartin09} suggested that NGC 3627 is a CT AGN candidate on the basis of its $L_{\rm 2-10}/L_{\rm [O III]}$ ratio using the X-ray luminosity estimated from \emph{XMM-Newton} observation (see also Figure 12). In contrast, using the same data, \cite{Brightman08} measured an ionized absorption column density of $\approx$5.0 $\times$ 10$^{21}$ cm$^{-2}$ toward the source, suggesting that it might actually be an unobscured AGN.

\subsubsection{X-ray observations and spectral fitting}

NGC 3627 has been observed multiple times in X-rays using, e.g., \emph{Chandra} and \emph{XMM-Newton}. We acquired a \emph{NuSTAR} observation of the galaxy in 2017 (2017-12-20; ObsID 60371003002) due to evidence from multiwavelength diagnostics that it could be CT (e.g., \citealp{GonzalezMartin09}; see also Section 5.1, and Figures 12 and 13). Our \emph{NuSTAR} data were fitted by \cite{Esparza20} using a simple model partial covering absorber component. Based on their analysis, they measured a CT column density of $N_{\rm H} =$ (1.8 $\pm$ 6.7) $\times$ 10$^{24}$ cm$^{-2}$. After correcting for this absorption, they estimated an intrinsic 2--10 keV luminosity of $L_{\rm 2-10, int} <$ 1.58 $\times$ 10$^{42}$ erg s$^{-1}$. When compared to its [O {\sc{iii}}] and 12$\mu$m luminosities, they found the source to be underluminous in X-ray, suggesting that the AGN is in the early fading stage of the AGN duty cycle. This is supported by \cite{Saade22} who also analysed the \emph{NuSTAR} data of the AGN together with \emph{Chandra} data to account for multiple brighter off-nuclear sources within the \emph{NuSTAR} beam (that could also heavily contaminate the \emph{XMM-Newton} data analysed by previous studies). In contrast however, they did not find evidence for obscuration due to the absence of a reflection component, and therefore did not provide a column density measurement. Given the significantly low intrinsic luminosity of the AGN in X-ray as compared to the mid-IR, they also suggest that the AGN is fading and recently deactivated at least $\sim$90 years ago. 

Here, we re-analyze the \emph{NuSTAR} data, together with archival \emph{Chandra} data to provide a measurement of its column density using physically motivated torus models. Due to the complexity of the central region of the galaxy, with multiple off-nuclear sources near the AGN, we took a different approach in analyzing the AGN spectra for this source. We extracted the \emph{NuSTAR} spectra using a circular region with 30$\arcsec$ radius, whilst for the \emph{Chandra} data, we isolated the AGN emission using a smaller circular region of 2$\arcsec$ radius. This is because the AGN emission starts to dominate over most of the off-nuclear sources in the energy in the high energy band of the \emph{Chandra} data. The \emph{NuSTAR} data also seem to be consistent with the \emph{Chandra} data if we were to extrapolate it into the \emph{NuSTAR} band (see \citealp{annuar20} for a similar case in NGC 660). Based on a simple modeling approach using an absorbed power-law model, we measured a column density of $N_{\rm H} =$ (1.3 $\pm$ 0.7) $\times$ 10$^{22}$ cm$^{-2}$. We therefore proceed with the MY{\sc{torus}} and B{\sc{orus}} models which indeed confirm a column density of this magnitude, indicating a mildly obscured AGN. These results broadly agree with the conclusion of \cite{Saade22} and argue against NGC 3627 harbouring a CTAGN. The 2--10 keV intrinsic luminosity of the AGN is $\sim$ 5 $\times$ 10$^{38}$ erg s$^{-1}$ (see Table 3).

\subsection{NGC 3628} 
 
NGC 3628 is an edge-on Sb spiral located at $D =$ 10 Mpc with distorted dust lanes due to its interaction with the other two galaxies in the Leo Triplet; i.e, NGC 3623 and NGC 3627. The nucleus is classified as LINER in the optical (e.g., GA09), and has been detected in radio (e.g., \citealp{Filho00}; \citealp{Nagar05}) and \emph{Chandra} \citep{GonzalezMartin09}. However, the \emph{Chandra} image shows a diffuse central source rather than a point source, which led \cite{GonzalezMartin09} to {infer} that the galaxy does not host an AGN. It is also not detected at 12$\mu$m using high angular resolution observations by VLT-VISIR, and \cite{Asmus14} therefore concluded that any AGN contribution to the mid-IR emission of the central $\sim$0.2 kpc is minor. However, GA09 {detected} the [Ne {\sc{v}}] line emission at a 3$\sigma$ significance level from the galaxy, indicating the presence of an AGN. The AGN has not been detected in the \emph{Swift}-BAT survey \citep{Oh18}. 

\subsubsection{X-ray observations and spectral fitting}

NGC 3628 has been observed multiple times in X-rays by {\it Chandra} and {\it XMM-Newton}. These data show several nearby, off-nuclear sources which dominate at low energies. There is a lack of a strong indication that it is highly obscured. However, motivated by the lack of reliable high-energy archival data, we obtained a {\it NuSTAR} observation of NGC 3628 in 2017 (2017-12-23; ObsID 60371004002). {The} \emph{NuSTAR} data {were} analysed by \cite{Esparza20} using a simple model which include{s} a partial covering absorber component. They found a column density of $\sim$ 2 $\times$ 10$^{23}$ cm$^{-2}$, indicating a heavily obscured AGN. \cite{Osorio22} also analysed the \emph{NuSTAR} data using reflection models (i.e., {\sc{pexrav}} and {\sc{pexmon}}). However, they did not {find} any signs of reflection in the spectrum, and measured a column density limit of $N_{\rm H} < 1.2 \times 10^{22}$ cm$^{-2}$, indicating that the AGN is most likely unobscured. 

\begin{figure}
\begin{center}
\includegraphics[scale=0.41]{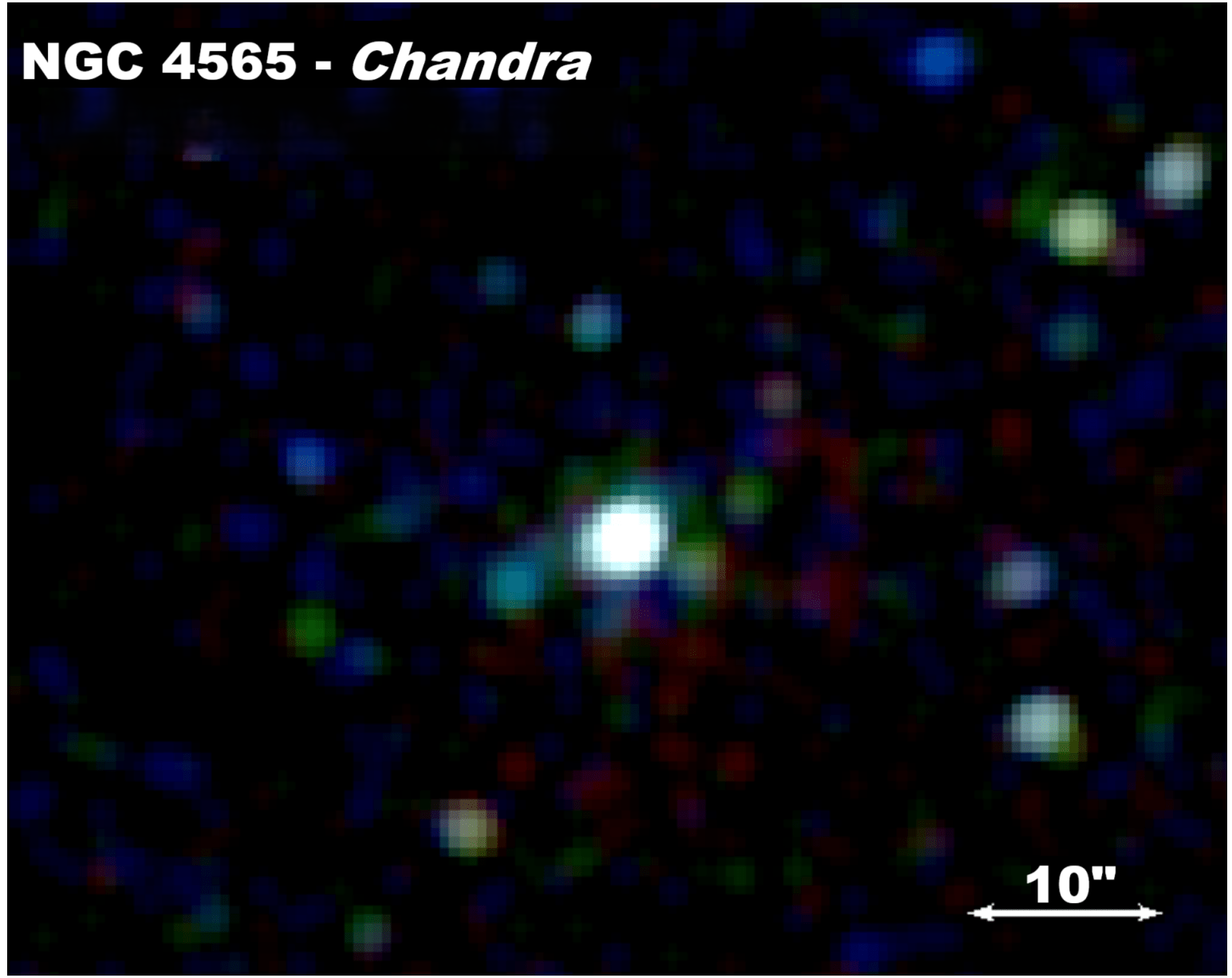}
\includegraphics[scale=0.31]{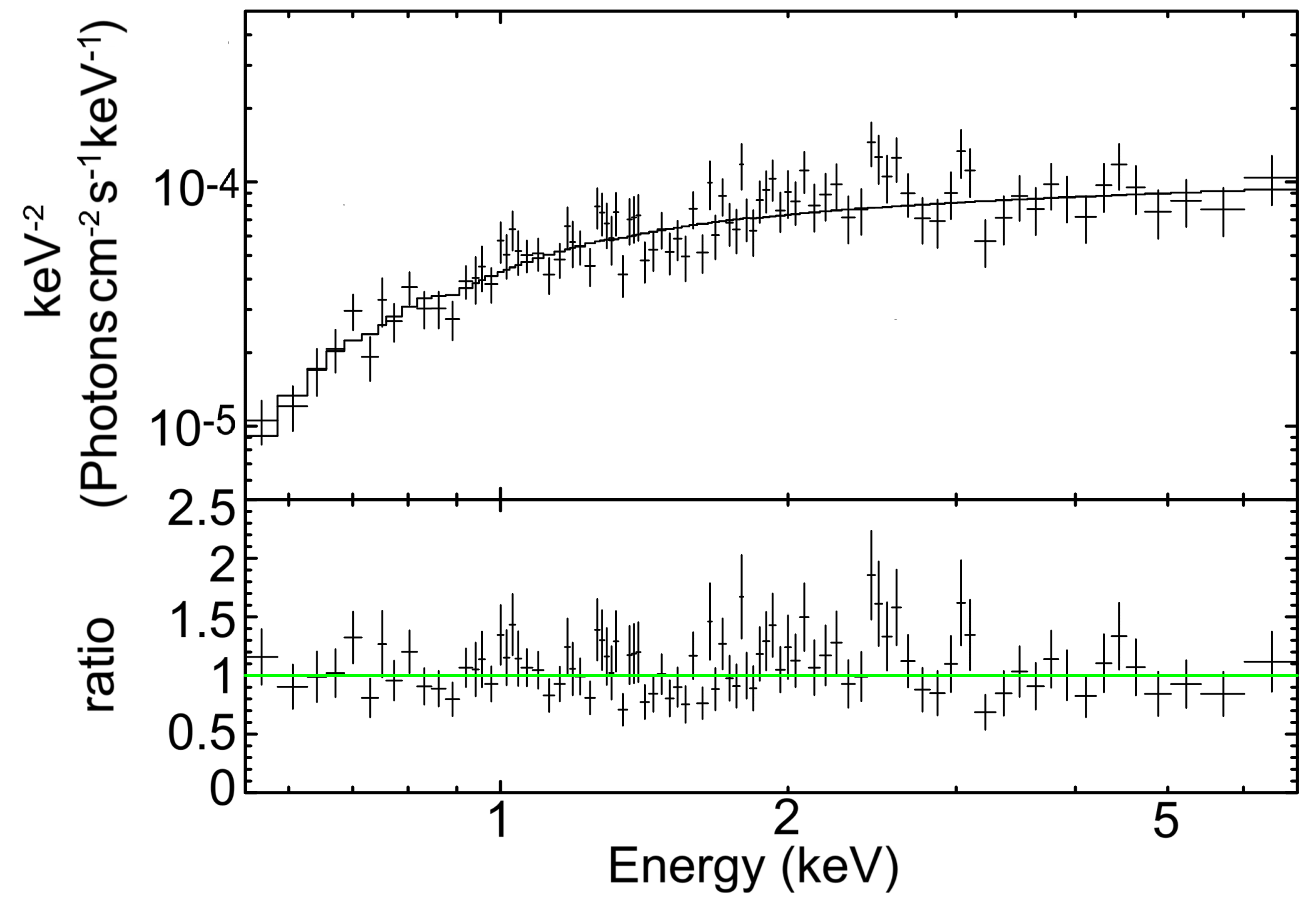}
\caption[
  NGC4565
]{
\emph{Top}: \textsl{Chandra} RGB image of NGC 4565. \emph{Bottom}: Best-fitting absorbed power-law model to the data. Figure description is the same as Figure 4.
}
\end{center}
\end{figure}

We analyze the \emph{NuSTAR} data again, together with its archival \emph{Chandra} data to {help account} for off-nuclear source contributions in the \emph{NuSTAR} spectrum that could significantly affect the results presented by both \cite{Esparza20} and \cite{Osorio22} due to the relatively large extraction region used in these studies (1{$\arcmin$} and 2$\arcmin$, respectively). The \emph{Chandra} data {reveal} that the AGN is embedded {in diffuse} emission with several off-nuclear point sources detected at 0.5--8 keV band within a 20$\arcsec$-radius of the AGN, of which two are detected in the 2--8 keV band (see Figure 7). Whilst one of these high-energy emitting off-nuclear sources is visible in the \emph{NuSTAR} data (ON 1), the AGN seems to be undetected. We firstly analyzed the source spectra extracted from a 20$\arcsec$-radius circular region using a simple absorbed power-law model. Based on this, we measured a column density of $N_{\rm H} =$ (1.3 $\pm$ 0.6) $\times$ 10$^{22}$ cm$^{-2}$, with $L_{\rm 2-10, int} =$ 6.92 $\times$ 10$^{39}$ erg s$^{-1}$. This is in agreement with that measured using just the \emph{Chandra} data with a smaller aperture region of 5$\arcsec$ {radius}, minimizing contamination from much of the off-nuclear sources; i.e., $N_{\rm H} =$ (7.4 $\pm$ 3.6) $\times$ 10$^{21}$ cm$^{-2}$, with $L_{\rm 2-10, int} =$ 5.24 $\times$ 10$^{38}$ erg s$^{-1}$.  As with NGC 3621, we {adopt} the results from the \emph{Chandra} data for this source throughout this paper as well as {they are} more reliable (see Table 3).

\begin{landscape} 
\begin{table}
  \caption{X-ray spectral fitting results.}
   \resizebox{\columnwidth}{!}{%
    \begin{tabular}{cccccccccc}
  \hline 
   Name & Facility & Model & Energy Band & $\Gamma$ & $\log N_{\rm H}$ & $\log{L_{\rm 2-10, obs}}$ &  $\log{L_{\rm 2-10, int}}$  & $\chi^{2}$ or C-stat / d.o.f & Ref. \\
             &   &   & [keV] & & [cm$^{-2}$] & [erg s$^{-1}$] & [erg s$^{-1}$] & &  \\
   (1) & (2)  & (3) & (4) & (5) & (6) & (7) & (8) & (9) & (10)  \\
   \hline
             &   &   &    & This work & & & & & \\
    \hline
   NGC 0613 & \it{C}  & ({\sc{zwabs}}*{\sc{zpow}})+{\sc{zgau}} & 0.5--8  & 2.24$^{+1.29}_{-1.06}$  
   & 23.58$^{+0.14}_{-0.16}$  
   & 40.47 & 41.05
   & 89/88 ($\chi^{2}$) 
   & ...  \\ 
           &    \it{C}     & B{\sc{orus}} & 0.5--8 & 1.8$^{f}$ & 23.52$^{+0.05}_{-0.04}$ & 40.49 & 41.05 & 80/74 ($\chi^{2}$) & ... \\
           &    \it{C}     & MY{\sc{torus}} & 0.5--8 & 2.03$^{+0.22}_{-0.28}$ & 23.54$^{+0.05}_{-0.04}$ & 40.47 & 41.12 & 105/82 ($\chi^{2}$) & ... \\
   NGC 1792$^{a}$ & \it{C} + \it{N}  & {\sc{zwabs(zcutoffpow)}}  &  0.5--24  & 1.80$^{f}$ & $\leq$22.39  & $\leq$38.80  & $\leq$38.80 & 99 / 129 (C-stat)  & ...   \\  
   NGC 3621 & {\it{XMM}} + {\it{N}}     & {\sc{zwabs(zcutoffpow)}}  & 0.5--24  & 1.60 $\pm$ 0.10  & 20.54$^{+0.19}_{-0.33}$  & 39.64 & 39.83 & 457 / 423 ($\chi^{2}$) & ... \\ 
            & {\it{C}} & {\sc{zwabs(zpow)}} & 0.5--8 & 1.80$^{f}$ & $\leq$ 21.25 & 37.53 & 37.53 & 20 / 21 (C-stat) & ... \\
   NGC 3627 & {\it{C}} + {\it{N}} & {\sc{zwabs(zpow)}}  & 0.5--24  & 1.80$^{f}$   & 22.10$^{+0.18}_{-0.24}$  & 38.64  & 38.66  & 22 / 13 ($\chi^{2}$) & ... \\ 
    & {\it{C}} + {\it{N}} & B{\sc{orus}}  & 0.5--24  & 1.80$^{f}$   & 22.71$^{+0.99}_{-0.42}$  & 38.70  & 38.72  & 16 / 12 ($\chi^{2}$) & ... \\ 
    & {\it{C}} + {\it{N}} & MY{\sc{torus}}  & 0.5--24  & 1.80$^{f}$   & 22.67$^{+1.02}_{-0.39}$  & 38.70  & 38.72  & 16 / 12 ($\chi^{2}$) & ... \\ 
   NGC 3628 & \it{C} + \it{N}  & {\sc{zwabs(zpow)}}  & 0.5--24  & 2.01$^{+0.29}_{-0.24}$ & 22.10$^{+0.16}_{-0.17}$ & 39.84 & 39.86 & 218 / 181 ($\chi^{2}$) & ... \\ 
& \it{C} & {\sc{zwabs(zpow)}}  & 0.5--8  & 2.00$^{+0.44}_{-0.39}$ & 21.87$^{+0.17}_{-0.19}$ & 38.72 & 38.73 & 12 / 12 ($\chi^{2}$) & ... \\ 
   NGC 4565 & \it{C}  & {\sc{zwabs(zpow)}}   &  0.5--8 & 1.88 $\pm$ 0.04 &  21.34$^{+0.08}_{-0.09}$ & 39.43  & 39.44 & 90/86 ($\chi^{2}$) & ... \\  
    \hline
             &   &   &    & Past studies & & & & & \\
    \hline
   Circinus & \it{C} + \it{XMM} + \it{N}   & MY{\sc{torus}} & 2--79  & 2.40 $\pm$ 0.03  & 24.87$^{+1.68}_{-0.20}$ &  40.45    & 42.57  & 2785 / 2637 ($\chi^{2}$)  & \cite{Arevalo14}  \\ 
   ESO121-G6 & \it{C} + \it{N}  & BN{\sc{torus}}  & 0.5--50  & 1.89$^{+0.11}_{-0.06}$  & 23.29 $\pm$ 0.02  & 40.53 & 41.01  & 368 / 317 ($\chi^{2}$) & \cite{annuar20}  \\ 
   NGC 0660$^{b}$ & \it{C} + \it{N}   & MY{\sc{torus}}  & 0.5--30 & 1.8$^{f}$  & 23.78$^{+0.18}_{-0.22}$  & 39.07 & 39.76 & 181 / 186 (C-stat) & \cite{annuar20} \\  
   NGC 1068 &  \it{C} + \it{XMM} + \it{N} + BAT &  MY{\sc{torus}}  & 2--195  & 2.10$^{+0.06}_{-0.07}$  & 25.00$^{+u}_{-0.01}$  &  41.21  & 43.30 & 1899.2 / 1666 ($\chi^{2}$) &  \cite{Bauer15}   \\ 
   NGC 1448 & \it{C} + \it{N}   & MY{\sc{torus}} & 0.6--40  & 1.9$^{f}$ & 24.65$^{+u}_{-0.22}$  & 38.95  & 40.88  & 429 / 440 (C-stat) & \cite{Annuar17} \\ 
   NGC 4051 & \it{N}   &  MY{\sc{torus}} &  2--70 & 2.33 $\pm$ 0.05  & 20.06$^{+0.05}_{-0.04}$ & 41.09   & 41.09 & 569/537 ($\chi^{2}$) & \cite{Turner17} \\ 
   NGC 4945 & \it{C} + \it{S} + \it{N}  & MY{\sc{torus}}  & 0.5--79 & 1.96 $\pm$ 0.07   & 24.54$^{+0.02}_{-0.01}$  & 39.48  & 42.76 & 1118 / 1055 ($\chi^{2}$) & \cite{Puccetti14} \\ 
   NGC 5033 & \it{XMM} + \it{N} + \it{BAT}  & {\sc{borus}} & 0.5--195  & 1.74 $\pm$ 0.02  & 20.00$^{+0.09}_{-u}$  & 41.13  & 41.13 & 1177/1058 ($\chi^{2}$) & \cite{Diaz23} \\
   NGC 5128 & \it{XMM} + \it{N}  & MY{\sc{torus}}    & 3--78   & 1.82 $\pm$ 0.01  & 23.04$^{+0.06}_{-0.01}$ &  41.73 & 42.00   & 1667 / 1536 ($\chi^{2}$) & \cite{Furst16}\\ 
   NGC 5194 & \it{C} + \it{N}  & MY{\sc{torus}}   & 0.6--50 & 1.8 $\pm$ 0.3  & 24.85$^{+0.15}_{-0.24}$ & 38.99  & 40.77  & 169.3 / 155 ($\chi^{2}$) & \cite{Xu16}\\ 
   NGC 5195 &  \it{C} + \it{N}   & {\sc{zwabs}}({\sc{zpow}})  & 0.5--24 & 2.12$^{+0.61}_{-0.23}$  & 22.07$^{+0.40}_{-0.81}$  &  38.80 & 38.82 &  226 / 243 (C-stat) & \cite{annuar20} \\ 
   NGC 5643 &  \it{C} + \it{XMM} + \it{N} + \it{BAT}  & MY{\sc{torus}} & 0.5--100  & 2.10$^{+0.04}_{-0.02}$   & 24.76$^{+u}_{-0.10}$  & 40.20 & 41.95 & 570 / 471 ($\chi^{2}$) & \cite{Annuar15}\\ 
   NGC 6300$^{c}$ & \it{N} & MY{\sc{torus}}  & 3--40 & 1.51$^{+0.04}_{-0.05}$  & 23.08$^{+0.01}_{-0.03}$ & 41.51 & 41.83 & 812 / 819 ($\chi^{2}$) & \cite{Jana20} \\
  \hline
     \end{tabular}
     }
  \begin{tablenotes}
   \small  
   \item \emph{Notes.} Column (1) AGN name; (2) X-ray facilities used in the analysis (BAT: \textsl{Swift}-BAT; C: \textsl{Chandra}; N: \textsl{NuSTAR}; XMM: \textsl{XMM-Newton}; XRT: \textsl{Swift}-XRT); (3) Best-fit models to the spectra; (4) Energy band used in the analysis in keV; (5) Best-fitting photon index; (6) Logarithm of the best-fitting line-of-sight column density measured in cm$^{-2}$; (7-8) Logarithm of the observed and absorption-corrected 2--10 keV luminosities, respectively, in erg s$^{-1}$; (9) Fit statistic results and approach; (10) Reference for the results.
   \item $^{f}$ fixed
    \item $^{a}$ The observed and intrinsic luminosity for NGC 1792 should be regarded as upper limits as the AGN was not detected in either of the \textsl{Chandra} or \textsl{NuSTAR} observations. However we also note that the AGN could be extremely CT, causing it to not be detected in both X-ray observations (see Section 4.2).
   \item $^{b}$ The column density and intrinsic luminosity for NGC 660 should be regarded as lower limits due to strong evidence for CT obscuration (see \citealp{annuar20}).
    \item $^{c}$ NGC 6300 has been demonstrated to be variable between different X-ray observations \citep{Jana20}. Here, we quote the results from the most recent \textsl{NuSTAR} data. We note, however, that the column density value has been consistent between all observations.
  \end{tablenotes}
\end{table} 
\end{landscape}

\subsection{NGC 4565}

NGC 4565 is an {edge-on} spiral galaxy (Sb) located at a distance of $D =$ 10 Mpc, with a nucleus classified as Seyfert 2 \citep{Ho97}. Despite the optical classification, the nuclear source was found to be unabsorbed by \cite{Chiaberge06} with a column density of $N_{\rm H} =$ (2.5 $\pm$ 0.6) $\times$ 10$^{21}$ cm$^{-2}$ and intrinsic luminosity of $L_{\rm 2-10, int} \approx$ 2.5 $\times$ 10$^{39}$ erg s$^{-1}$, measured using \emph{Chandra} data. The luminosity measured is comparable with that obtained by other studies using \emph{XMM-Newton} data (\citealp{Wu02}; \citealp{Cappi06}). In addition, \cite{Chiaberge06} found that the spectral energy distribution (SED) of the AGN shows no sign of a UV bump or thermal IR emission. The Eddington ratio determined for the AGN is also very low; i.e., $\sim$10$^{-6}$, and its position on the diagnostic planes for low-luminosity AGN suggests that the optical nucleus is disk-dominated, instead of jet-dominated. These pieces of collective evidences indicate that the AGN is undergoing low radiative efficiency accretion \citep{Chiaberge06}. The AGN has remained undetected in the \emph{Swift}-BAT survey \citep{Oh18}, and has not been observed at mid-IR using high angular resolution observations.

 \subsubsection{X-ray observations and spectral fitting}

The exact models used to fit the \emph{Chandra} data of NGC 4565 were not specified by \cite{Chiaberge06}. We therefore re-analysed the data for consistency with our modelling technique. The AGN has been observed by \emph{Chandra} twice, and in both data sets, the AGN was clearly detected, free from any contaminants, making its analysis straightforward. This source is one of the two AGN in our sample where we did not pursue a \emph{NuSTAR} observation (see Section 3). We combined the \emph{Chandra} data, and extracted the AGN spectrum from a 3$\arcsec$-radius circular aperture region. We then modelled it using a simple absorbed power-law model. The best-fit spectrum implies a column density of $N_{\rm H} \sim$ 2.2 $\times$ 10$^{21}$ cm$^{-2}$, in agreement with that obtained by \cite{Chiaberge06}, confirming that it is an unobscured AGN. The intrinsic luminosity measured is $L_{\rm 2-10, int} =$ 2.75 $\times$ 10$^{39}$ erg s$^{-1}$ (see Table 3). Figure 8 shows the \emph{Chandra} RGB images of the source and the best-fit spectrum.

\subsection{Summary}

In this section, we have presented the X-ray spectral analysis for the remaining six sources in the D $\leq$ 15 Mpc AGN sample, completing our X-ray analysis for the entire sample. Based on our analysis, we found that none of the six AGN is CT. Figure 9 shows the intrinsic 2--10 keV luminosity as a function of $N_{\rm{H}}$ for our sample and the \textsl{Swift}-BAT AGN sample. Based on the figure, we can deduce that all of our sources with $L_{\rm 2-10, int} <$ 10$^{40}$ erg s$^{-1}$ are unobscured or just mildly obscured; i.e., $N_{\rm H} \lesssim$ 10$^{22}$ cm$^{-2}$, with the exception of the CT AGN candidate, NGC 660, which has $L_{\rm 2-10, int} \gtrsim$ 5.8 $\times$ 10$^{39}$ erg s$^{-1}$ and $N_{\rm H} \gtrsim$ 6.0 $\times$ 10$^{23}$ cm$^{-2}$ (see Table 3 and \citealp{annuar20}). This may indicate that at the low-luminosity of $L_{\rm 2-10, int} <$ 10$^{40}$ erg s$^{-1}$, the AGN torus is not well developed, supporting findings by multiple past studies (e.g., \citealp{Elitzur06}; \citealp{Hoenig07}; \citealp{Elitzur09}; \citealp{Maoz05}; \citealp{Gonzalez-Martin17}). 

\begin{figure}
\centering
 \includegraphics[scale=0.52]{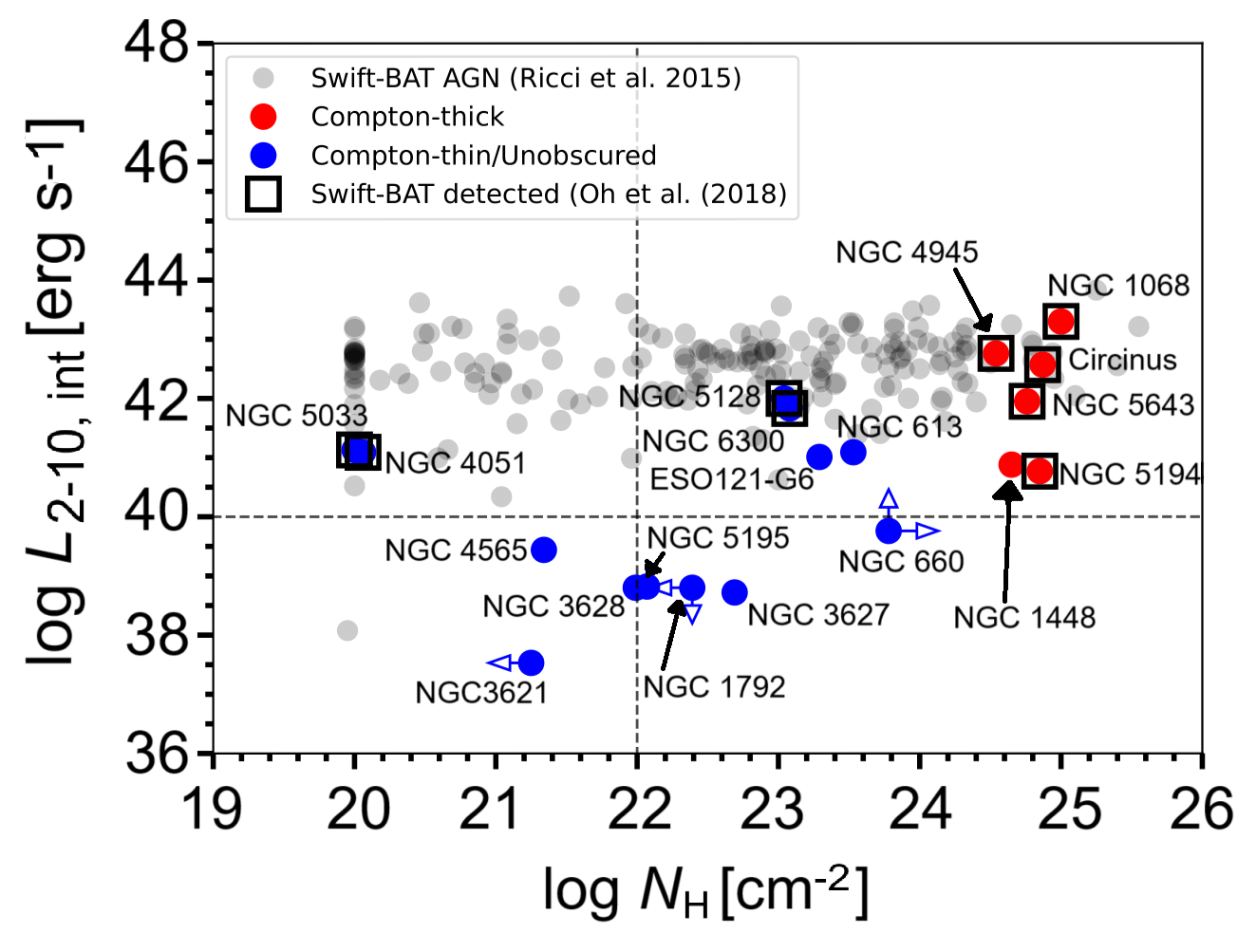}
\caption[
  Lxvsnh
]{
Intrinsic 2--10 keV luminosity vs. $N_{\rm{H}}$ for our sample and the \textsl{Swift}-BAT AGN at $D \leq$ 100 Mpc.\footref{footnote6} The dashed lines divide the low/high luminosity and obscured/unobscured AGN.
}
\end{figure}

\begin{figure}
\centering
\includegraphics[scale=0.53]{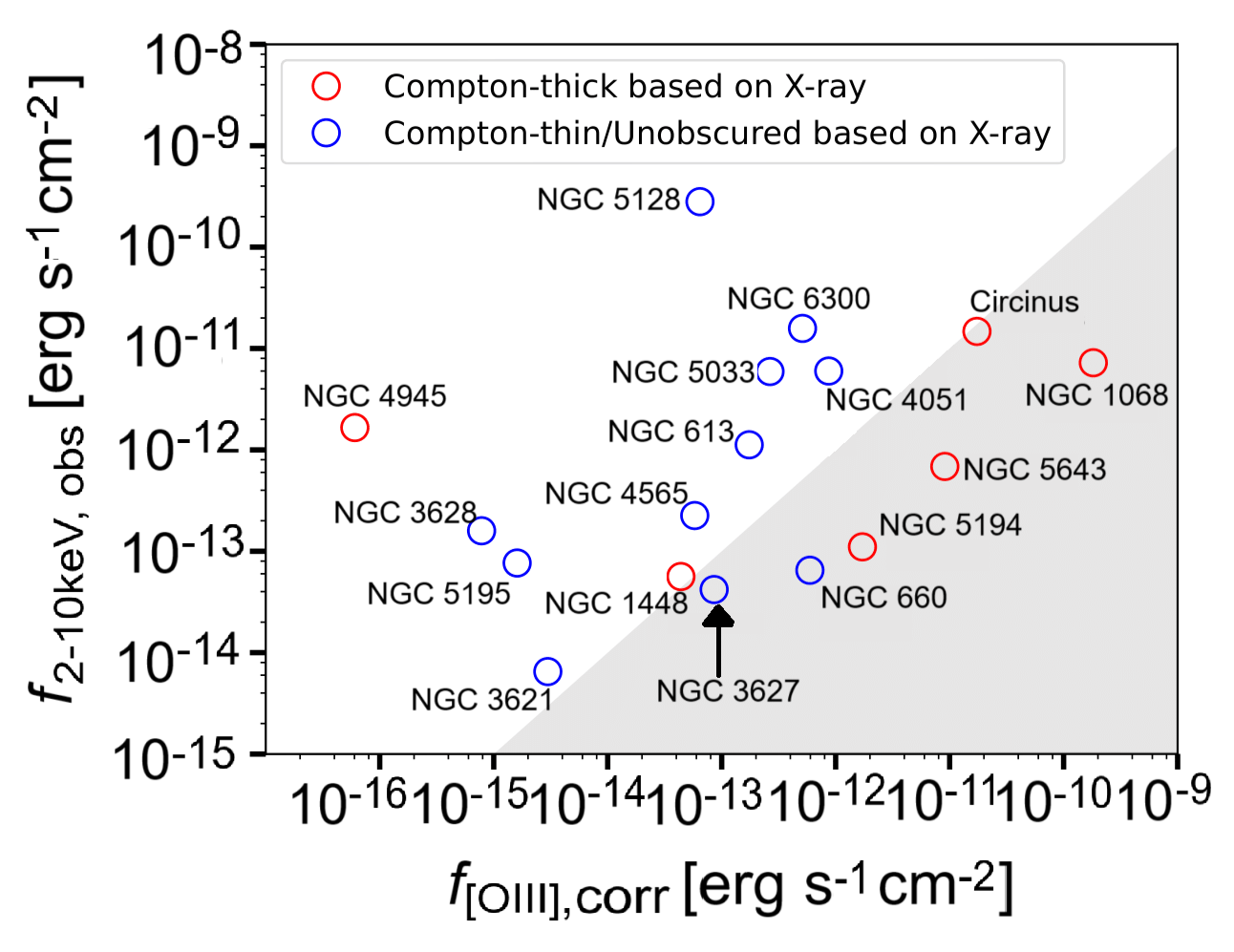}
\caption[
  oiii
]{
Observed 2--10 keV flux vs. {[O {\sc{iii}}]$\uplambda$5007\AA\ }flux corrected for the Balmer decrement plot for our sample. Red and blue circles marks Compton-thick and Compton-thin AGN on the basis of $N_{\rm{H}}$ measurements from X-ray spectroscopy, respectively. The grey area indicates a region where $f_{\rm 2-10, obs}$/$f_{\rm [OIII], corr}$ $\leq$1, which can be used as a CT AGN indicator \citep{Bassani99}. ESO 121-G6 and NGC 1792 are not plotted in the diagram as they lack good quality optical data.
}
\end{figure}

\section{Comparisons with multiwavelength data}

In this section, we discuss the multiwavelength properties of the AGN in our sample. We compare the observed and intrinsic 2--10 keV emission measured for the AGN from our X-ray analyses with their optical [O {\sc{iii}}]$\uplambda$5007\AA\ emission (Section 4.1), infrared 12$\micron$ and [Ne {\sc{v}}]$\uplambda$14.32$\micron$ luminosities (Section 4.2 and 4.3, respectively) in order to complement our X-ray analysis results.

\begin{figure*}
\centering
\includegraphics[scale=0.34]{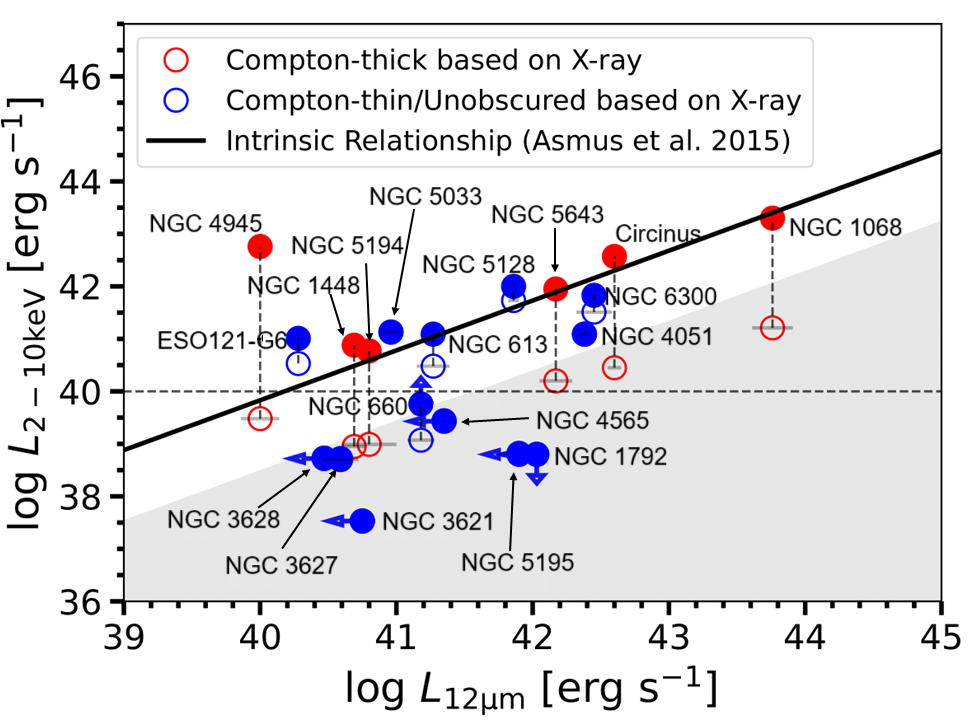} 
\includegraphics[scale=0.34]{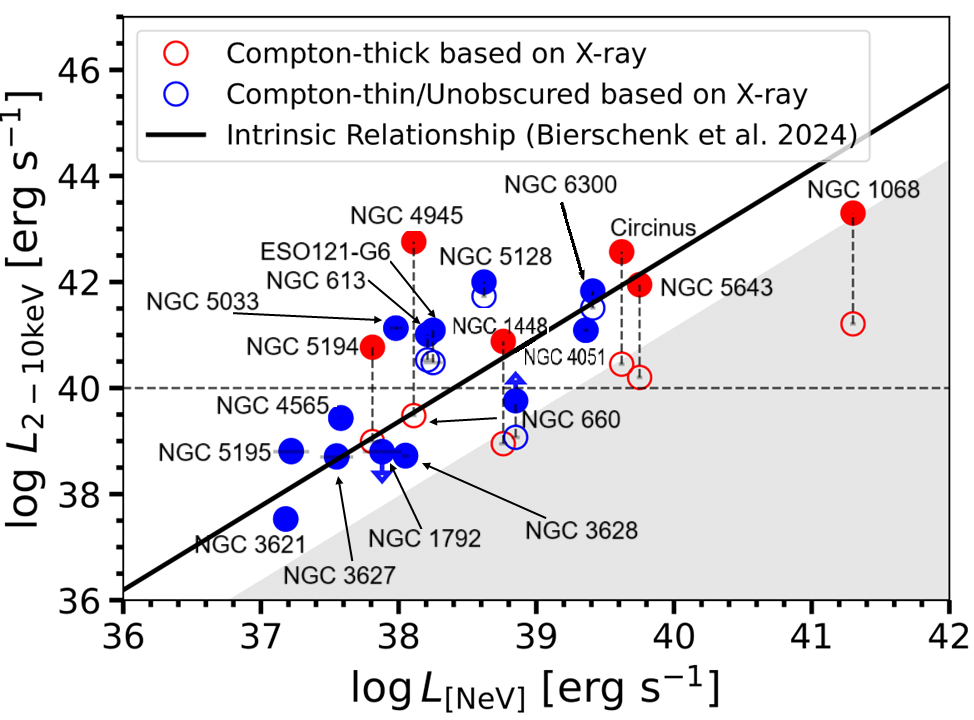}
\caption[
  Lxvsnev
]{
Observed 2--10 keV luminosities vs. 12$\mu$m (left) and {[Ne {\sc{v}}]} luminosities (right) for our sample. The solid lines for each plot correspond to the intrinsic relationships derived by \cite{Asmus15} (scatter, $\sigma \approx$ 0.3 dex) and \cite{Weaver10} (scatter, $\sigma \approx$ 0.5 dex), respectively. The symbols are the same as in Figure 10, with filled circles indicating the intrinsic 2--10 keV luminosities of the AGN. We identified sources which lie $>$ 25$\times$ below the relation (grey region) as CT AGN candidates, following detailed X-ray studies on CT AGN (e.g. \citealp{Iwasawa97}; \citealp{Matt00}; \citealp{Balokovic14}; \citealp{Annuar17}). The dashed lines divide the low/high luminosity AGN.
}
\end{figure*}

\subsection{Optical [O {\sc{iii}}]$\uplambda$5007\AA\ emission}

The [O {\sc{iii}}] emission-line in powerful AGN is mostly produced in the NLR due to photoionization from the central source, and is therefore considered to be a good indicator for the intrinsic flux of the AGN. Since the physical scale of this region extends beyond the torus, it does not suffer from nuclear obscuration like the X-ray emission. Optical emission from the NLR however, can be affected by extinction from the host-galaxy. Although in general this can be corrected for using the Balmer decrement (i.e., H$\alpha$/H$\beta$ flux ratio), in extreme cases, the host-galaxy obscuration can be so high that the optical Balmer decrement only provides a lower limit on the extinction (GA09). Figure 10 shows the comparison between the observed 2--10 keV ($f_{\rm 2-10, obs}$) and corrected [O {\sc{iii}}] flux ($f_{\rm [OIII], corr}$; Table 1) for our AGN sample. Using a sample of Seyfert 2 galaxies with good quality X-ray spectra and therefore reliable $N_{\rm{H}}$ measurements, \cite{Bassani99} found that all the CT AGN in their sample have an observed 2--10 keV and intrinsic [O {\sc{iii}}] flux ratio of $f_{\rm 2-10, obs}$/$f_{\rm [OIII], corr}$ $\lesssim$1.

Based on this diagnostic technique, we found that most of our X-ray identified CT AGN are selected as CT (4/6, 67$^{+33}_{-43}\%$), with the exception of NGC 1448 and NGC 4945. This is due to the fact that the optical emission from these two AGN suffers from significant absorption by their highly inclined host galaxies causing the Balmer decrement to underestimate the NLR extinction (see \citealp{Annuar17} and GA09, respectively). This consequently causes their [O {\sc{iii}}] luminosities to also be underestimated. Most of the X-ray identified Compton-thin or unobscured AGN (9/11, 82$^{+18}_{-39}\%$) lie outside the shaded region in Figure 10, suggesting that they are not CT, consistent with the results of their X-ray spectral analyses. The remaining sources; i.e., NGC 660 and NGC 3627, are selected as CT using this diagnostic. NGC 660 is found to be at least heavily obscured, and may be CT based upon our X-ray spectral analysis \citep{annuar20}. This therefore provides additional evidence in favor of this scenario. On the other hand, NGC 3627 has a Compton-thin column density and has been suggested to be a turned-off AGN (\citealp{Esparza20}; \citealp{Saade22}). 

\begin{figure}
\centering
 \includegraphics[scale=0.56]{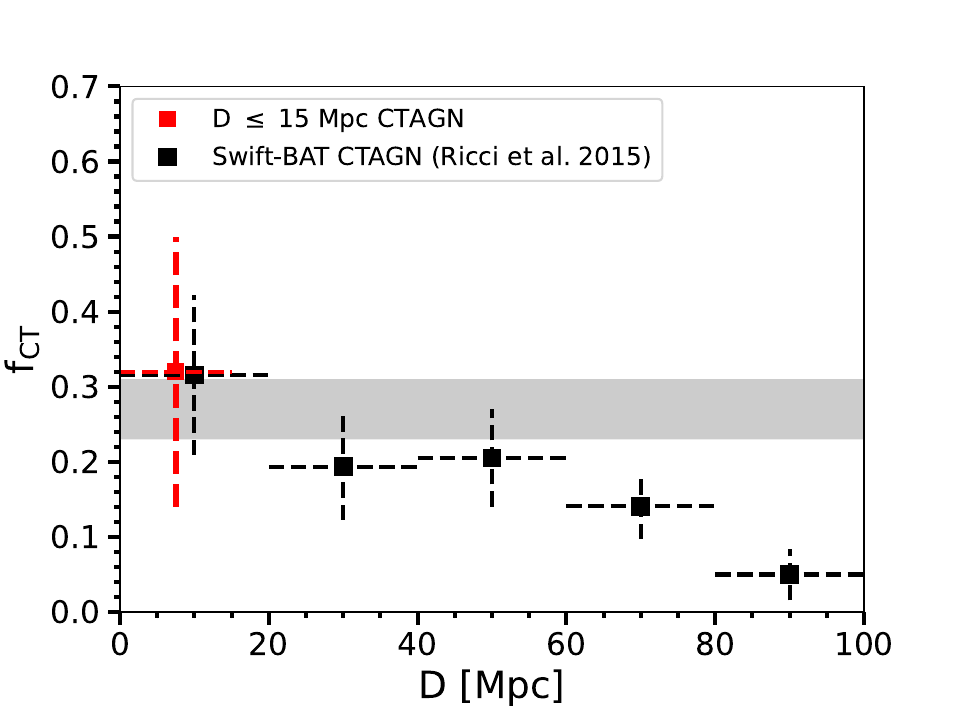}
 \includegraphics[scale=0.34]{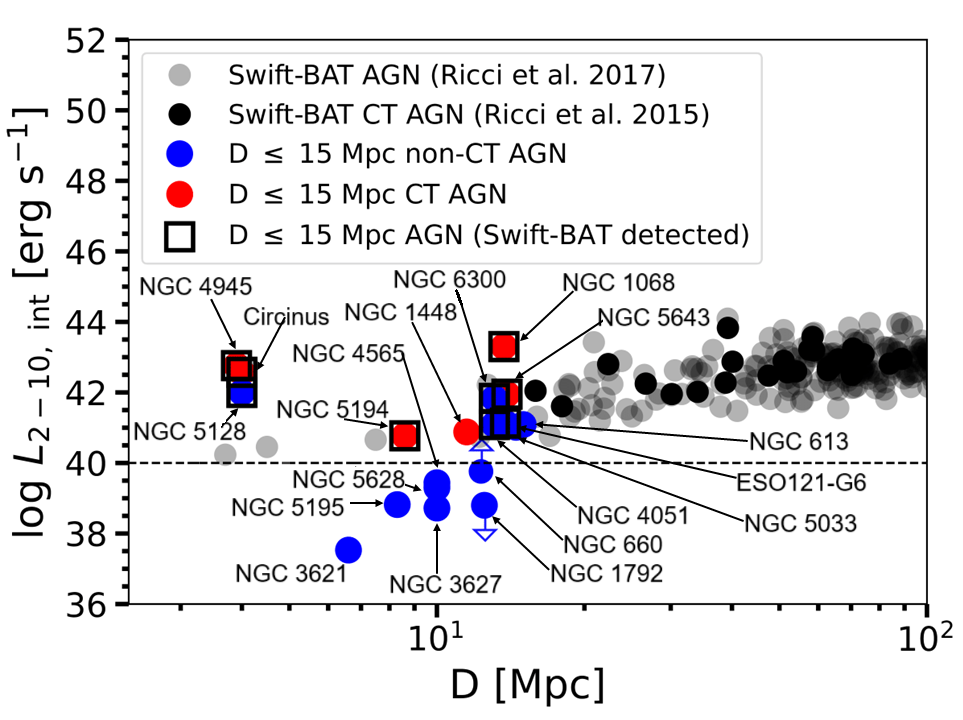}
\caption[
  Lxvsdct_all
]{
\emph{Top}: CT AGN fraction as a function of distance for our sample (red) and the \textsl{Swift}-BAT AGN located within 100 Mpc (black). The grey region shows the range of the intrinsic CT AGN fraction inferred using the whole \textsl{Swift}-BAT AGN sample (i.e., over the entire range of X-ray luminosity; log $L_{\rm 14-195} =$ 40--46 erg s$^{-1}$; \citealp{Ricci15}). \emph{Bottom}: Intrinsic 2--10 keV luminosities for the CT AGN in our sample\footref{footnote6} (red) and the entire \textsl{Swift}-BAT sample (grey) vs. distance up to $D =$ 100 Mpc. Squares indicate the CT AGN in our sample that are detected in the \textsl{Swift}-BAT survey. The dashed line divides the high/low luminosity AGN. {\it{Swift}}-BAT AGN within 15 Mpc that are not in our sample are discussed in Section 2.1.
}

\end{figure}

\subsection{Infrared 12{$\mu$m} emission}

We also compare the observed X-ray and 12 $\mu$m luminosities measured for our sources with the intrinsic X-ray:12$\mu$m correlation found by \cite{Asmus15}. The mid-IR continuum emission from AGN is produced by the obscuring torus (\citealp{Shi2014}) and/or polar dust structures, distributed along the ionization cones (\citealp{Asmus2016}; \citealp{Honig2019}). Therefore, it should provide a reliable estimate for the intrinsic luminosity of the AGN. The mid-IR emission can also be produced by dust around massive O-B stars; however, the emission from the AGN will typically dominate in this waveband and can be resolved, particularly when high-resolution imaging is used.

The \cite{Asmus15} X-ray:12$\mu$m intrinsic luminosities relationship has been shown to predict the intrinsic X-ray luminosity of AGN very well (see also \citealp{Horst08}; \citealp{Gandhi09}; \citealp{Mason12}). The relationship was derived using mid-IR data from high angular resolution mid-IR observations ($\sim$0$\farcs$4) of local Seyfert galaxies. As described earlier, for CT AGN, the X-ray emission that we observe is generally attributed to X-ray photons that are scattered or reflected from the back side of the torus or other circumnuclear material, which consists of just a few percent of the intrinsic power of the AGN in the 2--10 keV band (e.g. \citealp{Iwasawa97}; \citealp{Matt00}; \citealp{Balokovic14}; \citealp{Annuar17}). Therefore, we can use the \cite{Asmus15} X-ray:12$\mu$m relationship to identify AGN with observed 2--10 keV luminosities that deviate significantly from this intrinsic relationship, suggesting that they are likely to be CT  . We show the X-ray:12$\mu$m relationship by \citealp{Asmus15} in Figure 11 (left), with a grey region that we have adopted to select CT AGN candidates, representing a factor of 25$\times$ suppression of the X-ray flux (e.g., \citealp{Rovilos14}).

We obtain the majority of the 12 $\mu$m fluxes (15/19, 79$^{+21}_{-30}\%$) from high spatial resolution mid-IR observations (\citealp{Asmus14}; \citealp{Annuar17}; \citealp{annuar20}). The fluxes for the remaining four sources (NGC 1792, NGC 3621, NGC 4565 and NGC 5195) were obtained from {\it{WISE}} \citep{Wright10}, and are used as upper limits due to the lower angular resolution of {\it{WISE}}, which means the measurements can suffer significant contamination by the host-galaxy. Based on Figure 11, we find that most of our bona-fide CT AGN would be selected as CT on the basis of this diagnostic (5/6, 83$^{+17}_{-50}\%$), except for NGC 4945. This source is known to be an outlier, having fainter infrared emission than expected, which could be due to low torus covering factor (\citealp{Madejski2000}; \citealp{Done2003}; \citealp{Yaqoob2012}; \citealp{Marinucci2012}; \citealp{Puccetti14}; \citealp{Marchesi2019}; \citealp{Boorman2024}).

Six out of thirteen of our X-ray identified Compton-thin/unobscured AGN would also be selected as CT candidates based on this technique, although the majority are with upper limits for 12 $\mu$m measurements, making their classification uncertain on the basis of this method. We note that all of these sources are intrinsically low-luminosity AGN with $L_{\rm 2-10, int} <$ 10$^{40}$ erg s$^{-1}$. NGC 660, which may be CT based upon our X-ray spectral analysis \citep{annuar20} is located very near to the CT region. If we were to adopt the CT solution for this AGN with $L_{\rm 2-10, int} >$ 10$^{41}$ erg s$^{-1}$, it would be in good agreement with the intrinsic correlation \citep{annuar20}. These data provide further evidence that the AGN is most likely CT. On the other hand, NGC 3627, which is Compton-thin based on our X-ray analysis, is located at the edge of the CT region. However, this source has been suggested to be deactivated recently, contributing to the lower intrinsic X-ray luminosity measured from the corona, whilst the 12 $\mu$m emission from the much larger scale torus could continue to emit (\citealp{Esparza20}; \citealp{Saade22}). 

The fact that the sources that are misidentified as CT AGN have $L_{\rm 2-10, int} <$ 10$^{40}$ erg s$^{-1}$ (see Figure 11 -- left) might suggest a few scenarios: (1) the intrinsic correlations derived do not hold at lower AGN luminosities; or (2) flux contamination by non-AGN sources in the mid-infrared wavelength such as stellar activities; or (3) recent deactivation of AGN, contributing to the lower intrinsic X-ray luminosity measured from the corona, whilst the 12 $\mu$m emission from the much larger scale torus could continue to emit, such as the case for NGC 3627 (\citealp{Esparza20}; \citealp{Saade22}); or (4) the AGN are extremely heavily obscured (e.g., $N_{\rm H} \geq 10^{25}$ cm$^{-2}$) that their X-ray emission are significantly suppressed at all X-ray wavelengths, even the high energy, which could be the case for NGC 1792.

\subsection{Infrared [Ne {\sc{v}}]$\uplambda$14.32 $\mu$m emission}

Our parent sample from GA09 was derived using the detections of the high-ionisation [Ne {\sc{v}}] line as an unambiguous indicator of AGN activity in the galaxies. Thus, it should also be a reliable proxy for the intrinsic AGN power. As described earlier, the [Ne {\sc{v}}] line is also produced in the NLR, as with the [O {\sc{iii}}] emission-line. However, since it is produced in the mid-IR ($\lambda$ = 14.32 $\mu$m), it is less likely to be affected by extinction through the host-galaxy, unlike the optical [O {\sc{iii}}] line (see Figure 9 of GA09). Therefore, in addition to the two widely used CT AGN selection criteria described in previous sections, we also explore the use of the [Ne {\sc{v}}] line as an intrinsic AGN luminosity indicator and a tool to identify CT candidates.

We compared the observed 2--10 keV and [Ne {\sc{v}}] luminosities for our sample with the intrinsic correlation found by \cite{Bierschenk2024}, which was derived using the \textsl{Swift}-BAT AGN sample. Again, we classified those AGN with observed 2-10 keV luminosities lying more than 25$\times$ below this intrinsic relationship provided by \cite{Bierschenk2024} as likely to be CT. This is shown in Figure 11 (right). This technique managed to identify four out of six (67$^{+33}_{-43}\%$) of our confirmed CT AGN. The two CT AGN that were missed are NGC 4945 and NGC 5194. In fact, their observed X-ray luminosities are well in agreement with the intrinsic relationship, indicating that they would have been considered as unobscured if we were relying on this technique to estimate the AGN obscuration. Nevertheless, none of our Compton-thin/unobscured AGN are selected as CT using this method, including those with $L_{\rm 2-10, int} <$ 10$^{40}$ erg s$^{-1}$, with the exception of NGC 660. In fact for these sources, their luminosities are consistent with the established relationship, which could be used as evidence to support that the [Ne {\sc{v}}] emission detected by GA09 in this galaxies genuinely originate from AGN activities instead of other processes. Adopting the CT solution for NGC 660 would give an intrinsic 2--10 keV luminosity that is consistent with the established X-ray-[Ne {\sc{v}}] relationship, supporting this scenario.

\begin{figure*}
\centering
   \includegraphics[scale=0.54]{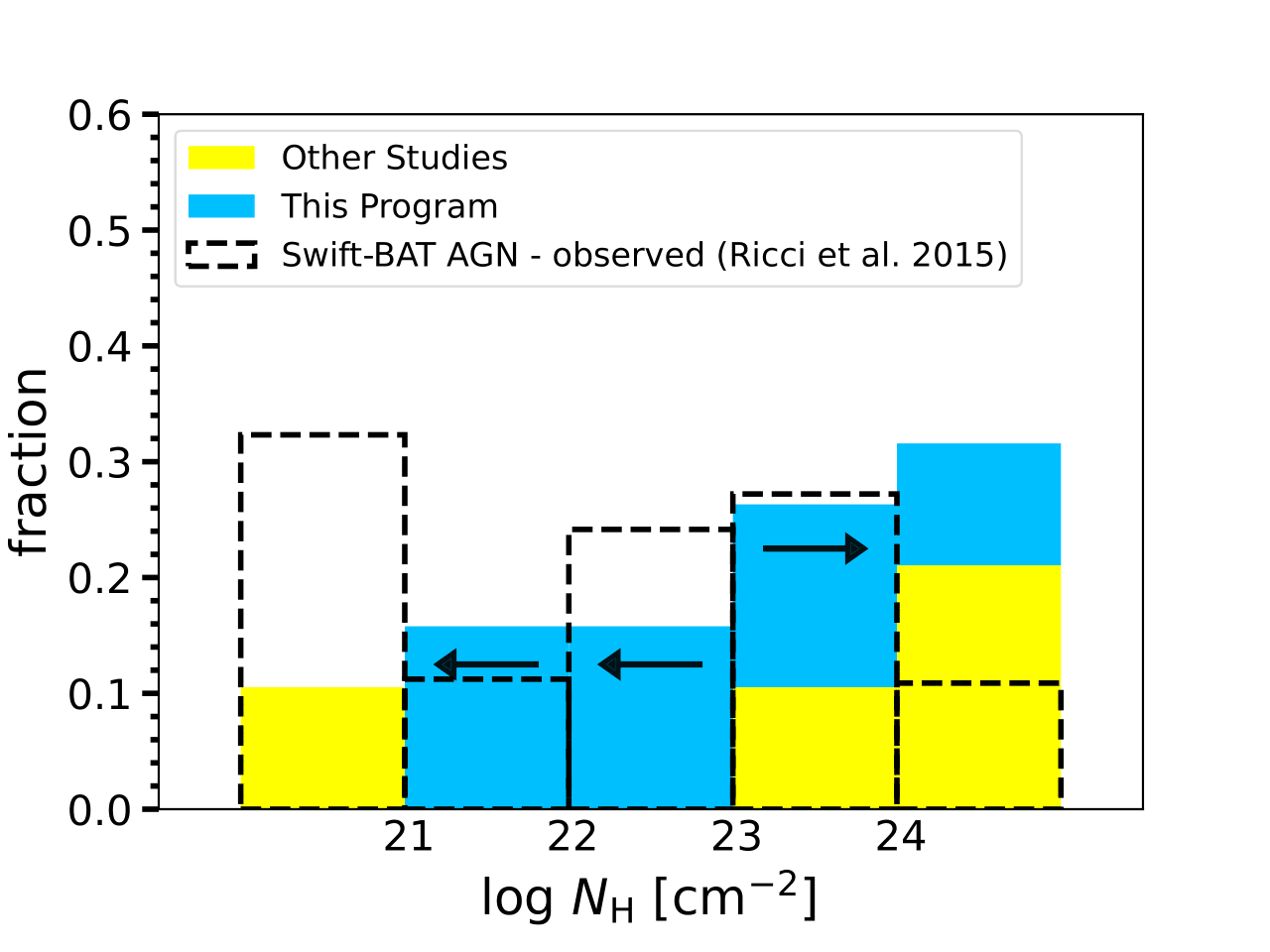}
    \includegraphics[scale=0.54]{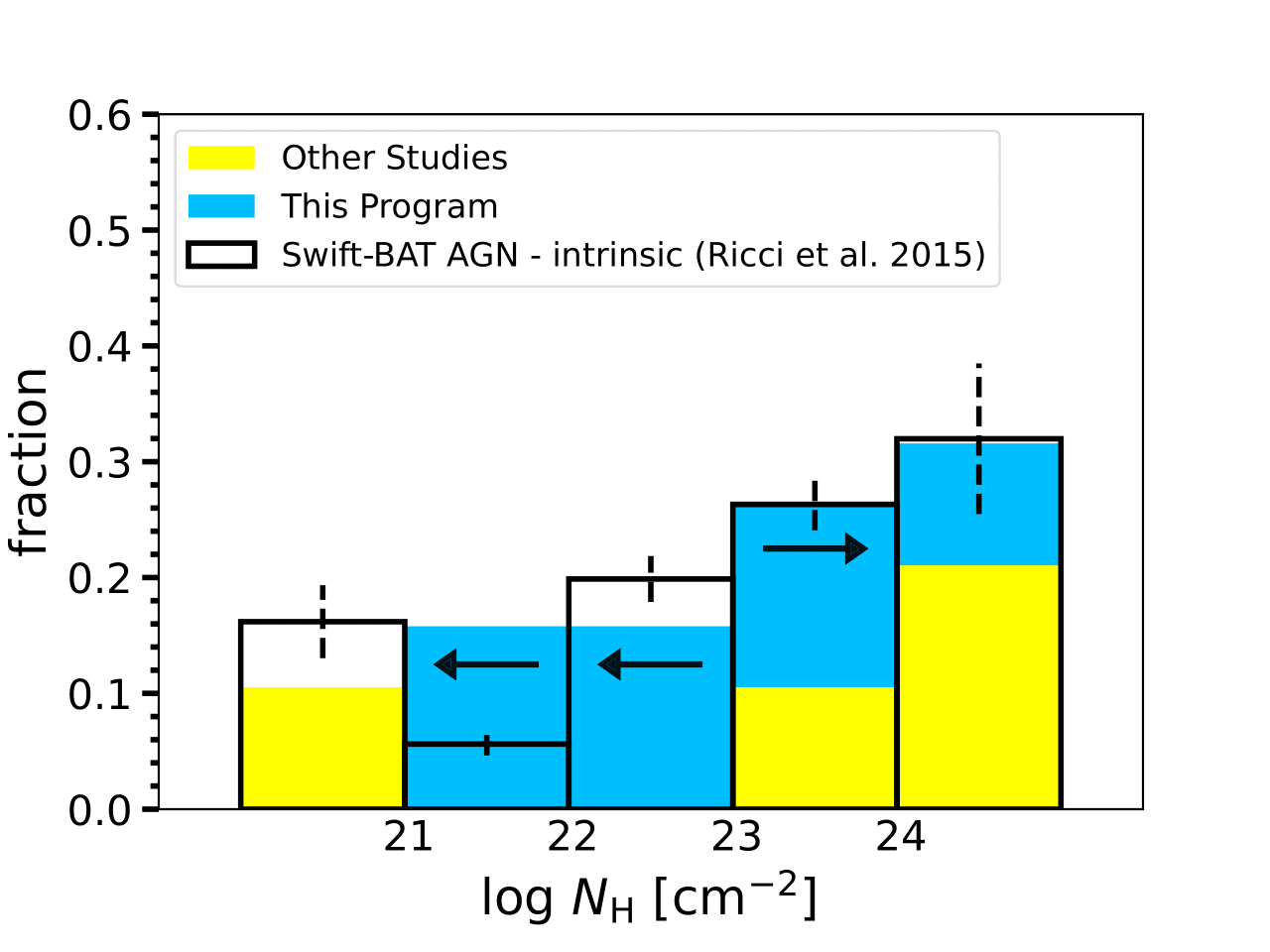}
\caption{The $N_{\rm{H}}$ distribution for our sample compared to the observed (left) and intrinsic (right) $N_{\rm{H}}$ distributions of the \textsl{Swift}-BAT AGN sample with log $L_{\rm 14-195} =$ 40.0--43.7 erg s$^{-1}$ \citep{Ricci15}. Yellow indicates those with $N_{\rm{H}}$ measurements from other studies, whilst blue indicates those that were measured as part of this program (\citealp{Annuar15}; \citealp{Annuar17}; \citealp{annuar20}; this work).}
\end{figure*}

\section{Compton-thick AGN Fraction and $N_{\rm H}$ Distribution}

Based on our results presented in Table 3, we find that the majority of our sources; i.e., 14/19 (74$^{+26}_{-29}\%$) are obscured, and 6/19 (i.e., 32$^{+30}_{-18}\%$) are CT (37$^{+33}_{-20}\%$ if including NGC 660; see \citealp{annuar20}). The CT AGN fraction is significantly higher than that directly observed in the entire \textsl{Swift}-BAT sample (i.e., 7.6$^{+1.1}_{-2.1}$$\%$ over the entire X-ray luminosity range of the \textsl{Swift}-BAT AGN sample; log $L_{\rm 14-195} =$ 40--46 erg s$^{-1}$; \citealp{Ricci15}), even after accounting for the uncertainties due to small-number statistics for our sample. Their inferred intrinsic fraction however (i.e., 27 $\pm$ 4 \%), is well in agreement with our value. Although the overall observed CT AGN fraction found for the \textsl{Swift}-BAT sample is significantly lower than the intrinsic fraction inferred, and also determined using our sample, \cite{Ricci15} show that this discrepancy is due to a bias against finding relatively faint CT AGN at larger distances.  This is demonstrated in Figure 12 (top). From this figure, we can see that at a smaller distance of 20 Mpc, \cite{Ricci15} found that the CT AGN population in the \textsl{Swift}-BAT sample is consistent with their inferred intrinsic fraction. Their CT fraction at $D \leq$ 20 Mpc is also in agreement with our fraction at $D \leq$ 15 Mpc. However, with our sample, we managed to identify additional lower luminosity CT AGN; i.e., NGC 1448 \citep{Annuar17} and possibly NGC 660 (\citealp{annuar20}; see Figure 12 -- bottom). \footnote{\label{footnote6}The 2--10 keV intrinsic luminosities for the \textsl{Swift}-BAT AGN were converted from their 14--195 keV luminosities using the scaling relation from \cite{Rigby09}.}

The same conclusion can be made when comparing our results with {\it{NuSTAR}} study of the \textsl{Swift}-BAT sample by \cite{Torres-Alba2021} which only directly measured a CT fraction of $\sim$8$\%$. However, when limiting down the sample to smaller redshift of $z \lesssim$ 0.01, this fraction increases to 20 $\pm$ 5$\%$, consistent with our findings. In addition, our fraction is also well in agreement with that directly found by the {\it{NuSTAR}} study of mid-IR (NuLANDS; \citealp{Boorman2024-arXiv}) and optical AGN \citep{Kammoun2020}, which measure CT fractions of $\sim$35$\%$ and $\sim$37--53$\%$, respectively. 

This consistency also holds when we compare our results with the CT AGN X-ray luminosity functions (XLFs) from previous studies, which estimate CT AGN fractions ranging from $\sim$17--24$\%$ (e.g., \citealp{Akylas2016}; \citealp{Ananna2022}; \citealp{Laloux2023}; \citealp{Georgantopoulos2025}), in agreement with our findings. The lower luminosity limit in these studies is typically log $L_{\rm X} = 41$ erg s$^{-1}$. \citet{Laloux2023} extended down to $L_{\rm X} = 40$ erg s$^{-1}$; however, the local CT XLF in that study remains an upper limit. In contrast, our work provides direct observational constraints down to this low luminosity limit.

\cite{Akylas2024} did a similar study to ours using a sample of infrared-selected AGN within 100 Mpc, but focusing on luminous AGN with $\log L _{12\mu\text{m}} > 42.3$ erg s$^{-1}$. They found a CT fraction of $25 \pm 5\%$, which is consistent with our overall fraction. This is also in agreement with our CT AGN fraction below this $12\mu\text{m}$ luminosity threshold; i.e., $19^{+30}_{-14}\%$ (3/16). However, above this threshold $(\log L _{12\mu\text{m}} > 42.3$ erg s$^{-1})$, we obtained a $100^{+0}_{-73}\%$ (3/3) CT population, which is much higher than that found by \cite{Akylas2024}. Nevertheless, within the large statistical uncertainty of our value, they agree with each other. We also found the same results when considering a similar 2–10 keV X-ray luminosity threshold of $L_{\rm 2-10, int} =$ 10$^{42}$ erg s$^{-1}$, where we found a CT AGN fraction of $19^{+30}_{-14}\%$ (3/16) below this limit. This indicates a moderate CT AGN population within this low-luminosity regime, consistent with those found at higher luminosities.

Figure 13 shows the $N_{\rm{H}}$ distribution of our AGN sample compared to the observed and intrinsic $N_{\rm{H}}$ distributions of the \textsl{Swift}-BAT AGN with comparable X-ray luminosities to our sample (i.e., log $L_{\rm 14-195} =$ 40.0--43.7 erg s$^{-1}$; \citealp{Ricci15}). Based on this figure, the $N_{\rm{H}}$ distribution of the AGN in our sample seems to be different from the observed $N_{\rm{H}}$ distribution of the \textsl{Swift}-BAT AGN, particularly at the lowest and highest (CT) regimes. However, it agrees very well with the intrinsic $N_{\rm{H}}$ distribution inferred after taking into account the \textsl{Swift}-BAT survey sensitivity limit. This demonstrates that using our sample, we are able to directly identify a higher fraction of CT AGN as compared to that directly observed in the \textsl{Swift}-BAT survey, exemplifying the bias against finding CT AGN even in a hard X-ray survey such as \textsl{Swift}-BAT. Based on this figure, we can also see how much our program has significantly improved the knowledge of X-ray properties of the nearest AGN down to lower luminosity. 

Based on Figure 9, if we apply a luminosity cut-off of $L_{\rm 2-10, int} =$ 10$^{40}$ erg s$^{-1}$, to separate low/high-luminosity AGN \citep{annuar20} and also to match the lower luminosity end of the entire \textsl{Swift}-BAT AGN sample, the CT AGN fraction calculated for our sample above this luminosity is 50$^{+49}_{-28}\%$ (6/12). This is higher than that observed and predicted using the \textsl{Swift}-BAT sample. This demonstrates the large uncertainty in determining an accurate $N_{\rm{H}}$ distribution of the AGN population at this high-obscuration regime. 
We note that, however, the lower limit fraction calculated is consistent with the intrinsic fraction inferred by the \textsl{Swift}-BAT survey if we take into account the uncertainties due to small number statistics (i.e., 50$^{+49}_{-28}\%$; see footnote 2).

\section{Discussion}

\subsection{Eddington Ratio}

\begin{figure}
\centering
  \includegraphics[scale=0.55]{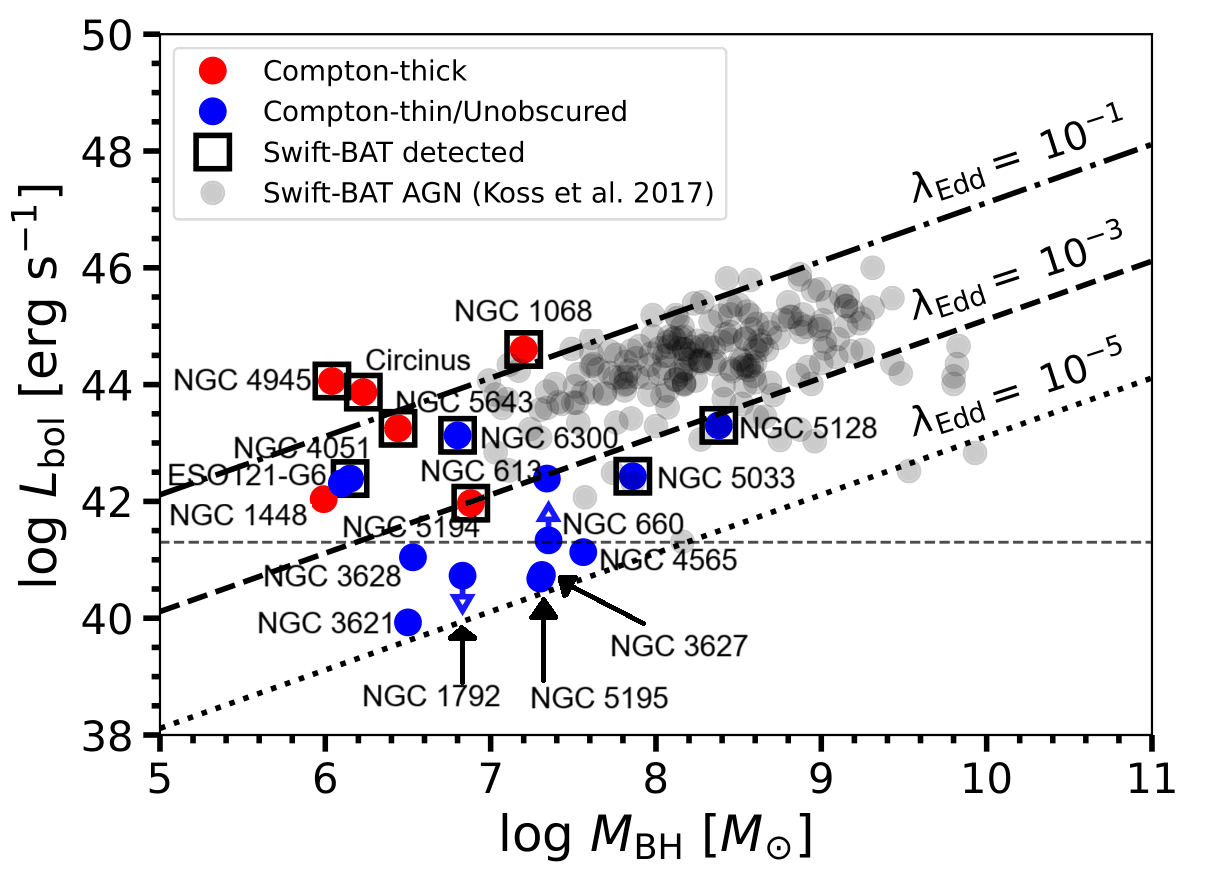}
\caption{The bolometric luminosity vs. $M_{\rm{BH}}$ for our sample and the \textsl{Swift}-BAT AGN sample (grey). Symbols are the same as in Figure 11. The dashed and dotted lines indicate constant Eddington ratios of $\lambda_{\rm Edd} =$ 10$^{-5}$, 10$^{-3}$ and 10$^{-1}$, respectively. Figure adapted from \protect\cite{Goulding10}.}
\end{figure}

In Figure 14 we show a plot of the AGN bolometric luminosity, $L_{\rm bol}$, as a function of black-hole mass, $M_{\rm{BH}}$, for our sample and the \textsl{Swift}-BAT AGN sample \citep{Koss17}. The bolometric luminosities for the AGN in our sample were calculated using their absorption-corrected 2--10 keV luminosities, assuming bolometric corrections of $\kappa \approx$ 20 for those with $L_{\rm 2-10, int} \geq$ 10$^{40}$ erg s$^{-1}$ (e.g., \citealp{Elvis94}; \citealp{Vasudevan10}), and $\kappa \approx$ 13 ($L_{\rm 2-10\ keV,\ int}$/10$^{41}$ erg s$^{-1}$)$^{-0.37}$) for the lower luminosity AGN \citep{Nemmen14}. The bolometric luminosities for the \textsl{Swift}-BAT AGN are given in \cite{Koss17} and were determined from their 14--195 keV luminosities. \cite{Koss17} measured $M_{\rm{BH}}$ for the \textsl{Swift}-BAT AGN sample using the velocity dispersion method, whilst for our AGN sample, the $M_{\rm{BH}}$ values were determined using a range of techniques (e.g., maser mapping, velocity dispersion and bulge luminosity; see GA09 for more details). 

The $M_{\rm{BH}}$ distribution of the two samples are significantly different from each other, with our sample dominating at lower $M_{\rm{BH}}$. Performing a Kolmogorov--Smirnov (KS) test on the $M_{\rm{BH}}$ distributions between the two samples produced a KS test probability of $P_{\rm KS} \sim$ 2 $\times$ $10^{-11}$, indicating that the two distributions are significantly different from each other. The mean $M_{\rm{BH}}$ calculated for our sample, log $<$$M_{\rm BH, 15 Mpc}$$> =$  6.88 $\pm$ 0.65 $M_{\odot}$, is {$\approx$ 1.5 dex}  lower than that determined for the \textsl{Swift}-BAT sample, log $<$$M_{\rm BH, BAT}$$> =$  8.32 $\pm$ 0.61 $M_{\odot}$.

Figure 14 shows that our sample has AGN bolometric luminosities in the range of $L_{\rm bol} \sim$ 10$^{39}$--10$^{45}$ erg s$^{-1}$, extending to lower luminosities than the \textsl{Swift}-BAT sample. The Eddington-ratio ranges probed by both samples however are similar, although the latter dominates at higher accretion rates of $\lambda_{\rm Edd} \geq$ 10$^{-3}$, and our sample is more evenly distributed between Eddington ratios below and above of this threshold. Interestingly, the majority of our sample at high Eddington ratios (i.e., $\lambda_{\rm Edd} \geq$ 10$^{-3}$) are CT, whereas the majority of AGN at lower Eddington ratios ($\lambda_{\rm Edd} <$ 10$^{-3}$) are unobscured or just mildly obscured. These results are consistent with that found by past studies (e.g., \citealp{Ho08}; \citealp{DraperBallantyne10}).

\subsection{Host-Galaxy Properties}

In this section we compare the host-galaxy and optical AGN  properties of our sample with those of the \textsl{Swift}-BAT AGN sample to investigate any potential differences between the two samples, given the unique parameter space that we probe using our $D \leq$ 15 Mpc AGN sample (i.e., lower AGN luminosities and black hole masses). 

\subsubsection{Stellar Mass}

Firstly, we investigated the stellar mass ($M_{\rm \ast}$) distributions for our sample and the \textsl{Swift}-BAT sample. We show the comparison between the two distributions in Figure 15. The stellar masses for our sample were determined using the $K$-band magnitude and the $J - K$ colour from the 2MASS Large Galaxy Atlas \citep{Jarrett03}. We calculated $M_{\rm \ast}$ for our sample using the relation derived by \cite{Westmeier11} between the $K$-band stellar mass-to-light ratio and the $J - K$ colour index. Meanwhile, $M_{\rm \ast}$ for the \textsl{Swift}-BAT sample was derived using spectral energy distribution fitting by \cite{Koss11}. For one of our AGN (NGC 4051) that has a measurement in \cite{Koss11}, we used the value given by that paper as we believe that it is more reliable. However, we note that the value that we calculated for the source using our method (log $M_{\rm \ast} = 9.60 \, M_{\odot}$) is in agreement with that derived by \cite{Koss11} (log $M_{\rm \ast} = 9.44 \, M_{\odot}$), demonstrating the reliability of our method. Additionally, NGC 4051 is a broad-line AGN, where the AGN contribution could significantly affect the optical-near-IR emission. This makes it a challenging case, and it is reassuring that our method provides a consistent stellar mass  with that derived by \cite{Koss11}. 

\begin{figure}
\centering
  \includegraphics[scale=0.5]{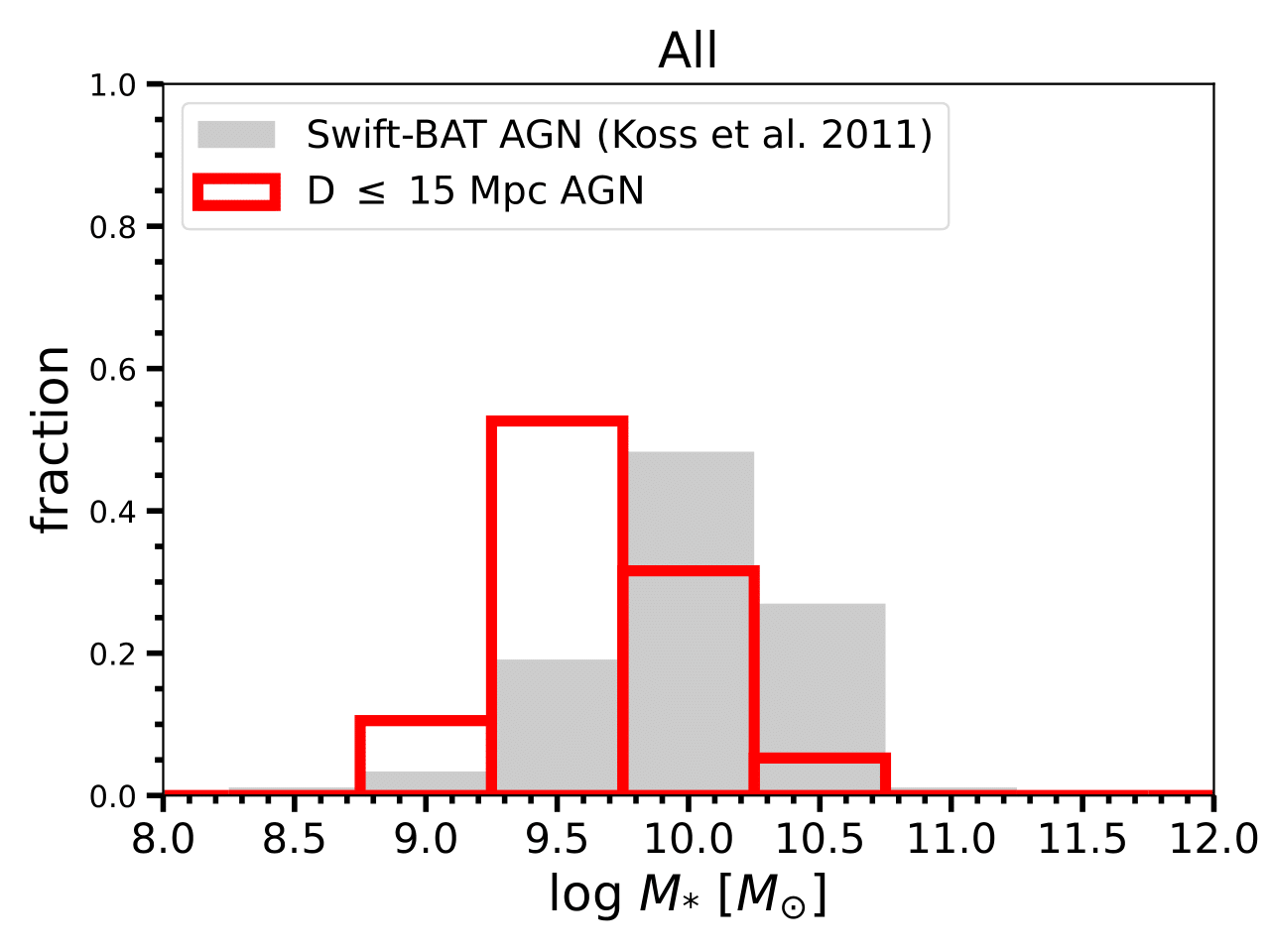}
\caption{$M_{\rm \ast}$ distributions for our sample (red) and the \textsl{Swift}-BAT sample (grey).}
\end{figure}

\begin{figure}
\centering
  \includegraphics[scale=0.35]{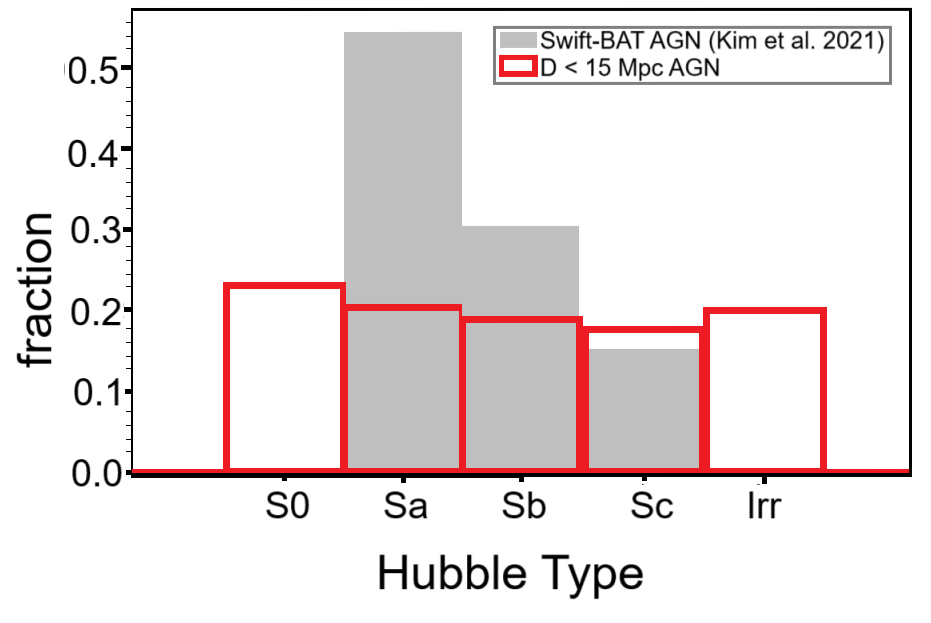}
\caption{The distribution of Hubble type for our sample (red) and the {\it{Swift-BAT}} sample (grey; \citealp{Kim2021})}
\end{figure}

Figure 15 shows that the galaxies in our sample generally have lower stellar masses as compared to the \textsl{Swift}-BAT sample. The mean stellar mass measured for our sample, log $<$$M_{\ast, \rm 15 Mpc}$$> =$  9.93 $\pm$ 0.31 $M_{\odot}$, is $\sim$ 0.3 dex lower than that calculated for the \textsl{Swift}-BAT sample; i.e., log $<$$M_{\ast, \rm BAT}$$> =$  10.24 $\pm$ 0.44 $M_{\circ}$. This can be attributed to the smaller $M_{\rm{BH}}$ and volume that we probe in our sample as compared to the \textsl{Swift}-BAT survey. Within the statistical uncertainties however, the stellar masses for the two samples are in agreement with each other. We performed a KS test between the two distributions and found that they are not significantly different from each other, with $P_{\rm KS} \sim$ 0.9. 

Comparing the $M_{\mathrm{BH}}$ (see Section 7.1) with the $M_{\ast}$ distributions, we can see a more significant difference in $M_{\mathrm{BH}}$ between the two samples as compared to $M_{\ast}$. A reason for this could be due to the \textsl{Swift}-BAT sample having more bulge-dominated systems as compared to our sample (see Figure 16). In these systems, the $M_{\mathrm{BH}}$ and $M_{\ast}$ ratio ($M_{\mathrm{BH}}$/$M_{\ast}$) is higher than for weak/bulgeless systems (e.g., \citealp{Reines2015}), which is the case for most of our sources.

\subsubsection{Star Formation Rate}

We also compare the star formation rate (SFR) distributions between our sample and the \textsl{Swift}-BAT sample. The two distributions are shown in Figure 17. We calculated the SFR for our sample using the far-IR luminosity from IRAS \citep{Sanders03} following \cite{Murphy2011}. For the \textsl{Swift}-BAT sample, \cite{Shimizu17} determined the SFR of the AGN using the SF luminosity obtained through spectral energy distribution decomposition, and then using the same $L_{\text{IR}}$-SFR conversion from \cite{Murphy2011} as our study. For overlapping sources in both samples (i.e., NGC 5033 and NGC 6300), we used the measurements from \cite{Shimizu17} as we believe that the technique is more reliable. We note that, the values calculated from our method for the two sources are comparable (agree within $\sim$0.3 dex) to those derived from \cite{Shimizu17} measurements. In both cases, our values are systematically lower than those determined by \cite{Shimizu17}. For NGC 6300, the difference is smaller ($\sim$0.1 dex), while for NGC 5033, our value is $\sim$0.3 dex lower. However, these differences do not significantly affect our results or the overall SFR distributions.
 
Overall, we found that the median SFR for our sample, log $<$SFR$>_{\rm 15 Mpc} =$ 0.25 $\pm$ 0.39 $M_{\odot}$/yr is consistent with that found for the \textsl{Swift}-BAT AGN sample within the statistical uncertainties; i.e., log $<$SFR$>_{\rm BAT} =$ 0.23$^{+0.58}_{-0.80}$ $M_{\odot}$/yr \citep{Shimizu17}. Performing a KS test on the SFR distributions between the two samples provide a KS test probability of $P_{\rm KS} \sim$ 0.9 and 0.8 when excluding and including sources with upper limit SFR in the \textsl{Swift}-BAT sample, respectively. This indicates that the two distributions are not significantly different from each other. These results indicate that the galaxies in our sample have similar SFR to the \textsl{Swift}-BAT sample, despite the lower $M_{\rm{BH}}$ and $M_{\rm \ast}$ probed in our sample. However, we caution that this may be attributed to the IR luminosity cut-off applied by GA09 of $L_{\rm IR} =$ 3 $\times$ 10$^{9}$ \lsun\ (corresponding to log SFR $= -$0.49 $M_{\odot}$/yr, assuming that all of the IR luminosity is due to star formation), to select the AGN sample which could bias the sample against selecting $D \leq$ 15 Mpc AGN with low SFRs.

\subsubsection{Optical Type}

Finally, we compare the distribution of the optical classifications of the AGN in our sample with that of the \textsl{Swift}-BAT AGN sample \citep{Oh22}. This is shown in Figure 18. This diagram shows that most of our AGN are classified as Seyfert 2s. We have significantly more LINERs and H{\sc{ii}} galaxies, but a lower fraction of Seyfert 1s  than \textsl{Swift}-BAT. The higher fractions of LINER and H{\sc{ii}} galaxies found in our sample show that we managed to identify relatively weaker AGN that are not identified in the optical. The \textsl{Swift}-BAT survey is limited by its sensitivity ($f_{\rm 14-195} \sim$ 10$^{-11}$ erg s$^{-1}$ cm$^{-2}$), and is therefore also biased against finding weak AGN. This may be the reason why our sample has a lower fraction of Seyfert 1s; i.e., due to the challenge in identifying the broad line components in the optical spectra of weak AGN since they generally have lower black hole masses, and the optical spectra can also be significantly contaminated by the host galaxies. It is also interesting to note that most of our H{\sc{ii}} nuclei (3/4, 75$^{+25}_{-55}\%$) are heavily obscured AGN with $N_{\rm H} >$ 10$^{23}$ cm$^{-2}$, {suggesting} that these nuclei suffer significant obscuration by both the host-galaxy and the AGN tori (see also GA09).

\begin{figure}
\centering
  \includegraphics[scale=0.5]{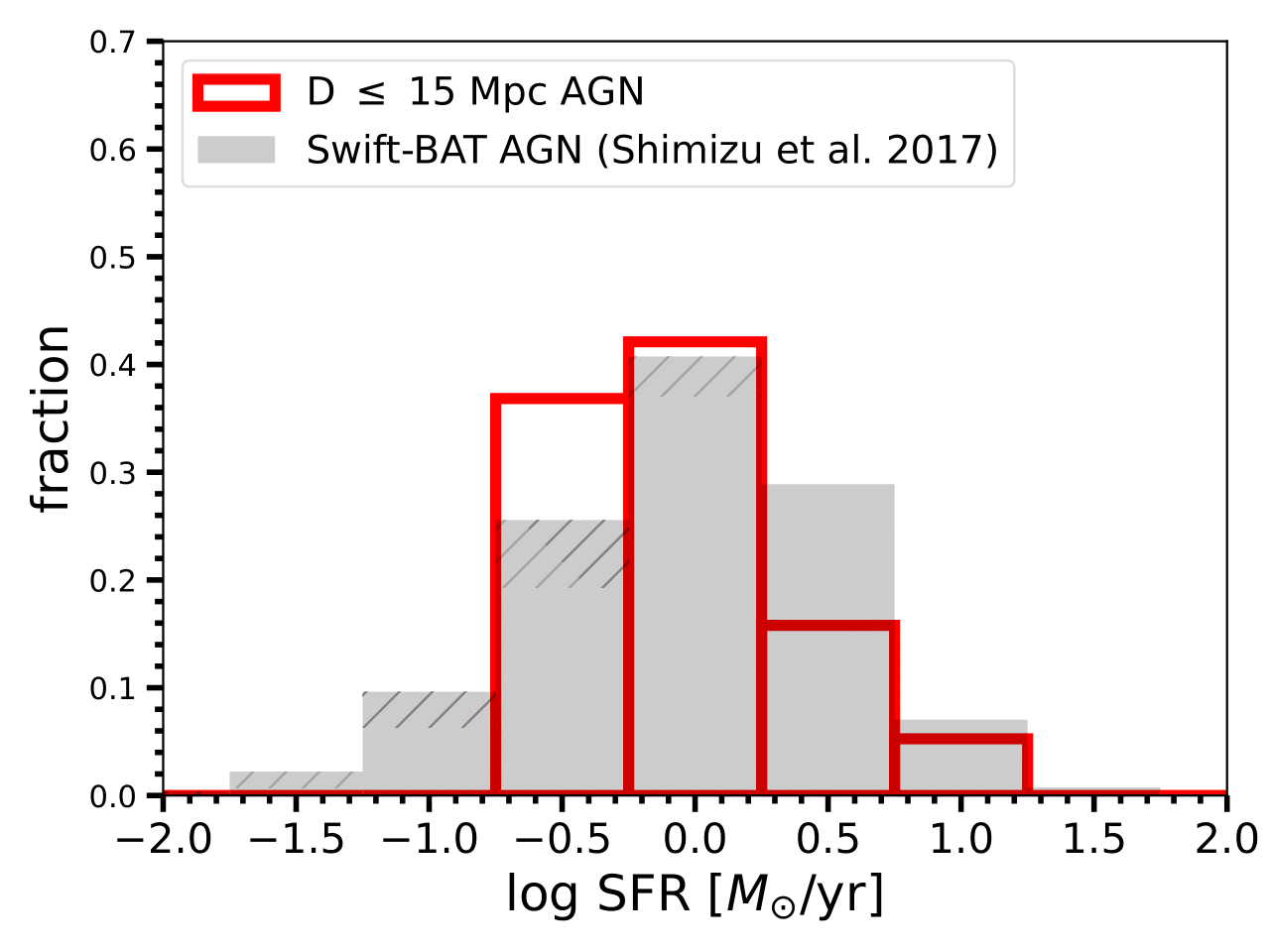}
\caption{SFR distributions for our sample (red) and the \textsl{Swift}-BAT sample to their (grey). Grey hatch indicates the \textsl{Swift}-BAT AGN which have upper limits SFRs.}
\end{figure}

\begin{figure}
\centering
  \includegraphics[scale=0.45]{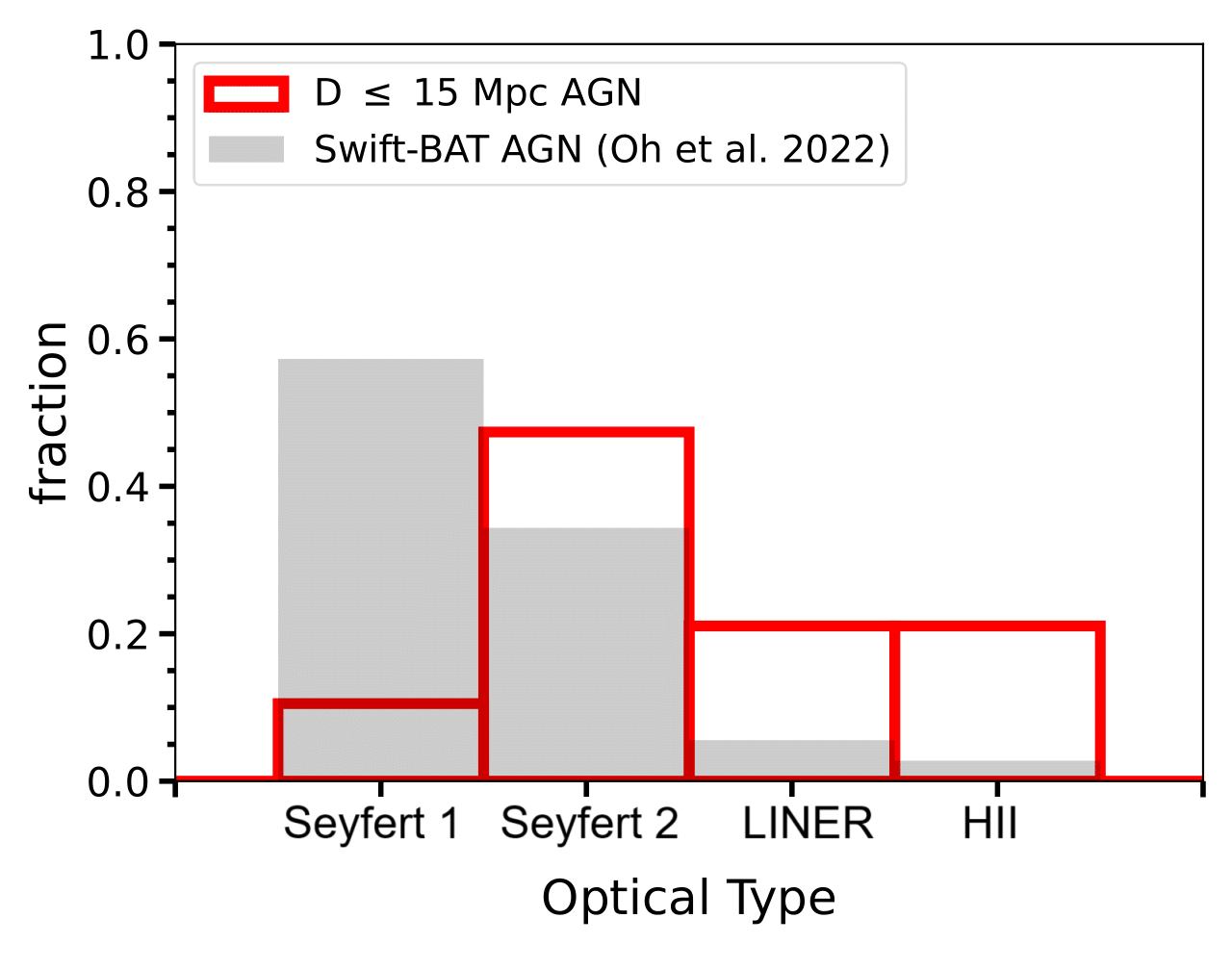}
\caption{The distribution of optical  nuclear classifications for our sample (red) and the \textsl{Swift}-BAT sample (grey).}
\end{figure}

\section{Conclusion}

We present a census of the CT AGN population and the $N_{\rm{H}}$ distribution of AGN in our cosmic backyard using an AGN sample within $D =$ 15 Mpc. We use $N_{\rm{H}}$ values measured directly from X-ray spectroscopic analysis using data from multiple observatories, mainly {\it{Chandra}} and {\it{NuSTAR}}. Our sample consists of AGN with intrinsic 2--10 keV luminosity in the range of $L_{\rm 2-10, int}$ = 10$^{37}$--10$^{43}$ erg s$^{-1}$, probing into a new parameter space that is not possible for more distant samples. Here is a summary of our main results:

\begin{itemize}
\item We found that 74$^{+26}_{-29}\%$ (14/19) of our sample is obscured AGN, with 32$^{+30}_{-18}\%$ (6/19) classified as Compton-thick (CT). This CT AGN fraction is significantly higher than that observed in the \textsl{Swift}-BAT survey but aligns with the fraction inferred after correcting for the survey's sensitivity limits. Our $N_{\rm{H}}$ distribution also agrees well with the corrected values. Applying a luminosity cut-off of $L_{\rm 2-10, int} = 10^{40}$ erg s$^{-1}$, we find a CT AGN fraction of 50$^{+49}_{-28}\%$ (6/12) above this threshold. This is significantly higher than that observed by \textsl{Swift}-BAT, but fully consistent with the inferred value within the uncertainties. Overall, our results provide direct observational evidence for AGN that are predicted by the \textsl{Swift}-BAT survey but remain undetected due to the survey's flux limitations.
\item If we consider only the typically defined low-luminosity regime of \( L_{2-10,\mathrm{int}} \leq 10^{42} \, \mathrm{erg} \, \mathrm{s}^{-1} \), we determined a CT AGN fraction of \( 19^{+30}_{-14}\% \) (3/16) within this relatively understudied domain. This is fully consistent with those observed at higher luminosities.
\item Overall, the multiwavelength data seem to support our X-ray analysis results, with the exception of a few cases, especially the intrinsically low-luminosity sources with $L_{\rm 2-10, int} \sim 10^{40}$ erg s$^{-1}$ in the X-ray:12 $\mu$m diagnostic diagram. These cases align with the typical limit of the intrinsic correlation derived. This suggests that such correlation may break down at lower AGN luminosities or that significant flux contamination from off-nuclear sources may affect these low-luminosity objects. Recent deactivation of the nuclear source, as well as extreme CT obscuration could also explain the discrepancies.
\item All of our sources with $L_{\rm 2-10, int} <$ 10$^{40}$ erg s$^{-1}$ (with the exception of NGC 660) are either unobscured or only mildly obscured, with $N_{\rm H} \lesssim 10^{22}$ cm$^{-2}$. This suggests that at such low luminosities, the AGN torus may be underdeveloped or absent, supporting previous studies that have proposed this scenario.
\item The Eddington-ratio range probed by our sample is similar to that of the \textsl{Swift}-BAT AGN sample, although the latter dominates at higher accretion rates. Majority of our sample at $\lambda_{\rm Edd} \geq$ 10$^{-3}$ tend to be CT AGN, while  AGN at $\lambda_{\rm Edd} <$ 10$^{-3}$ are mainly unobscured or just mildly obscured. On the other hand, our sample has lower black-hole masses, peaking at $\sim$10$^{6}$ \msun, $\sim$ 1.5 dex lower than the \textsl{Swift}-BAT sample, showing that our $D <$ 15 Mpc sample predominantly selects low-mass AGN.
\item In terms of basic host-galaxy properties, our sample shows star formation rates that are comparable to those in the \textsl{Swift}-BAT sample. However, most AGN in our sample are hosted by galaxies with lower stellar masses. This could be attributed to the smaller black hole masses ($M_{\rm{BH}}$) and the more-local volume we probe compared to the \textsl{Swift}-BAT survey. Additionally, our sample contains a significantly higher number of LINERs and H{\sc{ii}}-type optical nuclear spectra, which are largely missed by the \textsl{Swift}-BAT survey. This indicates that we have identified weaker AGN, or AGN that are heavily obscured by their host galaxies, causing them undetectable in optical wavelength.
\end{itemize}

Deeper observations with current facilities, or more sensitive and higher spatial resolution observations such as with the {\it{James Webb Space Telescope}} ({\it{JWST}}) mission, will be useful in constraining the 12$\mu$m and [Ne {\sc {v}}] nuclear emission which are currently not complete and might be contaminated by non-AGN emission. A future high-energy X-ray mission with higher sensitivity than {\it{NuSTAR}} will also be crucial in further constraining the X-ray properties of the low-luminosity AGN, including NGC 660 which is likely to be CT, and pushing this study out to a larger volume. Additionally, monitoring observations would also be useful to investigate whether variability or source deactivation could be the contributing cause for the multiwavelength discrepancies in the low-luminosity sources, such as the case for ESO 121-G6 \citep{annuar20} and NGC 3627 \citep{Saade22}, respectively. These will not only allow us to improve the AGN $N_{\rm{H}}$ distribution within the local volume, but also investigate the properties of low-luminosity AGN and test current theories on these sources; e.g., whether they have different physical structure than their more luminous counterparts.  

\section*{Acknowledgements}
A.A acknowledges financial support from {\it{Universiti Kebangsaan Malaysia}}'s {\it{Geran Universiti Penyelidikan}} grant code GUP-2023-033 and the Merdeka Award Grant for International Attachment 2021.  
D.M.A. acknowledges the Science and Technology Facilities Council (STFC) for support through grant code ST/T000244/1. 
M.N.R. acknowledges support from the MyBrainSc Scholarship program under the Ministry of Higher Education Malaysia (MoHE). 
CR acknowledges support from Fondecyt Regular grant 1230345 and ANID BASAL project FB210003. 
DRB is supported in part by NASA award 80NSSC24K0212 and NSF grants AST-2307278 and AST-2407658. 

\section*{Data Availability}

This work made use of data from the {\it{NuSTAR}} mission. {\it{NuSTAR}} is a project led by the California Institute of Technology (Caltech), managed by the Jet Propulsion Laboratory (JPL), and funded by the National Aeronautics and Space Administration (NASA). We thank the {\it{NuSTAR}} Operations, Software and Calibrations teams for support with these observations. This research has made use of the {\it{NuSTAR}} Data Analysis Software (NUSTARDAS) jointly developed by the ASI Science Data Center (ASDC, Italy) and Caltech (USA). The scientific results reported in this article are based on observations made by the {\it{Chandra X-ray Observatory}} and data obtained from the Chandra Data Archive. This research has made use of software provided by the Chandra X-ray Center (CXC) in the application packages CIAO. This work was also based on observations obtained with {\it{XMM-Newton}}, an ESA science mission with instruments and contributions directly funded by ESA Member States and NASA.

This research made use of various Python packages. We also used data obtained through the High Energy Astrophysics Science Archive Research Center (HEASARC) Online Service, provided by the NASA / Goddard Space Flight Center, and the NASA/IPAC extragalactic Database (NED) operated by JPL, Caltech under contract with NASA. 

Facilities: {\it{Chandra}}, {\it{NuSTAR}}, {\it{Swift}} and {\it{XMM–Newton}}.

\bibliographystyle{mnras}
\bibliography{d15mpc} 

\begin{thebibliography}{}
\makeatletter
\relax
\def\mn@urlcharsother{\let\do\@makeother \do\$\do\&\do\#\do\^\do\_\do\%\do\~}
\def\mn@doi{\begingroup\mn@urlcharsother \@ifnextchar [ {\mn@doi@} {\mn@doi@[]}}
\def\mn@doi@[#1]#2{\def\@tempa{#1}\ifx\@tempa\@empty \href {http://dx.doi.org/#2} {doi:#2}\else \href {http://dx.doi.org/#2} {#1}\fi \endgroup}
\def\mn@eprint#1#2{\mn@eprint@#1:#2::\@nil}
\def\mn@eprint@arXiv#1{\href {http://arxiv.org/abs/#1} {{\tt arXiv:#1}}}
\def\mn@eprint@dblp#1{\href {http://dblp.uni-trier.de/rec/bibtex/#1.xml} {dblp:#1}}
\def\mn@eprint@#1:#2:#3:#4\@nil{\def\@tempa {#1}\def\@tempb {#2}\def\@tempc {#3}\ifx \@tempc \@empty \let \@tempc \@tempb \let \@tempb \@tempa \fi \ifx \@tempb \@empty \def\@tempb {arXiv}\fi \@ifundefined {mn@eprint@\@tempb}{\@tempb:\@tempc}{\expandafter \expandafter \csname mn@eprint@\@tempb\endcsname \expandafter{\@tempc}}}

\bibitem[\protect\citeauthoryear{{Aalto} et~al.,}{{Aalto} et~al.}{2019}]{aalto2019}
{Aalto} S.,  et~al., 2019, \mn@doi [\aap] {10.1051/0004-6361/201935480}, \href {https://ui.adsabs.harvard.edu/abs/2019A&A...627A.147A} {627, A147}

\bibitem[\protect\citeauthoryear{{Akylas}, {Georgakakis}, {Georgantopoulos}, {Brightman}  \& {Nandra}}{{Akylas} et~al.}{2012}]{Akylas12}
{Akylas} A.,  {Georgakakis} A.,  {Georgantopoulos} I.,  {Brightman} M.,   {Nandra} K.,  2012, \mn@doi [\aap] {10.1051/0004-6361/201219387}, \href {http://adsabs.harvard.edu/abs/2012A%26A...546A..98A} {546, A98}

\bibitem[\protect\citeauthoryear{{Akylas}, {Georgantopoulos}, {Ranalli}, {Gkiokas}, {Corral}  \& {Lanzuisi}}{{Akylas} et~al.}{2016}]{Akylas2016}
{Akylas} A.,  {Georgantopoulos} I.,  {Ranalli} P.,  {Gkiokas} E.,  {Corral} A.,   {Lanzuisi} G.,  2016, \mn@doi [\aap] {10.1051/0004-6361/201628711}, \href {https://ui.adsabs.harvard.edu/abs/2016A&A...594A..73A} {594, A73}

\bibitem[\protect\citeauthoryear{{Akylas}, {Georgantopoulos}, {Gandhi}, {Boorman}  \& {Greenwell}}{{Akylas} et~al.}{2024}]{Akylas2024}
{Akylas} A.,  {Georgantopoulos} I.,  {Gandhi} P.,  {Boorman} P.,   {Greenwell} C.~L.,  2024, \mn@doi [\aap] {10.1051/0004-6361/202451906}, \href {https://ui.adsabs.harvard.edu/abs/2024A&A...692A.250A} {692, A250}

\bibitem[\protect\citeauthoryear{{Alexander}}{{Alexander}}{2001}]{Alexander01}
{Alexander} D.~M.,  2001, \mn@doi [\mnras] {10.1046/j.1365-8711.2001.04131.x}, \href {https://ui.adsabs.harvard.edu/abs/2001MNRAS.320L..15A} {320, L15}

\bibitem[\protect\citeauthoryear{{Ananna} et~al.,}{{Ananna} et~al.}{2019}]{Ananna2019}
{Ananna} T.~T.,  et~al., 2019, \mn@doi [\apj] {10.3847/1538-4357/aafb77}, \href {https://ui.adsabs.harvard.edu/abs/2019ApJ...871..240A} {871, 240}

\bibitem[\protect\citeauthoryear{{Ananna} et~al.,}{{Ananna} et~al.}{2022}]{Ananna2022}
{Ananna} T.~T.,  et~al., 2022, \mn@doi [\apjs] {10.3847/1538-4365/ac5b64}, \href {https://ui.adsabs.harvard.edu/abs/2022ApJS..261....9A} {261, 9}

\bibitem[\protect\citeauthoryear{{Annuar} et~al.,}{{Annuar} et~al.}{2015}]{Annuar15}
{Annuar} A.,  et~al., 2015, \mn@doi [\apj] {10.1088/0004-637X/815/1/36}, \href {http://adsabs.harvard.edu/abs/2015ApJ...815...36A} {815, 36}

\bibitem[\protect\citeauthoryear{{Annuar} et~al.,}{{Annuar} et~al.}{2017}]{Annuar17}
{Annuar} A.,  et~al., 2017, \mn@doi [\apj] {10.3847/1538-4357/836/2/165}, \href {http://adsabs.harvard.edu/abs/2017ApJ...836..165A} {836, 165}

\bibitem[\protect\citeauthoryear{{Annuar} et~al.,}{{Annuar} et~al.}{2020}]{annuar20}
{Annuar} A.,  et~al., 2020, \mn@doi [\mnras] {10.1093/mnras/staa1820}, \href {https://ui.adsabs.harvard.edu/abs/2020MNRAS.497..229A} {497, 229}

\bibitem[\protect\citeauthoryear{{Antonucci}}{{Antonucci}}{1993}]{Antonucci93}
{Antonucci} R.,  1993, \mn@doi [\araa] {10.1146/annurev.aa.31.090193.002353}, \href {http://adsabs.harvard.edu/abs/1993ARA%26A..31..473A} {31, 473}

\bibitem[\protect\citeauthoryear{{Ar{\'e}valo} et~al.,}{{Ar{\'e}valo} et~al.}{2014}]{Arevalo14}
{Ar{\'e}valo} P.,  et~al., 2014, \mn@doi [\apj] {10.1088/0004-637X/791/2/81}, \href {http://adsabs.harvard.edu/abs/2014ApJ...791...81A} {791, 81}

\bibitem[\protect\citeauthoryear{{Asmus}, {H{\"o}nig}, {Gandhi}, {Smette}  \& {Duschl}}{{Asmus} et~al.}{2014}]{Asmus14}
{Asmus} D.,  {H{\"o}nig} S.~F.,  {Gandhi} P.,  {Smette} A.,   {Duschl} W.~J.,  2014, \mn@doi [\mnras] {10.1093/mnras/stu041}, \href {http://adsabs.harvard.edu/abs/2014MNRAS.439.1648A} {439, 1648}

\bibitem[\protect\citeauthoryear{{Asmus}, {Gandhi}, {H{\"o}nig}, {Smette}  \& {Duschl}}{{Asmus} et~al.}{2015}]{Asmus15}
{Asmus} D.,  {Gandhi} P.,  {H{\"o}nig} S.~F.,  {Smette} A.,   {Duschl} W.~J.,  2015, \mn@doi [\mnras] {10.1093/mnras/stv1950}, \href {http://adsabs.harvard.edu/abs/2015MNRAS.454..766A} {454, 766}

\bibitem[\protect\citeauthoryear{{Asmus}, {H{\"o}nig}  \& {Gandhi}}{{Asmus} et~al.}{2016}]{Asmus2016}
{Asmus} D.,  {H{\"o}nig} S.~F.,   {Gandhi} P.,  2016, \mn@doi [\apj] {10.3847/0004-637X/822/2/109}, \href {https://ui.adsabs.harvard.edu/abs/2016ApJ...822..109A} {822, 109}

\bibitem[\protect\citeauthoryear{{Asmus} et~al.,}{{Asmus} et~al.}{2020}]{Asmus2020}
{Asmus} D.,  et~al., 2020, \mn@doi [\mnras] {10.1093/mnras/staa766}, \href {https://ui.adsabs.harvard.edu/abs/2020MNRAS.494.1784A} {494, 1784}

\bibitem[\protect\citeauthoryear{{Assef} et~al.,}{{Assef} et~al.}{2013}]{Assef13}
{Assef} R.~J.,  et~al., 2013, \mn@doi [\apj] {10.1088/0004-637X/772/1/26}, \href {http://adsabs.harvard.edu/abs/2013ApJ...772...26A} {772, 26}

\bibitem[\protect\citeauthoryear{{Audibert} et~al.,}{{Audibert} et~al.}{2019}]{Audibert19}
{Audibert} A.,  et~al., 2019, \mn@doi [\aap] {10.1051/0004-6361/201935845}, \href {https://ui.adsabs.harvard.edu/abs/2019A&A...632A..33A} {632, A33}

\bibitem[\protect\citeauthoryear{{Baldwin}, {Phillips}  \& {Terlevich}}{{Baldwin} et~al.}{1981}]{Baldwin81}
{Baldwin} J.~A.,  {Phillips} M.~M.,   {Terlevich} R.,  1981, \mn@doi [\pasp] {10.1086/130766}, \href {http://adsabs.harvard.edu/abs/1981PASP...93....5B} {93, 5}

\bibitem[\protect\citeauthoryear{{Balokovi{\'c}} et~al.,}{{Balokovi{\'c}} et~al.}{2014}]{Balokovic14}
{Balokovi{\'c}} M.,  et~al., 2014, \mn@doi [\apj] {10.1088/0004-637X/794/2/111}, \href {http://adsabs.harvard.edu/abs/2014ApJ...794..111B} {794, 111}

\bibitem[\protect\citeauthoryear{{Balokovi{\'c}} et~al.,}{{Balokovi{\'c}} et~al.}{2018}]{Balokovic18}
{Balokovi{\'c}} M.,  et~al., 2018, \mn@doi [\apj] {10.3847/1538-4357/aaa7eb}, \href {https://ui.adsabs.harvard.edu/abs/2018ApJ...854...42B} {854, 42}

\bibitem[\protect\citeauthoryear{{Barth}, {Strigari}, {Bentz}, {Greene}  \& {Ho}}{{Barth} et~al.}{2009}]{Barth09}
{Barth} A.~J.,  {Strigari} L.~E.,  {Bentz} M.~C.,  {Greene} J.~E.,   {Ho} L.~C.,  2009, \mn@doi [\apj] {10.1088/0004-637X/690/1/1031}, \href {https://ui.adsabs.harvard.edu/abs/2009ApJ...690.1031B} {690, 1031}

\bibitem[\protect\citeauthoryear{{Bassani}, {Dadina}, {Maiolino}, {Salvati}, {Risaliti}, {Della Ceca}, {Matt}  \& {Zamorani}}{{Bassani} et~al.}{1999}]{Bassani99}
{Bassani} L.,  {Dadina} M.,  {Maiolino} R.,  {Salvati} M.,  {Risaliti} G.,  {Della Ceca} R.,  {Matt} G.,   {Zamorani} G.,  1999, \mn@doi [\apjs] {10.1086/313202}, \href {http://adsabs.harvard.edu/abs/1999ApJS..121..473B} {121, 473}

\bibitem[\protect\citeauthoryear{{Bauer} et~al.,}{{Bauer} et~al.}{2015}]{Bauer15}
{Bauer} F.~E.,  et~al., 2015, \mn@doi [\apj] {10.1088/0004-637X/812/2/116}, \href {http://adsabs.harvard.edu/abs/2015ApJ...812..116B} {812, 116}

\bibitem[\protect\citeauthoryear{{Baumgartner}, {Tueller}, {Markwardt}, {Skinner}, {Barthelmy}, {Mushotzky}, {Evans}  \& {Gehrels}}{{Baumgartner} et~al.}{2013}]{Baumgartner13}
{Baumgartner} W.~H.,  {Tueller} J.,  {Markwardt} C.~B.,  {Skinner} G.~K.,  {Barthelmy} S.,  {Mushotzky} R.~F.,  {Evans} P.~A.,   {Gehrels} N.,  2013, \mn@doi [\apjs] {10.1088/0067-0049/207/2/19}, \href {http://adsabs.harvard.edu/abs/2013ApJS..207...19B} {207, 19}

\bibitem[\protect\citeauthoryear{{Bianchi}, {Matt}, {Fiore}, {Fabian}, {Iwasawa}  \& {Nicastro}}{{Bianchi} et~al.}{2002}]{Bianchi02}
{Bianchi} S.,  {Matt} G.,  {Fiore} F.,  {Fabian} A.~C.,  {Iwasawa} K.,   {Nicastro} F.,  2002, \mn@doi [\aap] {10.1051/0004-6361:20021414}, \href {http://adsabs.harvard.edu/abs/2002A%26A...396..793B} {396, 793}

\bibitem[\protect\citeauthoryear{{Bierschenk} et~al.,}{{Bierschenk} et~al.}{2024}]{Bierschenk2024}
{Bierschenk} M.,  et~al., 2024, \mn@doi [\apj] {10.3847/1538-4357/ad844a}, \href {https://ui.adsabs.harvard.edu/abs/2024ApJ...976..257B} {976, 257}

\bibitem[\protect\citeauthoryear{{Boorman} et~al.,}{{Boorman} et~al.}{2024a}]{Boorman2024-arXiv}
{Boorman} P.~G.,  et~al., 2024a, \mn@doi [arXiv e-prints] {10.48550/arXiv.2410.07339}, \href {https://ui.adsabs.harvard.edu/abs/2024arXiv241007339B} {p. arXiv:2410.07339}

\bibitem[\protect\citeauthoryear{{Boorman} et~al.,}{{Boorman} et~al.}{2024b}]{Boorman2024}
{Boorman} P.,  et~al., 2024b, in American Astronomical Society Meeting Abstracts. p. 175.02

\bibitem[\protect\citeauthoryear{{Brandt} \& {Alexander}}{{Brandt} \& {Alexander}}{2015}]{Brandt15}
{Brandt} W.~N.,  {Alexander} D.~M.,  2015, \mn@doi [\aapr] {10.1007/s00159-014-0081-z}, \href {http://adsabs.harvard.edu/abs/2015A%26ARv..23....1B} {23, 1}

\bibitem[\protect\citeauthoryear{{Brightman} \& {Nandra}}{{Brightman} \& {Nandra}}{2008}]{Brightman08}
{Brightman} M.,  {Nandra} K.,  2008, \mn@doi [\mnras] {10.1111/j.1365-2966.2008.13841.x}, \href {http://adsabs.harvard.edu/abs/2008MNRAS.390.1241B} {390, 1241}

\bibitem[\protect\citeauthoryear{{Brightman} \& {Nandra}}{{Brightman} \& {Nandra}}{2011}]{Brightman11}
{Brightman} M.,  {Nandra} K.,  2011, \mn@doi [\mnras] {10.1111/j.1365-2966.2011.18612.x}, \href {https://ui.adsabs.harvard.edu/abs/2011MNRAS.414.3084B} {414, 3084}

\bibitem[\protect\citeauthoryear{{Burlon}, {Ajello}, {Greiner}, {Comastri}, {Merloni}  \& {Gehrels}}{{Burlon} et~al.}{2011}]{Burlon11}
{Burlon} D.,  {Ajello} M.,  {Greiner} J.,  {Comastri} A.,  {Merloni} A.,   {Gehrels} N.,  2011, \mn@doi [\apj] {10.1088/0004-637X/728/1/58}, \href {http://adsabs.harvard.edu/abs/2011ApJ...728...58B} {728, 58}

\bibitem[\protect\citeauthoryear{{Cappi} et~al.,}{{Cappi} et~al.}{2006}]{Cappi06}
{Cappi} M.,  et~al., 2006, \mn@doi [\aap] {10.1051/0004-6361:20053893}, \href {http://adsabs.harvard.edu/abs/2006A%26A...446..459C} {446, 459}

\bibitem[\protect\citeauthoryear{{Cash}}{{Cash}}{1979}]{Cash79}
{Cash} W.,  1979, \mn@doi [\apj] {10.1086/156922}, \href {http://adsabs.harvard.edu/abs/1979ApJ...228..939C} {228, 939}

\bibitem[\protect\citeauthoryear{{Castangia}, {Panessa}, {Henkel}, {Kadler}  \& {Tarchi}}{{Castangia} et~al.}{2013}]{Castangia13}
{Castangia} P.,  {Panessa} F.,  {Henkel} C.,  {Kadler} M.,   {Tarchi} A.,  2013, \mn@doi [\mnras] {10.1093/mnras/stt1824}, \href {http://adsabs.harvard.edu/abs/2013MNRAS.436.3388C} {436, 3388}

\bibitem[\protect\citeauthoryear{{Chiaberge}, {Gilli}, {Macchetto}  \& {Sparks}}{{Chiaberge} et~al.}{2006}]{Chiaberge06}
{Chiaberge} M.,  {Gilli} R.,  {Macchetto} F.~D.,   {Sparks} W.~B.,  2006, \mn@doi [\apj] {10.1086/508131}, \href {http://adsabs.harvard.edu/abs/2006ApJ...651..728C} {651, 728}

\bibitem[\protect\citeauthoryear{{Cisternas} et~al.,}{{Cisternas} et~al.}{2013}]{Cisternas13}
{Cisternas} M.,  et~al., 2013, \mn@doi [\apj] {10.1088/0004-637X/776/1/50}, \href {https://ui.adsabs.harvard.edu/abs/2013ApJ...776...50C} {776, 50}

\bibitem[\protect\citeauthoryear{{Comastri}, {Gilli}, {Marconi}, {Risaliti}  \& {Salvati}}{{Comastri} et~al.}{2015}]{Comastri15}
{Comastri} A.,  {Gilli} R.,  {Marconi} A.,  {Risaliti} G.,   {Salvati} M.,  2015, \mn@doi [\aap] {10.1051/0004-6361/201425496}, \href {http://adsabs.harvard.edu/abs/2015A%26A...574L..10C} {574, L10}

\bibitem[\protect\citeauthoryear{{Diamond-Stanic}, {Rieke}  \& {Rigby}}{{Diamond-Stanic} et~al.}{2009}]{DiamondStanic09}
{Diamond-Stanic} A.~M.,  {Rieke} G.~H.,   {Rigby} J.~R.,  2009, \mn@doi [\apj] {10.1088/0004-637X/698/1/623}, \href {http://adsabs.harvard.edu/abs/2009ApJ...698..623D} {698, 623}

\bibitem[\protect\citeauthoryear{{Diaz} et~al.,}{{Diaz} et~al.}{2023}]{Diaz23}
{Diaz} Y.,  et~al., 2023, \mn@doi [\aap] {10.1051/0004-6361/202244678}, \href {https://ui.adsabs.harvard.edu/abs/2023A&A...669A.114D} {669, A114}

\bibitem[\protect\citeauthoryear{{Done}, {Madejski}, {{\.Z}ycki}  \& {Greenhill}}{{Done} et~al.}{2003}]{Done2003}
{Done} C.,  {Madejski} G.~M.,  {{\.Z}ycki} P.~T.,   {Greenhill} L.~J.,  2003, \mn@doi [\apj] {10.1086/374332}, \href {https://ui.adsabs.harvard.edu/abs/2003ApJ...588..763D} {588, 763}

\bibitem[\protect\citeauthoryear{{Draper} \& {Ballantyne}}{{Draper} \& {Ballantyne}}{2010}]{DraperBallantyne10}
{Draper} A.~R.,  {Ballantyne} D.~R.,  2010, \mn@doi [\apjl] {10.1088/2041-8205/715/2/L99}, \href {http://adsabs.harvard.edu/abs/2010ApJ...715L..99D} {715, L99}

\bibitem[\protect\citeauthoryear{{Dudik}, {Satyapal}, {Gliozzi}  \& {Sambruna}}{{Dudik} et~al.}{2005}]{Dudik05}
{Dudik} R.~P.,  {Satyapal} S.,  {Gliozzi} M.,   {Sambruna} R.~M.,  2005, \mn@doi [\apj] {10.1086/426856}, \href {https://ui.adsabs.harvard.edu/abs/2005ApJ...620..113D} {620, 113}

\bibitem[\protect\citeauthoryear{{Elitzur} \& {Ho}}{{Elitzur} \& {Ho}}{2009}]{Elitzur09}
{Elitzur} M.,  {Ho} L.~C.,  2009, \mn@doi [\apjl] {10.1088/0004-637X/701/2/L91}, \href {http://adsabs.harvard.edu/abs/2009ApJ...701L..91E} {701, L91}

\bibitem[\protect\citeauthoryear{{Elitzur} \& {Shlosman}}{{Elitzur} \& {Shlosman}}{2006}]{Elitzur06}
{Elitzur} M.,  {Shlosman} I.,  2006, \mn@doi [\apjl] {10.1086/508158}, \href {http://adsabs.harvard.edu/abs/2006ApJ...648L.101E} {648, L101}

\bibitem[\protect\citeauthoryear{{Elvis} et~al.,}{{Elvis} et~al.}{1994}]{Elvis94}
{Elvis} M.,  et~al., 1994, \mn@doi [\apjs] {10.1086/192093}, \href {http://adsabs.harvard.edu/abs/1994ApJS...95....1E} {95, 1}

\bibitem[\protect\citeauthoryear{{Esparza-Arredondo}, {Osorio-Clavijo}, {Gonz{\'a}lez-Mart{\'\i}n}, {Victoria-Ceballos}, {Haro-Corzo}, {Reyes-Amador}, {L{\'o}pez-S{\'a}nchez}  \& {Pasetto}}{{Esparza-Arredondo} et~al.}{2020}]{Esparza20}
{Esparza-Arredondo} D.,  {Osorio-Clavijo} N.,  {Gonz{\'a}lez-Mart{\'\i}n} O.,  {Victoria-Ceballos} C.,  {Haro-Corzo} S. A.~R.,  {Reyes-Amador} O.~U.,  {L{\'o}pez-S{\'a}nchez} J.,   {Pasetto} A.,  2020, \mn@doi [\apj] {10.3847/1538-4357/abc425}, \href {https://ui.adsabs.harvard.edu/abs/2020ApJ...905...29E} {905, 29}

\bibitem[\protect\citeauthoryear{{Falc{\'o}n-Barroso}, {Ramos Almeida}, {B{\"o}ker}, {Schinnerer}, {Knapen}, {Lan{\c{c}}on}  \& {Ryder}}{{Falc{\'o}n-Barroso} et~al.}{2014}]{Falcon-Barroso14}
{Falc{\'o}n-Barroso} J.,  {Ramos Almeida} C.,  {B{\"o}ker} T.,  {Schinnerer} E.,  {Knapen} J.~H.,  {Lan{\c{c}}on} A.,   {Ryder} S.,  2014, \mn@doi [\mnras] {10.1093/mnras/stt2189}, \href {https://ui.adsabs.harvard.edu/abs/2014MNRAS.438..329F} {438, 329}

\bibitem[\protect\citeauthoryear{{Feruglio}, {Maiolino}, {Piconcelli}, {Menci}, {Aussel}, {Lamastra}  \& {Fiore}}{{Feruglio} et~al.}{2010}]{Feruglio2010}
{Feruglio} C.,  {Maiolino} R.,  {Piconcelli} E.,  {Menci} N.,  {Aussel} H.,  {Lamastra} A.,   {Fiore} F.,  2010, \mn@doi [\aap] {10.1051/0004-6361/201015164}, \href {https://ui.adsabs.harvard.edu/abs/2010A&A...518L.155F} {518, L155}

\bibitem[\protect\citeauthoryear{{Filho}, {Barthel}  \& {Ho}}{{Filho} et~al.}{2000}]{Filho00}
{Filho} M.~E.,  {Barthel} P.~D.,   {Ho} L.~C.,  2000, \mn@doi [\apjs] {10.1086/313412}, \href {https://ui.adsabs.harvard.edu/abs/2000ApJS..129...93F} {129, 93}

\bibitem[\protect\citeauthoryear{{Fruscione} et~al.,}{{Fruscione} et~al.}{2006}]{Fruscione06}
{Fruscione} A.,  et~al., 2006, in Society of Photo-Optical Instrumentation Engineers (SPIE) Conference Series. p. 62701V, \mn@doi{10.1117/12.671760}

\bibitem[\protect\citeauthoryear{{F{\"u}rst} et~al.,}{{F{\"u}rst} et~al.}{2016}]{Furst16}
{F{\"u}rst} F.,  et~al., 2016, \mn@doi [\apj] {10.3847/0004-637X/819/2/150}, \href {http://adsabs.harvard.edu/abs/2016ApJ...819..150F} {819, 150}

\bibitem[\protect\citeauthoryear{{Gandhi}, {Horst}, {Smette}, {H{\"o}nig}, {Comastri}, {Gilli}, {Vignali}  \& {Duschl}}{{Gandhi} et~al.}{2009}]{Gandhi09}
{Gandhi} P.,  {Horst} H.,  {Smette} A.,  {H{\"o}nig} S.,  {Comastri} A.,  {Gilli} R.,  {Vignali} C.,   {Duschl} W.,  2009, \mn@doi [\aap] {10.1051/0004-6361/200811368}, \href {http://adsabs.harvard.edu/abs/2009A%26A...502..457G} {502, 457}

\bibitem[\protect\citeauthoryear{{Gehrels}}{{Gehrels}}{1986}]{Gehrels86}
{Gehrels} N.,  1986, \mn@doi [\apj] {10.1086/164079}, \href {http://adsabs.harvard.edu/abs/1986ApJ...303..336G} {303, 336}

\bibitem[\protect\citeauthoryear{{Georgantopoulos}, {Pouliasis}, {Ruiz}  \& {Akylas}}{{Georgantopoulos} et~al.}{2025}]{Georgantopoulos2025}
{Georgantopoulos} I.,  {Pouliasis} E.,  {Ruiz} A.,   {Akylas} A.,  2025, \mn@doi [\aap] {10.1051/0004-6361/202451282}, \href {https://ui.adsabs.harvard.edu/abs/2025A&A...695A.128G} {695, A128}

\bibitem[\protect\citeauthoryear{{Gilli}, {Comastri}  \& {Hasinger}}{{Gilli} et~al.}{2007}]{Gilli07}
{Gilli} R.,  {Comastri} A.,   {Hasinger} G.,  2007, \mn@doi [\aap] {10.1051/0004-6361:20066334}, \href {http://adsabs.harvard.edu/abs/2007A%26A...463...79G} {463, 79}

\bibitem[\protect\citeauthoryear{{Gliozzi}, {Satyapal}, {Eracleous}, {Titarchuk}  \& {Cheung}}{{Gliozzi} et~al.}{2009}]{Gliozzi09}
{Gliozzi} M.,  {Satyapal} S.,  {Eracleous} M.,  {Titarchuk} L.,   {Cheung} C.~C.,  2009, \mn@doi [\apj] {10.1088/0004-637X/700/2/1759}, \href {http://adsabs.harvard.edu/abs/2009ApJ...700.1759G} {700, 1759}

\bibitem[\protect\citeauthoryear{{Gonz{\'a}lez-Alfonso} \& {Sakamoto}}{{Gonz{\'a}lez-Alfonso} \& {Sakamoto}}{2019}]{Gonzalez-Alfonso19}
{Gonz{\'a}lez-Alfonso} E.,  {Sakamoto} K.,  2019, \mn@doi [\apj] {10.3847/1538-4357/ab3a32}, \href {https://ui.adsabs.harvard.edu/abs/2019ApJ...882..153G} {882, 153}

\bibitem[\protect\citeauthoryear{{Gonz{\'a}lez-Mart{\'{\i}}n}, {Masegosa}, {M{\'a}rquez}, {Guainazzi}  \& {Jim{\'e}nez-Bail{\'o}n}}{{Gonz{\'a}lez-Mart{\'{\i}}n} et~al.}{2009}]{GonzalezMartin09}
{Gonz{\'a}lez-Mart{\'{\i}}n} O.,  {Masegosa} J.,  {M{\'a}rquez} I.,  {Guainazzi} M.,   {Jim{\'e}nez-Bail{\'o}n} E.,  2009, \mn@doi [\aap] {10.1051/0004-6361/200912288}, \href {http://adsabs.harvard.edu/abs/2009A%26A...506.1107G} {506, 1107}

\bibitem[\protect\citeauthoryear{{Gonz{\'a}lez-Mart{\'{\i}}n} et~al.,}{{Gonz{\'a}lez-Mart{\'{\i}}n} et~al.}{2017}]{Gonzalez-Martin17}
{Gonz{\'a}lez-Mart{\'{\i}}n} O.,  et~al., 2017, \mn@doi [\apj] {10.3847/1538-4357/aa6f16}, \href {http://adsabs.harvard.edu/abs/2017ApJ...841...37G} {841, 37}

\bibitem[\protect\citeauthoryear{{Goulding} \& {Alexander}}{{Goulding} \& {Alexander}}{2009}]{Goulding09}
{Goulding} A.~D.,  {Alexander} D.~M.,  2009, \mn@doi [\mnras] {10.1111/j.1365-2966.2009.15194.x}, \href {http://adsabs.harvard.edu/abs/2009MNRAS.398.1165G} {398, 1165}

\bibitem[\protect\citeauthoryear{{Goulding}, {Alexander}, {Lehmer}  \& {Mullaney}}{{Goulding} et~al.}{2010}]{Goulding10}
{Goulding} A.~D.,  {Alexander} D.~M.,  {Lehmer} B.~D.,   {Mullaney} J.~R.,  2010, \mn@doi [\mnras] {10.1111/j.1365-2966.2010.16700.x}, \href {http://adsabs.harvard.edu/abs/2010MNRAS.406..597G} {406, 597}

\bibitem[\protect\citeauthoryear{{Goulding}, {Alexander}, {Mullaney}, {Gelbord}, {Hickox}, {Ward}  \& {Watson}}{{Goulding} et~al.}{2011}]{Goulding11}
{Goulding} A.~D.,  {Alexander} D.~M.,  {Mullaney} J.~R.,  {Gelbord} J.~M.,  {Hickox} R.~C.,  {Ward} M.,   {Watson} M.~G.,  2011, \mn@doi [\mnras] {10.1111/j.1365-2966.2010.17755.x}, \href {http://adsabs.harvard.edu/abs/2011MNRAS.411.1231G} {411, 1231}

\bibitem[\protect\citeauthoryear{{Greenhill}, {Kondratko}, {Lovell}, {Kuiper}, {Moran}, {Jauncey}  \& {Baines}}{{Greenhill} et~al.}{2003}]{Greenhill03}
{Greenhill} L.~J.,  {Kondratko} P.~T.,  {Lovell} J.~E.~J.,  {Kuiper} T.~B.~H.,  {Moran} J.~M.,  {Jauncey} D.~L.,   {Baines} G.~P.,  2003, \mn@doi [\apjl] {10.1086/367602}, \href {http://adsabs.harvard.edu/abs/2003ApJ...582L..11G} {582, L11}

\bibitem[\protect\citeauthoryear{{Harrison} et~al.,}{{Harrison} et~al.}{2013}]{Harrison13}
{Harrison} F.~A.,  et~al., 2013, \mn@doi [\apj] {10.1088/0004-637X/770/2/103}, \href {http://adsabs.harvard.edu/abs/2013ApJ...770..103H} {770, 103}

\bibitem[\protect\citeauthoryear{{Helou} \& {Walker}}{{Helou} \& {Walker}}{1988}]{Helou88}
{Helou} G.,  {Walker} D.~W.,  eds, 1988, {Infrared astronomical satellite (IRAS) catalogs and atlases. Volume 7: The small scale structure catalog}  NASA RP-1190 Vol. 7

\bibitem[\protect\citeauthoryear{{Hickox} \& {Alexander}}{{Hickox} \& {Alexander}}{2018}]{HickoxAlexander18}
{Hickox} R.~C.,  {Alexander} D.~M.,  2018, \mn@doi [\araa] {10.1146/annurev-astro-081817-051803}, \href {https://ui.adsabs.harvard.edu/abs/2018ARA&A..56..625H} {56, 625}

\bibitem[\protect\citeauthoryear{{Ho}}{{Ho}}{2008}]{Ho08}
{Ho} L.~C.,  2008, \mn@doi [\araa] {10.1146/annurev.astro.45.051806.110546}, \href {http://adsabs.harvard.edu/abs/2008ARA%26A..46..475H} {46, 475}

\bibitem[\protect\citeauthoryear{{Ho}, {Filippenko}  \& {Sargent}}{{Ho} et~al.}{1997}]{Ho97}
{Ho} L.~C.,  {Filippenko} A.~V.,   {Sargent} W.~L.~W.,  1997, \mn@doi [\apjs] {10.1086/313041}, \href {http://adsabs.harvard.edu/abs/1997ApJS..112..315H} {112, 315}

\bibitem[\protect\citeauthoryear{{H{\"o}nig}}{{H{\"o}nig}}{2019}]{Honig2019}
{H{\"o}nig} S.~F.,  2019, \mn@doi [\apj] {10.3847/1538-4357/ab4591}, \href {https://ui.adsabs.harvard.edu/abs/2019ApJ...884..171H} {884, 171}

\bibitem[\protect\citeauthoryear{{H{\"o}nig} \& {Beckert}}{{H{\"o}nig} \& {Beckert}}{2007}]{Hoenig07}
{H{\"o}nig} S.~F.,  {Beckert} T.,  2007, \mn@doi [\mnras] {10.1111/j.1365-2966.2007.12157.x}, \href {http://adsabs.harvard.edu/abs/2007MNRAS.380.1172H} {380, 1172}

\bibitem[\protect\citeauthoryear{{Hopkins}, {Hernquist}, {Cox}  \& {Kere{\v s}}}{{Hopkins} et~al.}{2008}]{Hopkins08}
{Hopkins} P.~F.,  {Hernquist} L.,  {Cox} T.~J.,   {Kere{\v s}} D.,  2008, \mn@doi [\apjs] {10.1086/524362}, \href {http://adsabs.harvard.edu/abs/2008ApJS..175..356H} {175, 356}

\bibitem[\protect\citeauthoryear{{Horst}, {Gandhi}, {Smette}  \& {Duschl}}{{Horst} et~al.}{2008}]{Horst08}
{Horst} H.,  {Gandhi} P.,  {Smette} A.,   {Duschl} W.~J.,  2008, \mn@doi [\aap] {10.1051/0004-6361:20078548}, \href {http://adsabs.harvard.edu/abs/2008A%26A...479..389H} {479, 389}

\bibitem[\protect\citeauthoryear{{Houck} et~al.,}{{Houck} et~al.}{2004}]{Houck04}
{Houck} J.~R.,  et~al., 2004, \mn@doi [\apjs] {10.1086/423134}, \href {http://adsabs.harvard.edu/abs/2004ApJS..154...18H} {154, 18}

\bibitem[\protect\citeauthoryear{{Hummel} \& {Jorsater}}{{Hummel} \& {Jorsater}}{1992}]{Hummel92}
{Hummel} E.,  {Jorsater} S.,  1992, \aap, \href {https://ui.adsabs.harvard.edu/abs/1992A&A...261...85H} {261, 85}

\bibitem[\protect\citeauthoryear{{Hummel}, {Jorsater}, {Lindblad}  \& {Sandqvist}}{{Hummel} et~al.}{1987}]{Hummel87}
{Hummel} E.,  {Jorsater} S.,  {Lindblad} P.~O.,   {Sandqvist} A.,  1987, \aap, \href {https://ui.adsabs.harvard.edu/abs/1987A&A...172...51H} {172, 51}

\bibitem[\protect\citeauthoryear{{Ishibashi} \& {Fabian}}{{Ishibashi} \& {Fabian}}{2016}]{Ishibashi2016}
{Ishibashi} W.,  {Fabian} A.~C.,  2016, \mn@doi [\mnras] {10.1093/mnras/stw2063}, \href {https://ui.adsabs.harvard.edu/abs/2016MNRAS.463.1291I} {463, 1291}

\bibitem[\protect\citeauthoryear{{Iwasawa}, {Fabian}  \& {Matt}}{{Iwasawa} et~al.}{1997}]{Iwasawa97}
{Iwasawa} K.,  {Fabian} A.~C.,   {Matt} G.,  1997, \mnras, \href {http://adsabs.harvard.edu/abs/1997MNRAS.289..443I} {289, 443}

\bibitem[\protect\citeauthoryear{{Iwasawa} et~al.,}{{Iwasawa} et~al.}{2011}]{Iwasawa2011}
{Iwasawa} K.,  et~al., 2011, \mn@doi [\aap] {10.1051/0004-6361/201015872}, \href {https://ui.adsabs.harvard.edu/abs/2011A&A...528A.137I} {528, A137}

\bibitem[\protect\citeauthoryear{{Jana}, {Chatterjee}, {Kumari}, {Nandi}, {Naik}  \& {Patra}}{{Jana} et~al.}{2020}]{Jana20}
{Jana} A.,  {Chatterjee} A.,  {Kumari} N.,  {Nandi} P.,  {Naik} S.,   {Patra} D.,  2020, \mn@doi [\mnras] {10.1093/mnras/staa2552}, \href {https://ui.adsabs.harvard.edu/abs/2020MNRAS.499.5396J} {499, 5396}

\bibitem[\protect\citeauthoryear{{Jarrett}, {Chester}, {Cutri}, {Schneider}  \& {Huchra}}{{Jarrett} et~al.}{2003}]{Jarrett03}
{Jarrett} T.~H.,  {Chester} T.,  {Cutri} R.,  {Schneider} S.~E.,   {Huchra} J.~P.,  2003, \mn@doi [\aj] {10.1086/345794}, \href {http://adsabs.harvard.edu/abs/2003AJ....125..525J} {125, 525}

\bibitem[\protect\citeauthoryear{{Kalberla}, {Burton}, {Hartmann}, {Arnal}, {Bajaja}, {Morras}  \& {P{\"o}ppel}}{{Kalberla} et~al.}{2005}]{Kalberla05}
{Kalberla} P.~M.~W.,  {Burton} W.~B.,  {Hartmann} D.,  {Arnal} E.~M.,  {Bajaja} E.,  {Morras} R.,   {P{\"o}ppel} W.~G.~L.,  2005, \mn@doi [\aap] {10.1051/0004-6361:20041864}, \href {http://adsabs.harvard.edu/abs/2005A%26A...440..775K} {440, 775}

\bibitem[\protect\citeauthoryear{{Kammoun} et~al.,}{{Kammoun} et~al.}{2020}]{Kammoun2020}
{Kammoun} E.~S.,  et~al., 2020, \mn@doi [\apj] {10.3847/1538-4357/abb29f}, \href {https://ui.adsabs.harvard.edu/abs/2020ApJ...901..161K} {901, 161}

\bibitem[\protect\citeauthoryear{{Kauffmann} et~al.,}{{Kauffmann} et~al.}{2003}]{Kauffmann03}
{Kauffmann} G.,  et~al., 2003, \mn@doi [\mnras] {10.1111/j.1365-2966.2003.07154.x}, \href {http://adsabs.harvard.edu/abs/2003MNRAS.346.1055K} {346, 1055}

\bibitem[\protect\citeauthoryear{{Kewley}, {Dopita}, {Sutherland}, {Heisler}  \& {Trevena}}{{Kewley} et~al.}{2001}]{Kewley01}
{Kewley} L.~J.,  {Dopita} M.~A.,  {Sutherland} R.~S.,  {Heisler} C.~A.,   {Trevena} J.,  2001, \mn@doi [\apj] {10.1086/321545}, \href {http://adsabs.harvard.edu/abs/2001ApJ...556..121K} {556, 121}

\bibitem[\protect\citeauthoryear{{Kim}, {Barth}, {Ho}  \& {Son}}{{Kim} et~al.}{2021}]{Kim2021}
{Kim} M.,  {Barth} A.~J.,  {Ho} L.~C.,   {Son} S.,  2021, \mn@doi [\apjs] {10.3847/1538-4365/ac133e}, \href {https://ui.adsabs.harvard.edu/abs/2021ApJS..256...40K} {256, 40}

\bibitem[\protect\citeauthoryear{{Kirkpatrick} et~al.,}{{Kirkpatrick} et~al.}{2013}]{Kirkpatrick13}
{Kirkpatrick} A.,  et~al., 2013, \mn@doi [\apj] {10.1088/0004-637X/763/2/123}, \href {http://adsabs.harvard.edu/abs/2013ApJ...763..123K} {763, 123}

\bibitem[\protect\citeauthoryear{{Kocevski} et~al.,}{{Kocevski} et~al.}{2015}]{Kocevski15}
{Kocevski} D.~D.,  et~al., 2015, \mn@doi [\apj] {10.1088/0004-637X/814/2/104}, \href {http://adsabs.harvard.edu/abs/2015ApJ...814..104K} {814, 104}

\bibitem[\protect\citeauthoryear{{Kondratko} et~al.,}{{Kondratko} et~al.}{2006}]{Kondratko06}
{Kondratko} P.~T.,  et~al., 2006, \mn@doi [\apj] {10.1086/498641}, \href {https://ui.adsabs.harvard.edu/abs/2006ApJ...638..100K} {638, 100}

\bibitem[\protect\citeauthoryear{{Koss}, {Mushotzky}, {Veilleux}, {Winter}, {Baumgartner}, {Tueller}, {Gehrels}  \& {Valencic}}{{Koss} et~al.}{2011}]{Koss11}
{Koss} M.,  {Mushotzky} R.,  {Veilleux} S.,  {Winter} L.~M.,  {Baumgartner} W.,  {Tueller} J.,  {Gehrels} N.,   {Valencic} L.,  2011, \mn@doi [\apj] {10.1088/0004-637X/739/2/57}, \href {http://adsabs.harvard.edu/abs/2011ApJ...739...57K} {739, 57}

\bibitem[\protect\citeauthoryear{{Koss} et~al.,}{{Koss} et~al.}{2017}]{Koss17}
{Koss} M.,  et~al., 2017, \mn@doi [\apj] {10.3847/1538-4357/aa8ec9}, \href {http://adsabs.harvard.edu/abs/2017ApJ...850...74K} {850, 74}

\bibitem[\protect\citeauthoryear{{Laloux} et~al.,}{{Laloux} et~al.}{2023}]{Laloux2023}
{Laloux} B.,  et~al., 2023, \mn@doi [\mnras] {10.1093/mnras/stac3255}, \href {https://ui.adsabs.harvard.edu/abs/2023MNRAS.518.2546L} {518, 2546}

\bibitem[\protect\citeauthoryear{{Lansbury} et~al.,}{{Lansbury} et~al.}{2014}]{Lansbury14}
{Lansbury} G.~B.,  et~al., 2014, \mn@doi [\apj] {10.1088/0004-637X/785/1/17}, \href {http://adsabs.harvard.edu/abs/2014ApJ...785...17L} {785, 17}

\bibitem[\protect\citeauthoryear{{Leitherer}, {Byler}, {Lee}  \& {Levesque}}{{Leitherer} et~al.}{2018}]{Leitherer18}
{Leitherer} C.,  {Byler} N.,  {Lee} J.~C.,   {Levesque} E.~M.,  2018, \mn@doi [\apj] {10.3847/1538-4357/aada84}, \href {https://ui.adsabs.harvard.edu/abs/2018ApJ...865...55L} {865, 55}

\bibitem[\protect\citeauthoryear{{Li} \& {Draine}}{{Li} \& {Draine}}{2001}]{Li01}
{Li} A.,  {Draine} B.~T.,  2001, \mn@doi [\apj] {10.1086/323147}, \href {http://adsabs.harvard.edu/abs/2001ApJ...554..778L} {554, 778}

\bibitem[\protect\citeauthoryear{{Luo} et~al.,}{{Luo} et~al.}{2013}]{Luo13}
{Luo} B.,  et~al., 2013, \mn@doi [\apj] {10.1088/0004-637X/772/2/153}, \href {http://adsabs.harvard.edu/abs/2013ApJ...772..153L} {772, 153}

\bibitem[\protect\citeauthoryear{{Madejski}, {{\.Z}ycki}, {Done}, {Valinia}, {Blanco}, {Rothschild}  \& {Turek}}{{Madejski} et~al.}{2000}]{Madejski2000}
{Madejski} G.,  {{\.Z}ycki} P.,  {Done} C.,  {Valinia} A.,  {Blanco} P.,  {Rothschild} R.,   {Turek} B.,  2000, \mn@doi [\apjl] {10.1086/312703}, \href {https://ui.adsabs.harvard.edu/abs/2000ApJ...535L..87M} {535, L87}

\bibitem[\protect\citeauthoryear{{Madsen} et~al.,}{{Madsen} et~al.}{2015}]{Madsen15}
{Madsen} K.~K.,  et~al., 2015, \mn@doi [\apjs] {10.1088/0067-0049/220/1/8}, \href {https://ui.adsabs.harvard.edu/abs/2015ApJS..220....8M} {220, 8}

\bibitem[\protect\citeauthoryear{{Magdziarz} \& {Zdziarski}}{{Magdziarz} \& {Zdziarski}}{1995}]{Magdziarz95}
{Magdziarz} P.,  {Zdziarski} A.~A.,  1995, \mn@doi [\mnras] {10.1093/mnras/273.3.837}, \href {https://ui.adsabs.harvard.edu/abs/1995MNRAS.273..837M} {273, 837}

\bibitem[\protect\citeauthoryear{{Maoz}, {Nagar}, {Falcke}  \& {Wilson}}{{Maoz} et~al.}{2005}]{Maoz05}
{Maoz} D.,  {Nagar} N.~M.,  {Falcke} H.,   {Wilson} A.~S.,  2005, \mn@doi [\apj] {10.1086/429795}, \href {http://adsabs.harvard.edu/abs/2005ApJ...625..699M} {625, 699}

\bibitem[\protect\citeauthoryear{{Marchesi} et~al.,}{{Marchesi} et~al.}{2019}]{Marchesi2019}
{Marchesi} S.,  et~al., 2019, \mn@doi [\apj] {10.3847/1538-4357/aafbeb}, \href {https://ui.adsabs.harvard.edu/abs/2019ApJ...872....8M} {872, 8}

\bibitem[\protect\citeauthoryear{{Marinucci}, {Risaliti}, {Wang}, {Nardini}, {Elvis}, {Fabbiano}, {Bianchi}  \& {Matt}}{{Marinucci} et~al.}{2012}]{Marinucci2012}
{Marinucci} A.,  {Risaliti} G.,  {Wang} J.,  {Nardini} E.,  {Elvis} M.,  {Fabbiano} G.,  {Bianchi} S.,   {Matt} G.,  2012, \mn@doi [\mnras] {10.1111/j.1745-3933.2012.01232.x}, \href {https://ui.adsabs.harvard.edu/abs/2012MNRAS.423L...6M} {423, L6}

\bibitem[\protect\citeauthoryear{{Mart{\'\i}n} et~al.,}{{Mart{\'\i}n} et~al.}{2016}]{Martin2016}
{Mart{\'\i}n} S.,  et~al., 2016, \mn@doi [\aap] {10.1051/0004-6361/201528064}, \href {https://ui.adsabs.harvard.edu/abs/2016A&A...590A..25M} {590, A25}

\bibitem[\protect\citeauthoryear{{Mart{\'\i}nez-Sansigre}, {Rawlings}, {Lacy}, {Fadda}, {Marleau}, {Simpson}, {Willott}  \& {Jarvis}}{{Mart{\'\i}nez-Sansigre} et~al.}{2005}]{Martinez-Sansigre2005}
{Mart{\'\i}nez-Sansigre} A.,  {Rawlings} S.,  {Lacy} M.,  {Fadda} D.,  {Marleau} F.~R.,  {Simpson} C.,  {Willott} C.~J.,   {Jarvis} M.~J.,  2005, \mn@doi [\nat] {10.1038/nature03829}, \href {https://ui.adsabs.harvard.edu/abs/2005Natur.436..666M} {436, 666}

\bibitem[\protect\citeauthoryear{{Mason} et~al.,}{{Mason} et~al.}{2012}]{Mason12}
{Mason} R.~E.,  et~al., 2012, \mn@doi [\aj] {10.1088/0004-6256/144/1/11}, \href {http://adsabs.harvard.edu/abs/2012AJ....144...11M} {144, 11}

\bibitem[\protect\citeauthoryear{{Matt} et~al.,}{{Matt} et~al.}{1997}]{Matt97}
{Matt} G.,  et~al., 1997, \aap, \href {http://adsabs.harvard.edu/abs/1997A%26A...325L..13M} {325, L13}

\bibitem[\protect\citeauthoryear{{Matt}, {Fabian}, {Guainazzi}, {Iwasawa}, {Bassani}  \& {Malaguti}}{{Matt} et~al.}{2000}]{Matt00}
{Matt} G.,  {Fabian} A.~C.,  {Guainazzi} M.,  {Iwasawa} K.,  {Bassani} L.,   {Malaguti} G.,  2000, \mn@doi [\mnras] {10.1046/j.1365-8711.2000.03721.x}, \href {http://adsabs.harvard.edu/abs/2000MNRAS.318..173M} {318, 173}

\bibitem[\protect\citeauthoryear{{McKernan}, {Ford}  \& {Reynolds}}{{McKernan} et~al.}{2010}]{McKernan10}
{McKernan} B.,  {Ford} K.~E.~S.,   {Reynolds} C.~S.,  2010, \mn@doi [\mnras] {10.1111/j.1365-2966.2010.17068.x}, \href {http://adsabs.harvard.edu/abs/2010MNRAS.407.2399M} {407, 2399}

\bibitem[\protect\citeauthoryear{{Miyamoto}, {Nakai}, {Seta}, {Salak}, {Nagai}  \& {Kaneko}}{{Miyamoto} et~al.}{2017}]{Miyamoto17}
{Miyamoto} Y.,  {Nakai} N.,  {Seta} M.,  {Salak} D.,  {Nagai} M.,   {Kaneko} H.,  2017, \mn@doi [\pasj] {10.1093/pasj/psx076}, \href {https://ui.adsabs.harvard.edu/abs/2017PASJ...69...83M} {69, 83}

\bibitem[\protect\citeauthoryear{{Moorwood} \& {Glass}}{{Moorwood} \& {Glass}}{1984}]{Moorwood84}
{Moorwood} A.~F.~M.,  {Glass} I.~S.,  1984, \aap, \href {http://adsabs.harvard.edu/abs/1984A%26A...135..281M} {135, 281}

\bibitem[\protect\citeauthoryear{{Mould} et~al.,}{{Mould} et~al.}{2000}]{Mould00}
{Mould} J.~R.,  et~al., 2000, \mn@doi [\apj] {10.1086/308304}, \href {http://adsabs.harvard.edu/abs/2000ApJ...529..786M} {529, 786}

\bibitem[\protect\citeauthoryear{{Murphy} et~al.,}{{Murphy} et~al.}{2011}]{Murphy2011}
{Murphy} E.~J.,  et~al., 2011, \mn@doi [\apj] {10.1088/0004-637X/737/2/67}, \href {https://ui.adsabs.harvard.edu/abs/2011ApJ...737...67M} {737, 67}

\bibitem[\protect\citeauthoryear{{Murphy} et~al.,}{{Murphy} et~al.}{2012}]{Murphy12}
{Murphy} E.~J.,  et~al., 2012, \mn@doi [\apj] {10.1088/0004-637X/761/2/97}, \href {http://adsabs.harvard.edu/abs/2012ApJ...761...97M} {761, 97}

\bibitem[\protect\citeauthoryear{{Nagar}, {Falcke}  \& {Wilson}}{{Nagar} et~al.}{2005}]{Nagar05}
{Nagar} N.~M.,  {Falcke} H.,   {Wilson} A.~S.,  2005, \mn@doi [\aap] {10.1051/0004-6361:20042277}, \href {https://ui.adsabs.harvard.edu/abs/2005A&A...435..521N} {435, 521}

\bibitem[\protect\citeauthoryear{{Nandra}, {O'Neill}, {George}  \& {Reeves}}{{Nandra} et~al.}{2007}]{Nandra2007}
{Nandra} K.,  {O'Neill} P.~M.,  {George} I.~M.,   {Reeves} J.~N.,  2007, \mn@doi [\mnras] {10.1111/j.1365-2966.2007.12331.x}, \href {https://ui.adsabs.harvard.edu/abs/2007MNRAS.382..194N} {382, 194}

\bibitem[\protect\citeauthoryear{{Negus}, {Comerford}, {S{\'a}nchez}, {Revalski}, {Riffel}, {Bundy}, {Nevin}  \& {Rembold}}{{Negus} et~al.}{2023}]{Negus2023}
{Negus} J.,  {Comerford} J.~M.,  {S{\'a}nchez} F.~M.,  {Revalski} M.,  {Riffel} R.~A.,  {Bundy} K.,  {Nevin} R.,   {Rembold} S.~B.,  2023, \mn@doi [\apj] {10.3847/1538-4357/acb772}, \href {https://ui.adsabs.harvard.edu/abs/2023ApJ...945..127N} {945, 127}

\bibitem[\protect\citeauthoryear{{Nemmen}, {Storchi-Bergmann}  \& {Eracleous}}{{Nemmen} et~al.}{2014}]{Nemmen14}
{Nemmen} R.~S.,  {Storchi-Bergmann} T.,   {Eracleous} M.,  2014, \mn@doi [\mnras] {10.1093/mnras/stt2388}, \href {http://adsabs.harvard.edu/abs/2014MNRAS.438.2804N} {438, 2804}

\bibitem[\protect\citeauthoryear{{Oh} et~al.,}{{Oh} et~al.}{2018}]{Oh18}
{Oh} K.,  et~al., 2018, \mn@doi [\apjs] {10.3847/1538-4365/aaa7fd}, \href {http://adsabs.harvard.edu/abs/2018ApJS..235....4O} {235, 4}

\bibitem[\protect\citeauthoryear{{Oh} et~al.,}{{Oh} et~al.}{2022}]{Oh22}
{Oh} K.,  et~al., 2022, \mn@doi [\apjs] {10.3847/1538-4365/ac5b68}, \href {https://ui.adsabs.harvard.edu/abs/2022ApJS..261....4O} {261, 4}

\bibitem[\protect\citeauthoryear{{Osorio-Clavijo}, {Gonz{\'a}lez-Mart{\'\i}n}, {S{\'a}nchez}, {Esparza-Arredondo}, {Masegosa}, {Victoria-Ceballos}, {Hern{\'a}ndez-Garc{\'\i}a}  \& {D{\'\i}az}}{{Osorio-Clavijo} et~al.}{2022}]{Osorio22}
{Osorio-Clavijo} N.,  {Gonz{\'a}lez-Mart{\'\i}n} O.,  {S{\'a}nchez} S.~F.,  {Esparza-Arredondo} D.,  {Masegosa} J.,  {Victoria-Ceballos} C.,  {Hern{\'a}ndez-Garc{\'\i}a} L.,   {D{\'\i}az} Y.,  2022, \mn@doi [\mnras] {10.1093/mnras/stab3752}, \href {https://ui.adsabs.harvard.edu/abs/2022MNRAS.510.5102O} {510, 5102}

\bibitem[\protect\citeauthoryear{{Panessa}, {Bassani}, {Cappi}, {Dadina}, {Barcons}, {Carrera}, {Ho}  \& {Iwasawa}}{{Panessa} et~al.}{2006}]{Panessa06}
{Panessa} F.,  {Bassani} L.,  {Cappi} M.,  {Dadina} M.,  {Barcons} X.,  {Carrera} F.~J.,  {Ho} L.~C.,   {Iwasawa} K.,  2006, \mn@doi [\aap] {10.1051/0004-6361:20064894}, \href {http://adsabs.harvard.edu/abs/2006A%26A...455..173P} {455, 173}

\bibitem[\protect\citeauthoryear{{Pereira-Santaella}, {Diamond-Stanic}, {Alonso-Herrero}  \& {Rieke}}{{Pereira-Santaella} et~al.}{2010}]{Pereira-Santaella10}
{Pereira-Santaella} M.,  {Diamond-Stanic} A.~M.,  {Alonso-Herrero} A.,   {Rieke} G.~H.,  2010, \mn@doi [\apj] {10.1088/0004-637X/725/2/2270}, \href {http://adsabs.harvard.edu/abs/2010ApJ...725.2270P} {725, 2270}

\bibitem[\protect\citeauthoryear{{Puccetti} et~al.,}{{Puccetti} et~al.}{2014}]{Puccetti14}
{Puccetti} S.,  et~al., 2014, \mn@doi [\apj] {10.1088/0004-637X/793/1/26}, \href {http://adsabs.harvard.edu/abs/2014ApJ...793...26P} {793, 26}

\bibitem[\protect\citeauthoryear{{Reines} \& {Volonteri}}{{Reines} \& {Volonteri}}{2015}]{Reines2015}
{Reines} A.~E.,  {Volonteri} M.,  2015, \mn@doi [\apj] {10.1088/0004-637X/813/2/82}, \href {https://ui.adsabs.harvard.edu/abs/2015ApJ...813...82R} {813, 82}

\bibitem[\protect\citeauthoryear{{Ricci}, {Ueda}, {Koss}, {Trakhtenbrot}, {Bauer}  \& {Gandhi}}{{Ricci} et~al.}{2015}]{Ricci15}
{Ricci} C.,  {Ueda} Y.,  {Koss} M.~J.,  {Trakhtenbrot} B.,  {Bauer} F.~E.,   {Gandhi} P.,  2015, \mn@doi [\apjl] {10.1088/2041-8205/815/1/L13}, \href {http://adsabs.harvard.edu/abs/2015ApJ...815L..13R} {815, L13}

\bibitem[\protect\citeauthoryear{{Ricci} et~al.,}{{Ricci} et~al.}{2017a}]{Ricci17}
{Ricci} C.,  et~al., 2017a, \mn@doi [\apjs] {10.3847/1538-4365/aa96ad}, \href {http://adsabs.harvard.edu/abs/2017ApJS..233...17R} {233, 17}

\bibitem[\protect\citeauthoryear{{Ricci} et~al.,}{{Ricci} et~al.}{2017b}]{Ricci2017}
{Ricci} C.,  et~al., 2017b, \mn@doi [\mnras] {10.1093/mnras/stx173}, \href {https://ui.adsabs.harvard.edu/abs/2017MNRAS.468.1273R} {468, 1273}

\bibitem[\protect\citeauthoryear{{Rieke} \& {Lebofsky}}{{Rieke} \& {Lebofsky}}{1985}]{Rieke85}
{Rieke} G.~H.,  {Lebofsky} M.~J.,  1985, \mn@doi [\apj] {10.1086/162827}, \href {https://ui.adsabs.harvard.edu/abs/1985ApJ...288..618R} {288, 618}

\bibitem[\protect\citeauthoryear{{Rigby}, {Diamond-Stanic}  \& {Aniano}}{{Rigby} et~al.}{2009}]{Rigby09}
{Rigby} J.~R.,  {Diamond-Stanic} A.~M.,   {Aniano} G.,  2009, \mn@doi [\apj] {10.1088/0004-637X/700/2/1878}, \href {http://adsabs.harvard.edu/abs/2009ApJ...700.1878R} {700, 1878}

\bibitem[\protect\citeauthoryear{{Risaliti}, {Maiolino}  \& {Salvati}}{{Risaliti} et~al.}{1999}]{Risaliti99}
{Risaliti} G.,  {Maiolino} R.,   {Salvati} M.,  1999, \mn@doi [\apj] {10.1086/307623}, \href {http://adsabs.harvard.edu/abs/1999ApJ...522..157R} {522, 157}

\bibitem[\protect\citeauthoryear{{Roche}, {Aitken}, {Smith}  \& {James}}{{Roche} et~al.}{1986}]{Roche86}
{Roche} P.~F.,  {Aitken} D.~K.,  {Smith} C.~H.,   {James} S.~D.,  1986, \mn@doi [\mnras] {10.1093/mnras/218.1.19P}, \href {https://ui.adsabs.harvard.edu/abs/1986MNRAS.218P..19R} {218, 19P}

\bibitem[\protect\citeauthoryear{{Rovilos} et~al.,}{{Rovilos} et~al.}{2014}]{Rovilos14}
{Rovilos} E.,  et~al., 2014, \mn@doi [\mnras] {10.1093/mnras/stt2228}, \href {http://adsabs.harvard.edu/abs/2014MNRAS.438..494R} {438, 494}

\bibitem[\protect\citeauthoryear{{Saade}, {Brightman}, {Stern}, {Malkan}  \& {Garc{\'\i}a}}{{Saade} et~al.}{2022}]{Saade22}
{Saade} M.~L.,  {Brightman} M.,  {Stern} D.,  {Malkan} M.~A.,   {Garc{\'\i}a} J.~A.,  2022, \mn@doi [\apj] {10.3847/1538-4357/ac88cf}, \href {https://ui.adsabs.harvard.edu/abs/2022ApJ...936..162S} {936, 162}

\bibitem[\protect\citeauthoryear{{Sakamoto}, {Aalto}, {Costagliola}, {Mart{\'\i}n}, {Ohyama}, {Wiedner}  \& {Wilner}}{{Sakamoto} et~al.}{2013}]{Sakamoto2013}
{Sakamoto} K.,  {Aalto} S.,  {Costagliola} F.,  {Mart{\'\i}n} S.,  {Ohyama} Y.,  {Wiedner} M.~C.,   {Wilner} D.~J.,  2013, \mn@doi [\apj] {10.1088/0004-637X/764/1/42}, \href {https://ui.adsabs.harvard.edu/abs/2013ApJ...764...42S} {764, 42}

\bibitem[\protect\citeauthoryear{{Sakamoto}, {Gonz{\'a}lez-Alfonso}, {Mart{\'\i}n}, {Wilner}, {Aalto}, {Evans}  \& {Harada}}{{Sakamoto} et~al.}{2021}]{Sakamoto21}
{Sakamoto} K.,  {Gonz{\'a}lez-Alfonso} E.,  {Mart{\'\i}n} S.,  {Wilner} D.~J.,  {Aalto} S.,  {Evans} A.~S.,   {Harada} N.,  2021, \mn@doi [\apj] {10.3847/1538-4357/ac2746}, \href {https://ui.adsabs.harvard.edu/abs/2021ApJ...923..206S} {923, 206}

\bibitem[\protect\citeauthoryear{{Sanders} \& {Mirabel}}{{Sanders} \& {Mirabel}}{1996}]{Sanders96}
{Sanders} D.~B.,  {Mirabel} I.~F.,  1996, \mn@doi [\araa] {10.1146/annurev.astro.34.1.749}, \href {http://adsabs.harvard.edu/abs/1996ARA%26A..34..749S} {34, 749}

\bibitem[\protect\citeauthoryear{{Sanders}, {Mazzarella}, {Kim}, {Surace}  \& {Soifer}}{{Sanders} et~al.}{2003}]{Sanders03}
{Sanders} D.~B.,  {Mazzarella} J.~M.,  {Kim} D.-C.,  {Surace} J.~A.,   {Soifer} B.~T.,  2003, \mn@doi [\aj] {10.1086/376841}, \href {http://adsabs.harvard.edu/abs/2003AJ....126.1607S} {126, 1607}

\bibitem[\protect\citeauthoryear{{Satyapal}, {Vega}, {Heckman}, {O'Halloran}  \& {Dudik}}{{Satyapal} et~al.}{2007}]{Satyapal07}
{Satyapal} S.,  {Vega} D.,  {Heckman} T.,  {O'Halloran} B.,   {Dudik} R.,  2007, \mn@doi [\apjl] {10.1086/519995}, \href {http://adsabs.harvard.edu/abs/2007ApJ...663L...9S} {663, L9}

\bibitem[\protect\citeauthoryear{{Schaerer} \& {Stasi{\'n}ska}}{{Schaerer} \& {Stasi{\'n}ska}}{1999}]{Schaerer99}
{Schaerer} D.,  {Stasi{\'n}ska} G.,  1999, \aap, \href {http://adsabs.harvard.edu/abs/1999A%26A...345L..17S} {345, L17}

\bibitem[\protect\citeauthoryear{{Setti} \& {Woltjer}}{{Setti} \& {Woltjer}}{1989}]{Setti89}
{Setti} G.,  {Woltjer} L.,  1989, \aap, \href {http://adsabs.harvard.edu/abs/1989A%26A...224L..21S} {224, L21}

\bibitem[\protect\citeauthoryear{{Shi}, {Rieke}, {Ogle}, {Su}  \& {Balog}}{{Shi} et~al.}{2014}]{Shi2014}
{Shi} Y.,  {Rieke} G.~H.,  {Ogle} P.~M.,  {Su} K.~Y.~L.,   {Balog} Z.,  2014, \mn@doi [\apjs] {10.1088/0067-0049/214/2/23}, \href {https://ui.adsabs.harvard.edu/abs/2014ApJS..214...23S} {214, 23}

\bibitem[\protect\citeauthoryear{{Shimizu}, {Mushotzky}, {Mel{\'e}ndez}, {Koss}, {Barger}  \& {Cowie}}{{Shimizu} et~al.}{2017}]{Shimizu17}
{Shimizu} T.~T.,  {Mushotzky} R.~F.,  {Mel{\'e}ndez} M.,  {Koss} M.~J.,  {Barger} A.~J.,   {Cowie} L.~L.,  2017, \mn@doi [\mnras] {10.1093/mnras/stw3268}, \href {http://adsabs.harvard.edu/abs/2017MNRAS.466.3161S} {466, 3161}

\bibitem[\protect\citeauthoryear{{Spoon}, {Keane}, {Tielens}, {Lutz}  \& {Moorwood}}{{Spoon} et~al.}{2001}]{Spoon01}
{Spoon} H.~W.~W.,  {Keane} J.~V.,  {Tielens} A.~G.~G.~M.,  {Lutz} D.,   {Moorwood} A.~F.~M.,  2001, \mn@doi [\aap] {10.1051/0004-6361:20000557}, \href {https://ui.adsabs.harvard.edu/abs/2001A&A...365L.353S} {365, L353}

\bibitem[\protect\citeauthoryear{{Spoon}, {Marshall}, {Houck}, {Elitzur}, {Hao}, {Armus}, {Brandl}  \& {Charmandaris}}{{Spoon} et~al.}{2007}]{Spoon07}
{Spoon} H.~W.~W.,  {Marshall} J.~A.,  {Houck} J.~R.,  {Elitzur} M.,  {Hao} L.,  {Armus} L.,  {Brandl} B.~R.,   {Charmandaris} V.,  2007, \mn@doi [\apjl] {10.1086/511268}, \href {https://ui.adsabs.harvard.edu/abs/2007ApJ...654L..49S} {654, L49}

\bibitem[\protect\citeauthoryear{{Spoon} et~al.,}{{Spoon} et~al.}{2022}]{Spoon22}
{Spoon} H.~W.~W.,  et~al., 2022, \mn@doi [\apjs] {10.3847/1538-4365/ac4989}, \href {https://ui.adsabs.harvard.edu/abs/2022ApJS..259...37S} {259, 37}

\bibitem[\protect\citeauthoryear{{Stern} et~al.,}{{Stern} et~al.}{2014}]{Stern14}
{Stern} D.,  et~al., 2014, \mn@doi [\apj] {10.1088/0004-637X/794/2/102}, \href {http://adsabs.harvard.edu/abs/2014ApJ...794..102S} {794, 102}

\bibitem[\protect\citeauthoryear{{Stierwalt} et~al.,}{{Stierwalt} et~al.}{2013}]{Stierwalt13}
{Stierwalt} S.,  et~al., 2013, \mn@doi [\apjs] {10.1088/0067-0049/206/1/1}, \href {https://ui.adsabs.harvard.edu/abs/2013ApJS..206....1S} {206, 1}

\bibitem[\protect\citeauthoryear{{Tombesi}, {Mel{\'e}ndez}, {Veilleux}, {Reeves}, {Gonz{\'a}lez-Alfonso}  \& {Reynolds}}{{Tombesi} et~al.}{2015}]{Tombesi2015}
{Tombesi} F.,  {Mel{\'e}ndez} M.,  {Veilleux} S.,  {Reeves} J.~N.,  {Gonz{\'a}lez-Alfonso} E.,   {Reynolds} C.~S.,  2015, \mn@doi [\nat] {10.1038/nature14261}, \href {https://ui.adsabs.harvard.edu/abs/2015Natur.519..436T} {519, 436}

\bibitem[\protect\citeauthoryear{{Torres-Alb{\`a}} et~al.,}{{Torres-Alb{\`a}} et~al.}{2021}]{Torres-Alba2021}
{Torres-Alb{\`a}} N.,  et~al., 2021, \mn@doi [\apj] {10.3847/1538-4357/ac1c73}, \href {https://ui.adsabs.harvard.edu/abs/2021ApJ...922..252T} {922, 252}

\bibitem[\protect\citeauthoryear{{Treister} \& {Urry}}{{Treister} \& {Urry}}{2012}]{Treister12}
{Treister} E.,  {Urry} C.~M.,  2012, \mn@doi [Advances in Astronomy] {10.1155/2012/516193}, \href {http://adsabs.harvard.edu/abs/2012AdAst2012E..16T} {2012, 516193}

\bibitem[\protect\citeauthoryear{{Treister}, {Urry}  \& {Virani}}{{Treister} et~al.}{2009}]{Treister09}
{Treister} E.,  {Urry} C.~M.,   {Virani} S.,  2009, \mn@doi [\apj] {10.1088/0004-637X/696/1/110}, \href {http://adsabs.harvard.edu/abs/2009ApJ...696..110T} {696, 110}

\bibitem[\protect\citeauthoryear{{Treister}, {Urry}, {Schawinski}, {Cardamone}  \& {Sanders}}{{Treister} et~al.}{2010}]{Treister10}
{Treister} E.,  {Urry} C.~M.,  {Schawinski} K.,  {Cardamone} C.~N.,   {Sanders} D.~B.,  2010, \mn@doi [\apjl] {10.1088/2041-8205/722/2/L238}, \href {http://adsabs.harvard.edu/abs/2010ApJ...722L.238T} {722, L238}

\bibitem[\protect\citeauthoryear{{Turner}, {Miller}, {Reeves}  \& {Braito}}{{Turner} et~al.}{2017}]{Turner17}
{Turner} T.~J.,  {Miller} L.,  {Reeves} J.~N.,   {Braito} V.,  2017, \mn@doi [\mnras] {10.1093/mnras/stx388}, \href {http://adsabs.harvard.edu/abs/2017MNRAS.467.3924T} {467, 3924}

\bibitem[\protect\citeauthoryear{{Ueda}, {Akiyama}, {Hasinger}, {Miyaji}  \& {Watson}}{{Ueda} et~al.}{2014}]{Ueda14}
{Ueda} Y.,  {Akiyama} M.,  {Hasinger} G.,  {Miyaji} T.,   {Watson} M.~G.,  2014, \mn@doi [\apj] {10.1088/0004-637X/786/2/104}, \href {http://adsabs.harvard.edu/abs/2014ApJ...786..104U} {786, 104}

\bibitem[\protect\citeauthoryear{{Urry} \& {Padovani}}{{Urry} \& {Padovani}}{1995}]{Urry95}
{Urry} C.~M.,  {Padovani} P.,  1995, \mn@doi [\pasp] {10.1086/133630}, \href {http://adsabs.harvard.edu/abs/1995PASP..107..803U} {107, 803}

\bibitem[\protect\citeauthoryear{{Vasudevan}, {Fabian}, {Gandhi}, {Winter}  \& {Mushotzky}}{{Vasudevan} et~al.}{2010}]{Vasudevan10}
{Vasudevan} R.~V.,  {Fabian} A.~C.,  {Gandhi} P.,  {Winter} L.~M.,   {Mushotzky} R.~F.,  2010, \mn@doi [\mnras] {10.1111/j.1365-2966.2009.15936.x}, \href {http://adsabs.harvard.edu/abs/2010MNRAS.402.1081V} {402, 1081}

\bibitem[\protect\citeauthoryear{{Veron-Cetty} \& {Veron}}{{Veron-Cetty} \& {Veron}}{1986}]{Veron86}
{Veron-Cetty} M.-P.,  {Veron} P.,  1986, \aaps, \href {http://adsabs.harvard.edu/abs/1986A%26AS...66..335V} {66, 335}

\bibitem[\protect\citeauthoryear{{V{\'e}ron-Cetty} \& {V{\'e}ron}}{{V{\'e}ron-Cetty} \& {V{\'e}ron}}{2006}]{Veron06}
{V{\'e}ron-Cetty} M.~P.,  {V{\'e}ron} P.,  2006, \mn@doi [\aap] {10.1051/0004-6361:20065177}, \href {https://ui.adsabs.harvard.edu/abs/2006A&A...455..773V} {455, 773}

\bibitem[\protect\citeauthoryear{{Weaver} et~al.,}{{Weaver} et~al.}{2010}]{Weaver10}
{Weaver} K.~A.,  et~al., 2010, \mn@doi [\apj] {10.1088/0004-637X/716/2/1151}, \href {http://adsabs.harvard.edu/abs/2010ApJ...716.1151W} {716, 1151}

\bibitem[\protect\citeauthoryear{{Weedman} et~al.,}{{Weedman} et~al.}{2005}]{Weedman2005}
{Weedman} D.~W.,  et~al., 2005, \mn@doi [\apj] {10.1086/466520}, \href {https://ui.adsabs.harvard.edu/abs/2005ApJ...633..706W} {633, 706}

\bibitem[\protect\citeauthoryear{{Werner} et~al.,}{{Werner} et~al.}{2004}]{Werner04}
{Werner} M.~W.,  et~al., 2004, \mn@doi [\apjs] {10.1086/422992}, \href {http://adsabs.harvard.edu/abs/2004ApJS..154....1W} {154, 1}

\bibitem[\protect\citeauthoryear{{Westmeier}, {Braun}  \& {Koribalski}}{{Westmeier} et~al.}{2011}]{Westmeier11}
{Westmeier} T.,  {Braun} R.,   {Koribalski} B.~S.,  2011, \mn@doi [\mnras] {10.1111/j.1365-2966.2010.17596.x}, \href {http://adsabs.harvard.edu/abs/2011MNRAS.410.2217W} {410, 2217}

\bibitem[\protect\citeauthoryear{{Wethers} et~al.,}{{Wethers} et~al.}{2024}]{Wethers2024}
{Wethers} C.~F.,  et~al., 2024, \mn@doi [\aap] {10.1051/0004-6361/202347207}, \href {https://ui.adsabs.harvard.edu/abs/2024A&A...683A..27W} {683, A27}

\bibitem[\protect\citeauthoryear{{Wright} et~al.,}{{Wright} et~al.}{2010}]{Wright10}
{Wright} E.~L.,  et~al., 2010, \mn@doi [\aj] {10.1088/0004-6256/140/6/1868}, \href {http://adsabs.harvard.edu/abs/2010AJ....140.1868W} {140, 1868}

\bibitem[\protect\citeauthoryear{{Wu}, {Xue}, {Xia}, {Deng}  \& {Mao}}{{Wu} et~al.}{2002}]{Wu02}
{Wu} H.,  {Xue} S.~J.,  {Xia} X.~Y.,  {Deng} Z.~G.,   {Mao} S.,  2002, \mn@doi [\apj] {10.1086/341862}, \href {https://ui.adsabs.harvard.edu/abs/2002ApJ...576..738W} {576, 738}

\bibitem[\protect\citeauthoryear{{Xu}, {Liu}, {Gou}  \& {Liu}}{{Xu} et~al.}{2016}]{Xu16}
{Xu} W.,  {Liu} Z.,  {Gou} L.,   {Liu} J.,  2016, \mn@doi [\mnras] {10.1093/mnrasl/slv148}, \href {http://adsabs.harvard.edu/abs/2016MNRAS.455L..26X} {455, L26}

\bibitem[\protect\citeauthoryear{{Yaqoob}}{{Yaqoob}}{2012}]{Yaqoob2012}
{Yaqoob} T.,  2012, \mn@doi [\mnras] {10.1111/j.1365-2966.2012.21129.x}, \href {https://ui.adsabs.harvard.edu/abs/2012MNRAS.423.3360Y} {423, 3360}

\bibitem[\protect\citeauthoryear{{da Silva}, {Menezes}  \& {Steiner}}{{da Silva} et~al.}{2020a}]{daSilva20a}
{da Silva} P.,  {Menezes} R.~B.,   {Steiner} J.~E.,  2020a, \mn@doi [\mnras] {10.1093/mnras/staa007}, \href {https://ui.adsabs.harvard.edu/abs/2020MNRAS.492.5121D} {492, 5121}

\bibitem[\protect\citeauthoryear{{da Silva}, {Menezes}, {Steiner}  \& {Fraga}}{{da Silva} et~al.}{2020b}]{daSilva20b}
{da Silva} P.,  {Menezes} R.~B.,  {Steiner} J.~E.,   {Fraga} L.,  2020b, \mn@doi [\mnras] {10.1093/mnras/staa1500}, \href {https://ui.adsabs.harvard.edu/abs/2020MNRAS.496..943D} {496, 943}

\makeatother
\end{thebibliography}

\bsp
\label{lastpage}
\end{document}